\documentclass[sigconf, screen, natbib=false]{acmart}
\AtBeginDocument{%
  }
\usepackage[utf8]{inputenc}
\usepackage{xcolor}
\usepackage{amsmath}
\usepackage{amsthm}
\usepackage{amsfonts}
\usepackage{enumitem}
\usepackage{makecell}
\usepackage[vlined,boxed,linesnumbered]{algorithm2e}
\usepackage{subcaption}
\usepackage{hyperref}
\usepackage{cleveref,thm-restate}
\usepackage[table]{xcolor} 
\usepackage[normalem]{ulem}
\usepackage[datamodel=acmdatamodel,style=acmnumeric,backend=biber,sortcites]{biblatex}
\bibliography{refs}

\usepackage{tikz} 

\usetikzlibrary{arrows,fit, shapes,snakes,backgrounds,decorations}

\tikzset{
    client/.style={circle,draw=blue!70,fill=blue!20,minimum size=6mm},
    includedClient/.style={circle,draw=green!90,thick,fill=blue!40,minimum size=6mm},
    newClient/.style={circle,draw=purple!70,fill=purple!20,minimum size=6mm},
    droppedClient/.style={circle,draw=red!90,thick,fill=gray!40,minimum size=6mm},
    agg/.style={rectangle,draw=red!70,fill=red!20,minimum width=14mm,minimum height=8mm},
    msg/.style={->,thick},
    msgBold/.style={->,very thick},
    msgdashed/.style={->,thick,dashed,gray!60},
    msgb/.style={<-,thick},
    lightmsg/.style={->,thick,dashed,gray}
    operation/.style={rectangle,draw=blue!70,fill=blue!20,minimum size=6mm},
    operationFade/.style={rectangle,draw=blue!20,fill=blue!5,minimum size=6mm},
}

\newtheorem{definition}{Definition}

\newcommand{\rememberlines}[1]{\expandafter\xdef\csname #1\endcsname{\number\value{AlgoLine}}}
\newcommand{\resumenumbering}[1]{\setcounter{AlgoLine}{\csname #1\endcsname}}

\let\oldnl\nl
\newcommand{\nonl}{\renewcommand{\nl}{\let\nl\oldnl}}

\SetAlgoVlined
\SetArgSty{upshape}

\SetKwHangingKw{Init}{init:}{}{}
\SetKwProg{Operation}{Operation}{ is}{}
\SetKwProg{WhenReceived}{When}{ is received do}{}
\SetKwProg{WhenReceivedFromServer}{When}{ is received from the server do}{}
\SetKwProg{WhenReceivedFromClients}{When}{ is received from clients do}{}
\SetKwProg{WhenReceivedFromClient}{When}{ is received from client $c_j$ do}{}
\SetKwProg{WhenReceivedFromAggregator}{When}{ is received from aggregator $a_j$ do}{}
\SetKwProg{When}{When}{ do}{}
\SetKwProg{WhenReceivedFrom}{When}{ is \receivedd from $p_j$ do}{}
\SetKwProg{InternalOperation}{Internal operation}{ is}{}
\SetKwProg{ForAll}{For all}{ do}{}
\SetKwIF{If}{ElseIf}{Else}{If}{then}{Else if}{Else}{}
\SetKwProg{Function}{Function}{ is}{}
\SetKwProg{InternalOperation}{Internal operation}{ is}{}
\SetKwProg{WaitClient}{Wait for}{ messages from clients then}{}
\SetKwProg{WaitUntil}{Wait until}{ then}{}

\newcommand{\cOr}{\textup{\textbf{or}}\xspace}
\newcommand{\cAnd}{\textup{\textbf{and}}\xspace}

\SetKwComment{Comment}{$\rhd$}{}
\SetCommentSty{textit}

\newcommand{\trainmsg}{\texttt{TRAIN}\xspace}
\newcommand{\updatemsg}{\texttt{UPDATE}\xspace}
\newcommand{\aggregatemsg}{\texttt{SUM-SHARES}\xspace}
\newcommand{\disseminatemsg}{\texttt{INTRA-CLUSTER-RECONSTRUCTION}\xspace}
\newcommand{\certifymsg}{\texttt{CERTIFY}\xspace}

\newcommand{\wastedmsg}{\texttt{WASTED}\xspace}
\newcommand{\autoblamingmsg}{\texttt{AUTO-BLAMING}\xspace}
\newcommand{\gathermsg}{\texttt{INTER-CLUSTER-SUM}\xspace}
\newcommand{\finalizemsg}{\texttt{FINALIZE}\xspace}
\newcommand{\clientunificationmsg}{\texttt{UNIFICATION}\xspace}
\newcommand{\pingmsg}{\texttt{PING}\xspace}

\newcommand{\broadcast}{\ensuremath{\mathsf{Broadcast}}\xspace}
\newcommand{\send}{\ensuremath{\mathsf{Send}}\xspace}

\newcommand{\wastedcnt}{\ensuremath{\texttt{wastedcnt}}}
\newcommand{\aggcnt}{\ensuremath{\texttt{aggcnt}}}
\newcommand{\EncSecretShares}{\ensuremath{\texttt{encShares}}}
\newcommand{\ParticipatingClients}{\ensuremath{\texttt{partClients}}}
\newcommand{\sigSet}{\ensuremath{\texttt{sigSet}}}
\newcommand{\MaskedUpdates}{\texttt{maskedUpdates}}
\newcommand{\SigMaskedUpdates}{\texttt{sigMaskedUpdates}}

\newcommand{\FinalSelection}{\texttt{finalSelec}}

\newcommand{\maxDelta}{\ensuremath{\Delta_{\texttt{max}}}}
\newcommand{\StoredSumShares}{\texttt{StoredSumShares}}
\newcommand{\StoredSumProofs}{\texttt{StoredSumProofs}}

\newcommand{\pinglist}{\texttt{pingList}}

\newcommand{\assign}{\ensuremath{\mathsf{Assign}}}
\newcommand{\assigned}{\ensuremath{\mathsf{Assigned}}}

\newcommand{\wasteddetection}{\ensuremath{\mathsf{WastedDetection}}}
\newcommand{\blaming}{\ensuremath{\mathsf{InclusionFraudDetection}}}
\newcommand{\select}{\ensuremath{\mathsf{Include}}}
\newcommand{\CreateMaskedModel}{\ensuremath{\mathsf{CreateMaskedModel}}}
\newcommand{\PrepareAggregation}{\ensuremath{\mathsf{PrepareAggregation}}}

\newcommand{\return}{\ensuremath{\mathsf{Return}}\xspace}

\newcommand{\PKSig}{\ensuremath{\mathsf{PKSig}}}
\newcommand{\PKThreSig}{\ensuremath{\mathsf{PKThreSig}}}

\newcommand{\SKEnc}{\ensuremath{\mathsf{SKEnc}}}
\newcommand{\SKSig}{\ensuremath{\mathsf{SKSig}}}
\newcommand{\SKThreSig}{\ensuremath{\mathsf{SKThreSig}}}

\newcommand{\Sign}{\ensuremath{\mathsf{DigSig.Sign}}\xspace}
\newcommand{\Verify}{\ensuremath{\mathsf{DigSig.Verify}}\xspace}
\newcommand{\SignThre}{\ensuremath{\mathsf{ThreSig.Sign}}\xspace}
\newcommand{\VerifyThre}{\ensuremath{\mathsf{ThreSig.Verify}}\xspace}
\newcommand{\CombineThre}{\ensuremath{\mathsf{ThreSig.Combine}}\xspace}
\newcommand{\VerifyCombThre}{\ensuremath{\mathsf{ThreSig.VerifyCombined}}\xspace}

\newcommand{\Encrypt}{\ensuremath{\mathsf{PKCrypt.Encrypt}}}
\newcommand{\Decrypt}{\ensuremath{\mathsf{PKCrypt.Decrypt}}}

\newcommand{\SSShare}{\ensuremath{\mathsf{SS.Share}}\xspace}
\newcommand{\SSReconstruct}{\ensuremath{\mathsf{SS.Recover}}\xspace}

\newcommand{\VSSReconstruct}{\ensuremath{\mathsf{PVAHSS.Recover}}\xspace}
\newcommand{\VSSVerify}{\ensuremath{\mathsf{PVAHSS.Verify}}\xspace}
\newcommand{\VSSPartialVerify}{\ensuremath{\mathsf{PVAHSS.PartialVerify}}\xspace}

\newcommand{\VSSProof}{\ensuremath{\texttt{PVAHSSProof}}\xspace}
\newcommand{\VSSPartialProof}{\ensuremath{\texttt{PVAHSSPartialProof}}\xspace}

\newcommand{\commit}{{\ensuremath{\mathsf{Commit}}\xspace}}
\newcommand{\open}{{\ensuremath{\mathsf{Open}}\xspace}}

\newcommand{\ttrue}{\textbf{True}}
\newcommand{\ffalse}{\textbf{False}}

\usepackage{blindtext}

\usepackage{xpatch}

\makeatletter
\xpatchcmd{\ps@firstpagestyle}{Manuscript submitted to ACM}{}{\typeout{First patch succeeded}}{\typeout{first patch failed}}
\xpatchcmd{\ps@standardpagestyle}{Manuscript submitted to ACM}{}{\typeout{Second patch succeeded}}{\typeout{Second patch failed}}    \@ACM@manuscriptfalse
\makeatother

\renewcommand\footnotetextcopyrightpermission[1]{} 
\setcopyright{none}
\acmConference[]{}{}{}
\title{Privacy-Preserving Federated Averaging with Byzantine Aggregators in Asynchronous Networks}

\begin{document}
\author{Antonella Del Pozzo}
\affiliation{%
  \institution{Université Paris Saclay, CEA LIST}
\country{France}
}

\author{Achille Desreumaux}
\affiliation{%
  \institution{Université Paris Saclay, CEA LIST}
\country{France}
}
\author{Mathieu Gestin}
\affiliation{%
  \institution{Université Paris Saclay, CEA LIST}
\country{France}
}

\author{Alexandre Rapetti}
\affiliation{%
  \institution{Université Paris Saclay, CEA LIST}
\country{France}
}
\author{Sara Tucci}
\affiliation{%
  \institution{Université Paris Saclay, CEA LIST}
  \country{France}}

\begin{abstract}
Federated Learning requires secure aggregation to prevent gradient leakage, yet 
existing protocols suffer from key limitations: they assume synchrony, 
require heavy peer-to-peer coordination, and do not 
tolerate aggregators that halt or omit messages. These constraints make current 
secure aggregation schemes impractical in large-scale, unreliable distributed 
settings.

To overcome these limitations, we introduce a new secure aggregation protocol that 
operates in fully asynchronous networks — where messages may be arbitrarily delayed — and 
tolerates \emph{fully Byzantine} aggregators that are capable of arbitrary deviations 
including premature halting. Our design combines several key mechanisms: clustering 
clients under verifiable coordinators, lightweight LWE-based masking (with masking 
components distributed across aggregators), and differential privacy applied to both 
intermediary and final aggregated models. We further propose verifiable shuffling of 
clients across clusters, which prevents any client from being systematically excluded 
by a Byzantine coordinator, and a fair inclusion mechanism that ensures the inclusion 
of straggling clients whose messages are late. The protocol eliminates all client-to-client communication, 
and its communication overhead scales only with the number of aggregators—never with 
the number of clients. It also ensures equal representation of clients across rounds, 
avoiding bias and preventing unbalanced privacy risks among clients. 

Overall, our protocol provides the first secure aggregation primitive that is both 
privacy-preserving and robust to fully Byzantine behavior in asynchronous networks, 
closing the gap between prior secure aggregation assumptions and real-world 
distributed systems.
\end{abstract}



\keywords{Asynchronous communications, Federated averaging, Fairness} 
  
\maketitle

\section{Introduction}

Federated Learning (FL) is a paradigm where multiple clients collaboratively train a machine learning model while keeping their data local. A central aggregator collects, at each training round, the local model updates, called \textit{gradients}, to compute a global model that is then redistributed to clients for the next round. Training proceeds through successive rounds until convergence.

The main advantage of FL lies in preserving data privacy, since raw data never leaves the clients. However, recent studies have revealed that gradient reconstruction attacks \cite{YGFJ18,ZLH19,YMAAKM21}, where an adversary corrupts the central aggregator or intercepts gradients exchanged in the network, can reconstruct original data samples used during training.

To mitigate such privacy risks, several approaches have been proposed, including local differential privacy \cite{ACGMMTZ16}, homomorphic encryption \cite{DGLLNW16,EPI19}, and secure aggregation \cite{BIKMMPRSS17}. Each of these techniques, however, comes with important limitations.
Local differential privacy injects noise into gradients, but when applied at the granularity of individual updates, the amount of noise required to ensure meaningful privacy substantially degrades global model accuracy.
Homomorphic encryption enables computation over encrypted gradients, yet its computational and communication overhead remains prohibitive for large-scale models or resource-constrained clients.
Secure aggregation avoids these costs because masking is lightweight and ciphertexts retain the size of the input. Still, most secure aggregation protocols \cite{BIKMMPRSS17,bell2020secure,SSVRCN22} require extensive peer-to-peer coordination among clients, increasing per-client message complexity and creating mutual dependencies. Approaches that reduce this communication burden continue to rely on server-side computations that constitute a single point of failure \cite{flexible24}.
Indeed, to the best of our knowledge, existing secure aggregation protocols do not account for aggregators that may experience availability faults and halt the protocol at arbitrary points.
Network assumptions pose a further limitation. Many protocols assume synchrony—i.e., bounded message delays—or do not explicitly model the network. In contrast, realistic deployments must operate in asynchronous settings, where message delays may be arbitrary, possibly under adversarial influence. In such settings, slow or straggling clients may be systematically excluded from the aggregation, while fast clients may become disproportionately represented, potentially biasing the aggregated model (especially in the presence of non-independent and identically distributed (non-i.i.d.) data) or exposing fast clients to increased privacy risks.

In this paper, we present a new secure aggregation protocol that extends the state of the art to operate in asynchronous networks and tolerate fully Byzantine aggregators. Here, \textit{fully} means that aggregators may deviate in truly arbitrary ways: not only by attempting to illicitly reveal private information or corrupt the aggregated model (as in standard secure aggregation), but also by halting during protocol execution or selectively omitting messages.

Importantly, our goal is to achieve a synchronous federated learning scheme, in which training proceeds in rounds. Although a fully asynchronous federated learning scheme might appear more natural in an asynchronous network~\cite{Spyker24}, it is not attractive in our setting. Prior work has shown that asynchronous FL can reduce model utility and accuracy, especially in the presence of stale updates, straggler clients, non-i.i.d.\ data, and heterogeneous participants~\cite{zhou2022towards,mohammadi2025empirical}. Moreover, recent empirical evidence suggests that asynchronous FL may also exacerbate privacy and fairness disparities across clients~\cite{mohammadi2025empirical}. We therefore adopt a different approach: although the network is fully asynchronous, training is still organized into well-defined rounds. In other words, our approach aims to preserve the benefits of synchronous FL while operating over asynchronous communications.

This approach leverages a distributed architecture, where in each round clients are partitioned into clusters, and each cluster is verifiably assigned to a specific proxy aggregator, called a \emph{coordinator}. Each client prepares its model update by masking it using lightweight Learning With Errors (LWE) techniques~\cite{SSVRCN22} and generating shares of the mask, one for each aggregator. Each client then encrypts these shares under the aggregators’ keys and sends the encrypted shares to its coordinator, which redistributes them to the corresponding aggregators. The aggregators then collaboratively perform verifiable secure aggregation using the received shares: each coordinator securely computes the aggregated model for its cluster, assisted by the shares previously distributed to other aggregators. In addition, each client contributes a small differential privacy (DP) noise component that remains after aggregation and provides privacy guarantees for the aggregated model.

In our design, clients do not communicate with each other, and the communication overhead is proportional mainly to the number of aggregators—a quantity we expect to be significantly smaller than the number of clients.

A key aspect of our solution is the verifiable shuffling of clusters: at every round, clients are randomly reassigned to coordinators. This prevents any client from being systematically paired with a Byzantine coordinator that might drop its updates or omit returning the aggregated model. Moreover, because each cluster has a single coordinator that serializes the input shares for that cluster, there is no need to run a Byzantine Fault-Tolerant (BFT) consensus protocol~\cite{CL99,DRZ18}, which would otherwise be required when client shares are sent directly to multiple potentially Byzantine aggregators~\cite{ben1993asynchronous}. Avoiding BFT consensus is essential both to prevent high communication overhead and to ensure that our protocol operates correctly in an asynchronous network. Importantly, periodic shuffling is also key to convergence: by reassigning clients to different aggregators at each round, updates are gradually mixed across clusters, allowing the global model to align even though aggregators do not run consensus to produce a unique model at each round.

The second key aspect of our solution is the \emph{inclusion of straggling clients}. Our inclusion mechanism at the aggregator side increases the likelihood of choosing clients that have been silent in previous rounds, ensuring that every participant is equally represented in expectation. Crucially, both the shuffling and inclusion mechanisms are designed to be non-manipulable by Byzantine aggregators.

Overall, our solution enables, for the first time, privacy preserving and fair federated learning in asynchronous networks with potentially fully Byzantine aggregators, closing a long-standing gap between the assumptions made in secure aggregation and the realities of large-scale, unreliable distributed systems.

The paper is organized as follows. \Cref{sec:background}, presents building blocks. \Cref{sec:relatedwork}, studies state-of-the-art privacy preserving federated learning protocols. \Cref{sec:model} details the model and the formal problem statement of this paper. \Cref{sec:overview}, presents an overview of our protocol. \Cref{sec:description} details the working of the protocol. Finally, performances of our protocol are evaluated and compared to the state of the art in \Cref{sec:exp:analyzes}.

\section{Background} \label{sec:background}

\paragraph{Federated Learning with FedAvg}
We base our protocol on FedAvg~\cite{MMRHA17} as, in the context of heterogeneous data, it is the most efficient building block to create a privacy preserving Federated Learning (FL) mechanism  \cite{LNV2024,QWH2024}. FedAvg works by letting a set of $n_c$ clients collaboratively train a global model $w \in \mathbb{R}^d$ under the coordination of a central server, without sharing their raw data. Each client $c_i$ holds a private dataset of size $m_i$ and defines a local objective function $f_i:\mathbb{R}^d \rightarrow \mathbb{R}$. The global learning task is to minimize the aggregated objective:
\begin{equation*}
    \min_{w} f(w) 
    \;=\;
    \sum_{i=1}^{n_c} \frac{m_i}{M}\, f_i(w),
    \qquad
    M = \sum_{i=1}^{n_c} m_i.
\label{eq:fl-objective}
\end{equation*}

FL proceeds in communication rounds. At each round $\tau$, the server selects a subset $\mathcal{C}^{\tau}$ of $k \le n_c$ clients and sends them the current global model~$w^{\tau}$. Each selected client $c_i$ performs $E$ local epochs of stochastic gradient descent (SGD) on its private data. 
During SGD, the client processes its dataset in mini-batches of size $B$, where $B$ denotes the number of samples used to compute one stochastic gradient estimate.

After running SGD, client $i$ obtains an updated local model $w_i^\tau$ and sends the update $g^\tau_i = w^{\tau} - w_i^\tau$ to the server, which aggregates the received updates using the \emph{FedAvg} rule:
\begin{equation*}
    w^{\tau+1}
    =
    w^\tau - \sum_{c_i \in \mathcal{C}^\tau} \frac{m_i}{m_{\mathcal{C}^\tau}}\, g^\tau_i,
    \qquad
    m_{\mathcal{C}^\tau} = \sum_{c_i \in \mathcal{C}^\tau} m_i.
\label{eq:fedavg}
\end{equation*}

\paragraph{Differential Privacy}
Differential Privacy (DP) \cite{D06} makes it possible to publish statistics about a dataset while preserving individuals' privacy.
DP relies on the notion of adjacent datasets. Two datasets are said to be adjacent if their data differ by exactly one element.
The idea is to build a mechanism that transforms a dataset's statistics in a way that two adjacent datasets' statistics are indistinguishable, up to 
an $\epsilon$ factor, usually by injecting controlled noise. 
The original DP definition, called $\epsilon$-DP, was given for a parameter $\epsilon$, a mechanism $\mathsf{M}: \mathcal{R} \rightarrow \mathcal{S}$ and two adjacent datasets $R_1$ and $R_2$ as:
    \begin{equation*}
        Pr[\mathsf{M}(R_1) \in S] \le e^{\epsilon} Pr[\mathsf{M}(R_2) \in S],\ \ \ \forall S \in \mathcal{S}.
    \end{equation*}
In this framework, the parameter $\epsilon$ is interpreted as a \emph{privacy budget}: it quantifies the maximum privacy loss that the mechanism is allowed to incur. A smaller budget corresponds to stronger privacy guarantees, while a larger budget permits more information leakage. When a DP mechanism is applied repeatedly, as is the case in federated learning, where such mechanisms run across multiple training rounds, the privacy loss accumulates over time and must remain within the overall budget. 

Privacy preserving FL systems use a mechanism based on Gaussian noise to provide privacy. However, proving $\epsilon$-DP guarantees of this mechanism is deemed impractical.
For this reason, federated learning systems do not rely on plain $\epsilon$-DP for privacy accounting, but instead use  $(\alpha, \epsilon)$-Rényi DP \cite{M17} or $(\alpha, \epsilon)$-RDP.
\footnote{RDP measures privacy loss using the Rényi divergence, parameterized by an order 
$\alpha>1$ that controls the sensitivity of the divergence. Importantly, RDP guarantees 
bound privacy loss in expectation, while   $\epsilon$-DP mechanism provides a uniform guarantee.}
In the RDP framework, each invocation of a mechanism incurs a privacy cost $\epsilon(\alpha)$ at a given Rényi order $\alpha$. This cost plays the role of a \emph{privacy budget for that round}. Because RDP composes additively, the total privacy budget consumed after $T$ rounds is simply the sum of the individual costs. Moreover,  when privacy is enforced through \emph{Gaussian noise} with variance $\sigma^{2}$ , RDP cost admits a simple closed-form expression for $T$ rounds:$\epsilon(\alpha) = \alpha \cdot T /2\sigma^{2}.$

\paragraph{Homomorphic commitments}
A commitment scheme is a cryptographic scheme that makes it possible to create an element $C$ from a value $v$ that is binding to $v$ while hiding the value. It has an operation $\commit(v)$ which outputs the commitment $C$ to the value $v$.
The scheme also supports an operation $\open(C,v)$ that outputs $1$ if $C$ is a commitment to $v$.
We require that these commitments are additionally homomorphic, i.e., we have an operation $\oplus$ such that $\commit(v) \oplus \commit(v') = \commit(v+v')$. Such commitment schemes can be built using modular exponentiation in groups where solving the discrete logarithm is hard (as in \cite{S91}), or discrete logarithm-based polynomial commitments \cite{KZG10}.

\paragraph{Secret Sharing}
Secret Sharing (SS) is a mechanism used to share a value in a privacy preserving manner. A $(x,y)$-SS scheme has $2$ operations: \SSShare and \SSReconstruct. \SSShare takes an input $z$ and outputs $x$ ``shares''. An adversary that is given at most $y-1$ shares cannot learn anything about $z$. However, given any $y\leq x$ shares, $z$ can be reconstructed using the \SSReconstruct operation. %
We use additionally homomorphic SS schemes. They make it possible to sum shares and reconstruct the sum of those shares without revealing the individual unsummed values to any participant. We use the $\oplus$ operator to denote this homomorphic addition.

Furthermore, in \Cref{sec:certification}, we use a Publicly Verifiable, Additionally Homomorphic SS (PVAHSS) scheme. A PVAHSS scheme is an SS scheme with verifiability properties. We require \emph{any actor} to be able to use an algorithm \VSSVerify to verify that the reconstruction is executed honestly. The \VSSVerify algorithm takes as input a proof built during the creation of the shares and outputs $\ttrue$ if the reconstruction of the secret value was lawful. We further require that the proof is additionally homomorphic to enable verification of the reconstruction of the sum of shares. Finally, we require the ability to detect a misbehaving actor that sums wrong shares thanks to a \VSSPartialVerify function. Some prior works proposed some of these features \cite{TM20,CP22, HZ20,CGHJ25, P92}, but they either do not provide all features, or are designed for any type of computations rather than additions, and are thus too complex for our usage. To solve this problem, we propose to use
additionally homomorphic commitment schemes for both proofs.

\section{Related Work}\label{sec:relatedwork}

\paragraph{Preamble.}
In the FL literature, the term \emph{Byzantine} is often used with different scopes. 
For the sake of clarity, in our review of related work, we distinguish four notions. 
In secure aggregation, adversaries target confidentiality; we call them \emph{privacy-Byzantine} adversaries. 
They may deviate by colluding to recover individual updates or by manipulating masking shares. 
In verifiable secure aggregation, adversaries may also target integrity; we call them \emph{integrity-Byzantine} adversaries. 
For instance, they may deviate by submitting malformed vectors or computing incorrect aggregation shares. 
When adversaries target the model by submitting syntactically well-formed updates that nevertheless sabotage convergence, e.g., poisoned gradients or models, we call them \emph{model-Byzantine} adversaries. 
\emph{Fully Byzantine} adversaries, on the other hand, strictly subsume the above notions and also capture attacks against availability.  They may deviate arbitrarily from the protocol, including by halting, omitting or equivocating messages. 
When the context is clear, we simply write \emph{Byzantine}.

\paragraph{Secure Aggregation} 
Bonawitz et al.~\cite{BIKMMPRSS17} introduced secure aggregation 
for FL via pairwise random masks that cancel upon aggregation.
The protocol assumes a synchronous network and provides privacy against 
an honest-but-curious or privacy-Byzantine server, 
while considering clients to be potentially privacy-Byzantine or dropping out.  
To handle dropouts, the protocol employs a double-masking mechanism: 
each client also secret-shares the seed of an additional 
mask with all other clients, allowing the server to reconstruct 
the seed of any missing participant, which limits the number 
of tolerated dropouts. Subsequent works show that replacing the 
fully connected client graph with a logarithmic-degree 
k-regular graph significantly reduces communication costs \cite{bell2020secure,SBR22}.
These works focus on secure aggregation but do not apply 
differential privacy to safeguard the published aggregate.

FLDP \cite{SSVRCN22} addresses this limitation by introducing a differentially private
aggregation protocol based on Learning With Errors (LWE)\cite{R09} reducing the complexity of Bonawitz et al.~\cite{BIKMMPRSS17}. Each client masks its gradient using an LWE-based scheme, then sends it to the server. Clients further reconstruct the sum of their masks and transmit it to the server, which recovers the global sum by subtracting this aggregated mask from the sum of masked gradients.

All the previous solutions assume a single server. This implies that clients must collaborate to support mask removal and/or to tolerate dropouts, which means that there is no client-to-client independence. 
\begin{table*}[t]
\centering
\setlength{\tabcolsep}{5pt}
\renewcommand{\arraystretch}{1.3}
\resizebox{\textwidth}{!}{
\begin{tabular}{lccccc}
\toprule
\textbf{Property} & 
\textbf{Bonawitz}~\cite{BIKMMPRSS17} & 
\textbf{Bell}~\cite{bell2020secure} & 
\textbf{FLDP}~\cite{SSVRCN22} & 
\textbf{FlexScaAgg}~\cite{flexible24} & 
\textbf{This paper} \\
\midrule

\rowcolor{black!7}
\multicolumn{6}{l}{\textbf{Security \& Privacy}} \\
\midrule
Final and intermediary model privacy & $\times$ & $\times$ & $\checkmark$ & $\times$ & $\checkmark$ \\

\midrule
\rowcolor{black!7}
\multicolumn{6}{l}{\textbf{Robustness \& Liveness}} \\
\midrule
Model-Byzantine aggregator tolerance & $\times$ & $\times$ & $\times$ & $\checkmark$ & $\checkmark$ \\
Aggregator omission fault tolerance & $\times$ & $\times$ & $\times$ & $\times$ & $\checkmark$ \\

\midrule
\rowcolor{black!7}
\multicolumn{6}{l}{\textbf{Fairness \& Participation}} \\
\midrule
Straggler management (slow vs fast clients) & $\times$ & $\times$ & $\times$  & $\times$ & $\checkmark$ \\

\midrule
\rowcolor{black!7}
\multicolumn{6}{l}{\textbf{Architecture \& Efficiency}} \\
\midrule
Client-to-client independence & $\times$ & $\times$ & $\times$ & $\checkmark$ & $\checkmark$ \\
One-shot clients & $\times$ & $\times$ & $\times$ & $\times$ & $\checkmark$ \\
Communication assumptions & Sync & Unspecified & Unspecified & Unspecified & Async \\
\midrule
\rowcolor{black!7}
\multicolumn{6}{l}{\textbf{Communication \& computation complexity }} \\
\midrule
Client bits complexity 
    & $\mathcal{O}(N_g + n_{c})$ 
    & $\mathcal{O}(N_g + \log n_c)$ 
    & $\mathcal{O}(N_g + N_s + n_{c})$ 
    & $\mathcal{O}(N_{g}n_a)$ 
    & $\mathcal{O}(N_{g} + N_s n_{a})$ \\
Server bits complexity 
    & $\mathcal{O}(n_c(N_g+ n_c))$ 
    & $\mathcal{O}(n_c(N_g + \log n_c))$ 
    & $\mathcal{O}(N_g n_{c} + N_s)$ 
    & $\mathcal{O}(N_{g}(n_{c}+n_a))$ 
    & $\mathcal{O}(N_g (n_{c} + n_{a}) + n_{c}( N_{s} + n_a))$ \\
Client message complexity 
    & $\mathcal{O}(n_c)$ 
    & $\mathcal{O}(\log n_c)$ 
    & $\mathcal{O}(n_c)$ 
    & $\mathcal{O}(n_a)$ 
    & $\mathcal{O}(n_a)$ \\
Server message complexity 
    & $\mathcal{O}(n_c)$ 
    & $\mathcal{O}(n_c)$ 
    & $\mathcal{O}(n_c)$ 
    & $\mathcal{O}(n_a+n_c)$ 
    & $\mathcal{O}(n_a + n_c)$ \\
Client computational complexity 
    & $\mathcal{O}(n_c( N_g + n_c))$ 
    & $\mathcal{O}(\log n_c(N_g + \log n_c))$ 
    & $\mathcal{O}(n_c\log n_c + N_g N_s)$ 
    & $\mathcal{O}(N_g n_a)$ 
    & $\mathcal{O}(N_s n_a + N_g N_s)$ \\
Server computational complexity 
    & $\mathcal{O}(N_g n_c^2)$ 
    & $\mathcal{O}(n_c(\log^2 n_c + N_g \log n_c))$ 
    & $\mathcal{O}(N_g (n_c + N_s) + n_c\log n_c)$ 
    & $\mathcal{O}(N_g(n_a+n_c))$ 
    & $\mathcal{O}(N_g(n_c + n_a) + N_s (n_c n_a + N_g))$ \\
\bottomrule
\end{tabular}
}
\caption{Comparison of our solution with state-of-the-art secure aggregation protocols. 
We denote by $n_a$ the number of aggregators, by $n_c$ the number of clients, and by $N_g$ the size of an update. 
Note that $n_a$ is on the order of $\log(n_c)$.}
\label{tab:comparisonIntro}

\end{table*}
Prio \cite{prio17} (and its evolution Prio+ \cite{prio+22}), like our approach, addresses this problem by replicating the server. 
Unlike our protocol, all servers are assumed to be honest.
In a similar vein, Elsa \cite{elsa23} considers two servers, allowing one server and some clients to be Byzantine. In the case of a Byzantine server, the inputs of honest clients are guaranteed to remain private.

\paragraph{Verifiable Secure Aggregation} 
FlexScaAgg \cite{flexible24} proposes a solution that includes verifiable aggregation, enabling clients to check the integrity of the computation. For secure aggregation, the protocol uses two non-colluding entities—the server and the initiator—each generating a random seed for every client. Clients expand these seeds into two independent masks. Neither the server nor the initiator can recover individual updates on their own; only by collaborating can they compute the true sum. 
Unlike our solution, that tolerates only honest-but-curious clients, this mechanism can tolerate integrity-Byzantine clients, up to $n-2$.
Its main limitation, however, is availability: if either server or initiator fails, the protocol halts.
Besides \cite{flexible24}, few schemes address verifiable aggregation, allowing users to check the correctness of aggregation results, such as VerifyNet \cite{verifynet19} and VeriFL \cite{verifl20}. In those works, clients perform verification themselves; in contrast, our design delegates verification to the aggregators, leaving clients with the lightweight task of checking a quorum of aggregator signatures.

\paragraph{Comparison}
Table~\ref{tab:comparisonIntro} compares the works most closely related to ours along four axes.
(i)~Both FLDP~\cite{SSVRCN22}, and our work ensure that no intermediate aggregate is 
ever revealed without DP noise already embedded in it, a strictly stronger privacy 
property than SA alone~\cite{BIKMMPRSS17, bell2020secure, flexible24}.
(ii)~Our protocol tolerates both integrity-Byzantine aggregators --- as does 
FlexScaAgg~\cite{flexible24} --- and fully Byzantine ones, including omission faults, 
a liveness guarantee absent from all baselines.
(iii)~Clients neither communicate with each other (client-to-client independence) 
nor engage in multiple rounds (one-shot clients),  as mask removal is delegated 
entirely to aggregators. 
Furthermore, our protocol operates in an asynchronous communication model, which naturally 
accommodates straggler management --- preventing fast clients from disproportionate 
privacy exposure --- a concern unaddressed by all compared works.
(iv)~Under the practical assumptions $n_a = \mathcal{O}(\log n_c)$ and 
$N_s \log n_c = o(N_g)$, our protocol achieves state-of-the-art complexities 
for communication metrics: $\mathcal{O}(N_g)$ and $\mathcal{O}(N_g n_c)$ in client 
and server communication bits, $\mathcal{O}(\log n_c)$ and $\mathcal{O}(n_c)$ 
in message complexity. For computation, our client pays $\mathcal{O}(N_g N_s)$ 
and our server $\mathcal{O}(N_g n_c + N_g N_s)$, the inherent cost of the LWE approach, shared with FLDP.
A detailed analysis is provided in Appendix~\ref{app:complexity}.

\section{Model and problem statement}\label{sec:model}

\paragraph{Threat and network model.}
We consider a set of $n_c$ honest-but-curious clients, among which up to $t_c$ may \textit{crash}.
Crashed clients stop participating indefinitely but never behave maliciously.
The remaining $n_c - t_c$ clients are said to be \textit{correct}.
We also consider a set of $n_a$ \textit{aggregators} that collectively assume the role traditionally played by the central server in classical federated learning schemes.
Among these, up to $t_a$ may be \textit{fully Byzantine}.
The remaining $n_a - t_a$ aggregators are considered \textit{correct}, in the distributed-systems sense that they follow the protocol as specified.
However, from a security standpoint, they are modeled as \textit{honest-but-curious} (non-colluding semi-honest) entities: they do not deviate from the protocol but may attempt to infer information from the data they process.

The proposed solution operates in a \textit{reliable asynchronous network}, meaning that there is no known upper bound on message delays, yet messages sent by correct participants are eventually delivered and never lost.
However, clients may permanently drop out of the computation if they crash, and faulty aggregators too. 
To cope with this adversarial setting, we require that $n_a > 3t_a$. This threshold ensures protection from equivocation of Byzantine processes, thus enforcing the client's privacy (c.f. \cref{sec:inclusion}). We also require that $n_c >4t_c$. This threshold, higher than the theoretical crash based distributed computation theory, is required to ensure that clients can be included by aggregators in an unbiased manner (c.f. \cref{sec:inclusion}).

\paragraph{Data distribution} We consider that clients hold heterogeneous data.
Such heterogeneity impacts the protocol design, as asynchronous communication with no known bound on message delays may introduce bias in the resulting model: if “fast” clients contribute more frequently than “slow” ones, and their data are biased, the global model may also become biased.

\paragraph{Problem statement} We consider the following optimization problem: the clients and the aggregators must collaborate to find $\tilde{w}^*$, an estimation of the optimal value 
\begin{equation}
    w^* = \min_w \sum_{c_i \in \mathcal{C}} \frac{m_i}{M} F_{c_i}(w)= \min_w \sum_{c_i \in \mathcal{C}^*} \frac{m_i}{M} F_{c_i}(w), \qquad M = \sum_{i=1}^{n_c} m_i.
\end{equation} 

Where $\mathcal{C}^*$ is any subset of $n_c-2t_c$ clients in $\mathcal{C}$, and where $F_{c_i}$ is a smooth and strongly convex function $\forall i \in \{1, \cdots, n_c\}$. 

\emph{The goal for each correct aggregator} $a_i$ is to find its own estimation $\tilde{w}^*_i$ of $w^*$, where $||\tilde{w}^*_i-w^*|| \le \delta$, for a predefined $\delta$. However, $\tilde{w}^*_i$ can be different from $\tilde{w}^*_j$, for two different aggregators $a_i$ and $a_j$. 

Importantly, our objective is not only to solve the above optimization problem, but to solve it under the semantics of a \emph{synchronous federated learning scheme}. More precisely, our problem is to emulate a synchronous round-based FL scheme over asynchronous communications within the fault model described above.

\section{High-Level System and Protocol Description}\label{sec:overview}

This section provides a high-level overview of the \textit{privacy-preserving and fair federated averaging protocol with asynchronous communication and Byzantine aggregators.}

\begin{figure}[!ht]
\centering
\begin{tikzpicture}[>=stealth,scale=0.65, transform shape]

\node[client] (c1) at (0,0) {$C_1$};
\node[client] (c2) at (-0.5,-1) {$C_2$};
\node[client] (c3) at (0.5,-1) {$C_3$};
\node[draw=blue!50,dashed,rounded corners,fit=(c1)(c2)(c3),inner sep=6pt,label=below:{\scriptsize Cluster 1}] (cluster1) {};

\node[client] (d1) at (0,-3) {$C_4$};
\node[client] (d2) at (-0.5,-4) {$C_5$};
\node[client] (d3) at (0.5,-4) {$C_6$};
\node[draw=blue!50,dashed,rounded corners,fit=(d1)(d2)(d3),inner sep=6pt,label=below:{\scriptsize Cluster 2}] (cluster2) {};

\node[client] (e1) at (0,-6) {$C_7$};
\node[client] (e2) at (-0.5,-7) {$C_{8}$};
\node[client] (e3) at (0.5,-7) {$C_{9}$};
\node[draw=blue!50,dashed,rounded corners,fit=(e1)(e2)(e3),inner sep=6pt,label=below:{\scriptsize Cluster 3}] (cluster3) {};

\node[agg] (a1) at (5,-0.5) {$A_1$};
\node[agg] (a2) at (5,-3.5) {$A_2$};
\node[agg] (a3) at (5,-6.5) {$A_3$};

\node[newClient] (nc1) at (10,0) {$C_5$};
\node[newClient] (nc2) at (9.5,-1) {$C_{8}$};
\node[newClient] (nc3) at (10.5,-1) {$C_2$};
\node[draw=green!50,dashed,rounded corners,fit=(nc1)(nc2)(nc3),inner sep=6pt,label=below:{\scriptsize Cluster 1'}] (ncluster1) {};

\node[newClient] (nd1) at (10,-3) {$C_1$};
\node[newClient] (nd2) at (9.5,-4) {$C_7$};
\node[newClient] (nd3) at (10.5,-4) {$C_{9}$};
\node[draw=green!50,dashed,rounded corners,fit=(nd1)(nd2)(nd3),inner sep=6pt,label=below:{\scriptsize Cluster 2'}] (ncluster2) {};

\node[newClient] (ne1) at (10,-6) {$C_3$};
\node[newClient] (ne2) at (9.5,-7) {$C_{4}$};
\node[newClient] (ne3) at (10.5,-7) {$C_6$};
\node[draw=green!50,dashed,rounded corners,fit=(ne1)(ne2)(ne3),inner sep=6pt,label=below:{\scriptsize Cluster 3'}] (ncluster3) {};

\node (gear1) at (1.25,-0.5) {\includegraphics[width=13pt]{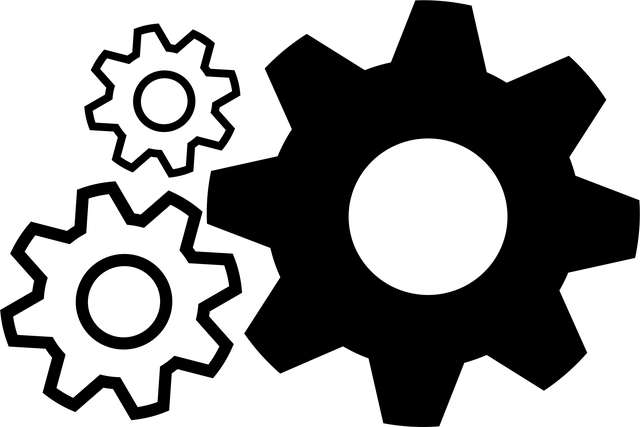}};
\node (gear2) at (1.25,-3.5) {\includegraphics[width=13pt]{./icons/gear.png}};
\node (gear3) at (1.25,-6.5) {\includegraphics[width=13pt]{./icons/gear.png}};

\draw[msgBold,blue!70] (a1.west) to[bend right=25] (gear1);
\path (a1.west) to[bend right=25] node[pos=0.5,above,inner sep=1pt,yshift=4pt] {\includegraphics[width=14pt]{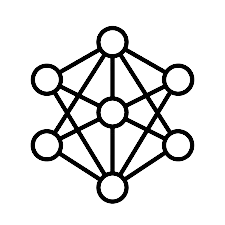}} (gear1);
\path (a1.west) to[bend right=25] node[pos=0.5,above,inner sep=1pt,scale=0.6,blue,yshift=1pt] {\textbf{Round $\tau\!-\!1$}} (gear1);

\node[fill=white,circle,draw=black,inner sep=2pt,font=\footnotesize\bfseries] at (2.95,0.8) {1};

\draw[msgBold,blue!70] (a2.west) to[bend right=25] (gear2);
  
\draw[msgBold,blue!70] (a3.west) to[bend right=25] (gear3);

\node[fill=white,circle,draw=black,inner sep=2pt,font=\footnotesize\bfseries] at (1.2,0) {2};
\draw[msgBold,green!70] (gear1) to[bend right=25] (a1.west);
\path (gear1) to[bend right=25] node[pos=0.5,below,inner sep=1pt] {
  \begin{tikzpicture}
    \node at (0,-0.15) {\includegraphics[width=13pt]{./icons/model.png}};
    \node at (0,0) {\includegraphics[width=11pt]{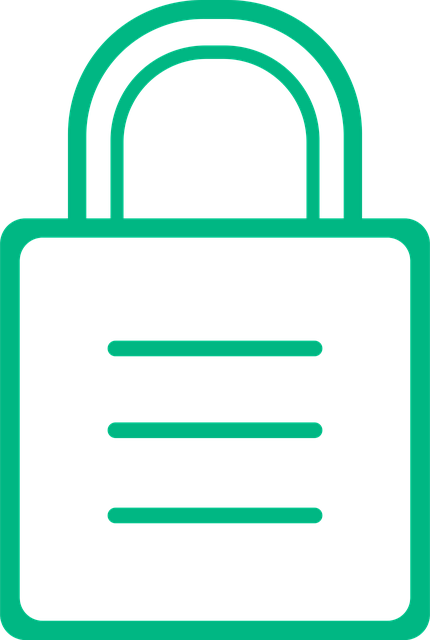}};
  \end{tikzpicture}
} (a1.west);

\node[fill=white,circle,draw=black,inner sep=2pt,font=\footnotesize\bfseries] at (2.95,-1.85) {3};

\draw[msgBold,green!70] (gear2) to[bend right=25] (a2.west);
  
\draw[msgBold,green!70] (gear3) to[bend right=25] (a3.west);



\node[fill=white,circle,draw=black,inner sep=2pt,font=\footnotesize\bfseries] at (5,-1.9) {4};

\draw[msgdashed,red!70] (a1) to[bend left=15] (a2);

\draw[msgdashed,red!70] (a2) to[bend left=15] (a3);

\draw[msgdashed,red!70] (a1) to[bend left=25] (a3);

\draw[msgdashed,red!70] (a2) to[bend left=15] (a1);

\draw[msgdashed,red!70] (a3) to[bend left=15] (a2);

\draw[msgdashed,red!70] (a3) to[bend left=25] (a1);

\node[draw=orange!70,fill=orange!15,rounded corners,line width=1.5pt,align=center,font=\scriptsize] 
  at (5,1) {
  \includegraphics[width=13pt]{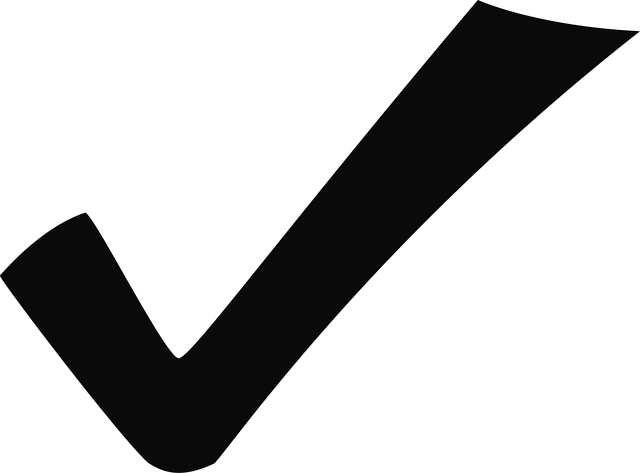} \\[-1pt]
  \tiny Generation of \\[-2pt]
  \tiny verified model \\[-2pt]
  \tiny Round $\tau$
};

\node[fill=white,circle,draw=black,inner sep=2pt,font=\footnotesize\bfseries] at (5.5,1.5) {5};

\draw[msgBold,purple!70] (a1.east) -- (ncluster1.west);
\path (a1.east) -- node[pos=0.5,above,inner sep=1pt] {\includegraphics[width=14pt]{./icons/model.png}} (ncluster1.west);
\path (a1.east) -- node[pos=0.5,below,inner sep=1pt,scale=0.6,purple] {\textbf{Round $\tau$}} (ncluster1.west);
  
\node[fill=white,circle,draw=black,inner sep=2pt,font=\footnotesize\bfseries] at (7.25,0.2) {6};

\draw[msgBold,purple!70] (a2.east) -- (ncluster2.west);
  
\draw[msgBold,purple!70] (a3.east) -- (ncluster3.west);

\node[align=center,font=\scriptsize] at (0,1.5) {
  \textbf{Round $\tau$} \\
  Current clusters
};

\node[align=center,font=\scriptsize] at (10,1.5) {
  \textbf{Round $\tau+1$} \\
  New clusters
};

\end{tikzpicture}
\caption{Overview of one training round.
(1) Client receives the global model from the previous round, with a proof of correct aggregation. 
(2) Client verifies the proof and compute a local model update. 
(3) It produces a privacy-preserving update by adding a mask and calibrated DP noise, and sends it to its coordinator. 
Aggregators collect masked updates from their cluster and, once enough contributions are received, include $\rho$ clients to mitigate delay bias and reduce the DP noise budget.
(4)  They aggregate the included masked updates, collaboratively remove the masks on the cluster-level sum, and perform a second inter-cluster aggregation to obtain the global model for the next round. 
(5) Aggregators produce a threshold signature as a proof of correct aggregation, determine their new cluster and (6) send the global model to the clients.}

\label{fig:overall:overview}
\end{figure}


 \newcommand{\msgAColor}{blue}
 \newcommand{\msgBColor}{red}
 \tikzset{
     ncbar angle/.initial=90,
     ncbar/.style={
         to path=(\tikztostart)
         -- ($(\tikztostart)!#1!\pgfkeysvalueof{/tikz/ncbar angle}:(\tikztotarget)$)
         -- ($(\tikztotarget)!($(\tikztostart)!#1!\pgfkeysvalueof{/tikz/ncbar angle}:(\tikztotarget)$)!\pgfkeysvalueof{/tikz/ncbar angle}:(\tikztostart)$)
         -- (\tikztotarget)
     },
     ncbar/.default=0.5cm,
 }

 \tikzset{square left brace/.style={ncbar=0.15cm}}
 \tikzset{square right brace/.style={ncbar=-0.15cm}}
 \tikzset{round left paren/.style={ncbar=0.5cm,out=120,in=-120}}
 \tikzset{round right paren/.style={ncbar=0.5cm,out=60,in=-60}}
 
 \begin{figure*}

     \centering
 \resizebox{\textwidth}{!}{
 \begin{tikzpicture}%
 \tikzstyle{operation}=[rectangle, rounded corners=3, draw=blue!70,fill=blue!20,minimum width=5mm,minimum height=5mm, text centered];
 \tikzstyle{operationFade}=[rectangle, rounded corners=3, draw=blue!30,fill=blue!5,minimum width=5mm,minimum height=5mm];

     \node[fill=white,circle,draw=black,inner sep=2pt,font=\footnotesize\bfseries] at (2,4) {1};
     \node[fill=white,circle,draw=black,inner sep=2pt,font=\footnotesize\bfseries] at (6,4) {2};
     \node[fill=white,circle,draw=black,inner sep=2pt,font=\footnotesize\bfseries] at (11,4) {3};
     \node[fill=white,circle,draw=black,inner sep=2pt,font=\footnotesize\bfseries] at (18,4) {4};
     \node[fill=white,circle,draw=black,inner sep=2pt,font=\footnotesize\bfseries] at (22.5,4) {5};
      \node[fill=white,circle,draw=black,inner sep=2pt,font=\footnotesize\bfseries] at (26,4) {6};

     \draw[-latex] (0,3) node[left,xshift=-4.5]{$a_1$} -- (28, 3) ;
     \draw[-latex] (0,2) node[left,xshift=-4.5]{$a_2$} -- (28, 2) ;
     \draw[-latex] (0,1) node[left,xshift=-4.5]{$a_3$} -- (28, 1) ;

     \node[operation, text width=3.5cm] (aggregation) at (2,3) {\texttt{SHARES COLLECTION AND PREPARATION}};
     \node[operation] (aggregate1) at (6,1) {\aggregatemsg};
     \node[operation] (aggregate2) at (6,2) {\aggregatemsg};

     \draw[msg,blue!70] (aggregation.east) to (aggregate1.west);
     \draw[msg,blue!70] (aggregation.east) to (aggregate2.west);
    
     \node[operation] (disseminate3) at (11,3) {\disseminatemsg};
     \node[operationFade] (disseminate2) at (11,2) { \textcolor{black!30}{\disseminatemsg}};
     \node[operationFade] (disseminate1) at (11,1) { \textcolor{black!30}{\disseminatemsg}};

     \draw[msg,blue!70] (aggregate1.east) to (disseminate3.west);
     \draw[msg,blue!70] (aggregate2.east) to (disseminate3.west);
    
     \node[operation] (gather3) at (18,3) {\gathermsg};
     \node[operation] (gather2) at (18,2) {\gathermsg};
     \node[operation] (gather1) at (18,1) {\gathermsg};

     \draw[msg,blue!70] (disseminate3.east) to (gather2.west);
     \draw[msg,blue!70] (disseminate3.east) to (gather1.west);
     \draw[msg,blue!20] (disseminate2.east) to (gather3.west);
     \draw[msg,blue!20] (disseminate2.east) to (gather1.west);
     \draw[msg,blue!20] (disseminate1.east) to (gather3.west);
     \draw[msg,blue!20] (disseminate1.east) to (gather2.west);

     \node[operation] (certify2) at (22.5,2) {\certifymsg};
     \node[operation] (certify1) at (22.5,1) {\certifymsg};

     \draw[msg,blue!70] (gather3.east) to (certify1.west);
     \draw[msg,blue!70] (gather3.east) to (certify2.west);

     \node[operation, text width=2cm] (finalize) at (26,3) {\centering \texttt{START NEW ROUND}};

     \draw[msg,blue!70] (certify1.east) to (finalize.west);
     \draw[msg,blue!70] (certify2.east) to (finalize.west);
    
    \draw[decorate,decoration={brace,amplitude=10pt}] (0,4.2) -- (14,4.2) ;
    \node[] at (7,4.8) {intra-cluster-aggregation stage};
    \draw[decorate,decoration={brace,amplitude=10pt}] (16,4.2) -- (28,4.2) ;
    \node[] at (22,4.8) {inter-cluster-aggregation stage};
 \end{tikzpicture}
 }
     \caption{Communication flow for aggregator $a_1$. $a_1$ waits for enough updates, then includes $\rho$ out of $k$ updates (1). It sends each share of the $\rho$ updates to the concerned aggregators which aggregate the shares and produce a proof of non-tampering (2). Each aggregator sends back the aggregated shares and proofs to $a_1$, which verifies the proof, reconstructs the mask, unmasks the cluster-level aggregated update, and aggregates the proofs of non-tampering (3). $a_1$ broadcasts the proof and the aggregate to the aggregators, who do the same thing with their own aggregates (4). Once enough intra-cluster aggregates and proofs are received, $a_1$ shares them with the aggregators for them to verify and sign the sum of the intra-cluster aggregate (5). Once $a_1$ receives enough such signatures, it begins a new round (6).}
    \label{fig:agg:comm}

 \end{figure*}

\paragraph{Learning protocol in a nutshell} 
Our protocol proceeds in rounds.
Each round follows a two-stage aggregation structure.

Clients first compute a local \textit{model update} using gradient descent on their private data, which is then sent to the aggregators in a privacy-preserving way.
To distribute the load across multiple aggregators, clients are organized into clusters of equal size~$k$, 
with each cluster managed by a dedicated aggregator, referred to as the \emph{cluster coordinator}.

In the first stage (\emph{intra-cluster aggregation}), each cluster coordinator acts as a proxy for the clients, who remain the actual dealers of their secrets—that is, their model updates.
Clients produce shares of their updates for all aggregators, encrypting each share with the recipient aggregator’s key.
The coordinator collects the encrypted update shares from its cluster and, once enough valid shares have been received, redistributes them to all aggregators.
It then initiates the reconstruction phase of the cluster-level aggregate.
Other aggregators collaborate in this process by computing the partial sums on their shares and contributing to the reconstruction of the cluster-level aggregate.

In the second stage (\emph{inter-cluster aggregation}), cluster-level aggregates are further combined by coordinators to produce the global model, which is then distributed back to clients for the next round.

\Cref{fig:overall:overview} illustrates this structure and the main steps executed by clients and aggregators within one round, while \Cref{fig:agg:comm} details communications among aggregators during the $4th$ step in \Cref{fig:overall:overview}.
In the following, we further detail the main protocol mechanisms.

\paragraph{Privacy-preserving method}
Client privacy is achieved through a combination of masking and differential privacy (DP).
Each client masks its model update using a random vector~$s_i$ that is secret-shared among all aggregators, and adds random noise~$e_i$ calibrated to satisfy the DP guarantees.
The noise provides DP protection at the \emph{cluster-level aggregated model}, whereas the masking mechanism hides each individual model update before aggregation.
To make masking efficient, we adapt a Learning-With-Errors (LWE)–based Verifiable Secret Sharing (VSS) scheme~\cite{SSVRCN22}, which produces compact shares on the mask rather than on the full model update.
Unlike~\cite{SSVRCN22}, in our protocol, the mask is collaboratively removed by the aggregators—rather than by the clients—to avoid inter-client communication and improve scalability.
This unmasking is performed only on the cluster-level aggregated sum of client updates, never on individual updates, 
ensuring that no single update is ever exposed in clear form.
Importantly, privacy is guaranteed only when updates are aggregated in sets of at least~$\rho$ clients,
where~$\rho$ denotes the minimum aggregation size required to satisfy the DP guarantees. This mechanism allows each client to only add a noise which follows the distribution law $\mathcal{N}(0,({\sigma^2}/{\rho})\mathbf{I})$, where $\sigma^2 = ({T\cdot C^2 \cdot  \alpha})/({2\cdot \epsilon_{\texttt{max}}}) \Leftrightarrow \epsilon_{\texttt{max}}(\alpha) =  ({T\cdot C^2 \cdot  \alpha})/({2\cdot \sigma^2})$, with $T$ the maximum number of times a client's update is added to a global model, $\alpha$ and $\epsilon_{\texttt{max}}$ the necessary RDP parameters required to reach the desired privacy guarantees, and $C$ the clipping parameter.\footnote{The clipping step restricts the gradient's norm to at most $C$. It is used to enable the DP analysis of privacy preserving learning schemes \cite{ACGMMTZ16}.}

\paragraph{Client-to-cluster assignment}
In our solution, the assignment of clients to clusters plays a crucial role. 
First, to preserve privacy, clusters must form a partition of the clients into disjoint subsets. 
Indeed, if two intra-cluster aggregates overlap on some clients, a Byzantine aggregator could compute a linear combination—e.g., subtract one aggregate from another—to isolate a smaller subset of updates and compromise privacy. 
For this reason, our assignment mechanism enforces disjoint clusters of size~$k > \rho$, 
ensuring that any linear combination of aggregates involves at least~$\rho$ distinct clients 
and cannot reveal information about smaller subsets. 

Second, our assignment mechanism reshuffles clients between clusters from one round to the next. 
A static cluster assignment could bias the learning process: clients assigned to a Byzantine or crashed coordinator might never contribute, and clusters with too many failed clients (which could prevent reaching the $\rho$ threshold) might be ignored by correct coordinators. 
Moreover, coordinators are not guaranteed to compute exactly the same global model at the end of a round. 
This divergence is unavoidable in an asynchronous system with faults, where each aggregator can wait for other intra-cluster aggregates from only $n_a - t_a$ other aggregators during the  inter-cluster aggregation. 
In this setting, a static assignment would therefore hinder convergence of the global model. 

To address these issues, we employ a verifiable deterministic shuffling algorithm that takes the round number as input and uses it as a random seed in a public hash function to create a new partition of the clients (disjoint sets) at each round. 
This ensures that the client-to-cluster assignment is independently verifiable by all participants and that, in expectation, each client is assigned to a uniformly random cluster in every round.

\paragraph{Client inclusion and bias mitigation}
Once a coordinator has received enough updates from its assigned clients, it must include a subset of $\rho < k$ clients whose updates will be included in the intra-cluster aggregation.
This inclusion process is critical to mitigate the bias induced by non-uniform communication delays among clients.
Without such regulation, faster clients could be over-represented in the training process, leading to biased model updates.

Our inclusion mechanism operates by letting each aggregator include from its current cluster the $\rho$ clients it has included least frequently in previous rounds.
To maximize the likelihood of collecting the largest possible number of contributions, including those from slower clients, a signaling mechanism is employed to estimate the number of correct clients within each cluster. 
We demonstrate experimentally that this procedure yields an approximately uniform inclusion distribution for the $n_c - 2t_c$ fastest clients, even when communication delays are not uniformly distributed.

Importantly, this inclusion mechanism also reduces the amount of noise that clients must inject to satisfy a given DP budget.
In a setting with non-uniform delays, a fast client could be included in every round and would therefore need to add more noise to maintain its DP guarantees.
With our inclusion mechanism, each client only needs to calibrate its noise proportionally to~${\rho}/{k}$, 
since it is included, in expectation, only a fraction~${\rho}/{k}$ of the rounds.
This significantly improves the utility of the aggregated model for a given privacy level.

\paragraph{Aggregate verification and global model certification}
Because coordinators may be Byzantine, each coordinator must produce a proof that the global model has been computed in compliance with the protocol and is derived from the model updates of the clients in the clusters.
These proofs enable clients to verify the global model they receive at the beginning of each round.
Certification, on the other hand, cannot be obtained by comparing coordinators' results, as is classical in systems using consensus,
because, as already mentioned, due to asynchrony, each coordinator may end up with a different inter-cluster aggregate.

To address this issue, we use a PVAHSS scheme. This mechanism enables the verification of the intra- and inter-cluster aggregates (or sums), ensuring that each aggregate is consistent with the committed values.
The verifications are performed by the aggregators themselves at the end of each round.
Each coordinator collects at least $n_a-t_a$ signed verifications for its global model, which together form the certificate associated with that model.

\section{Detailed protocol description}\label{sec:description}
This section incrementally presents the detailed learning protocol for asynchronous networks with Byzantine aggregators. Due to space constraints the complete algorithm for the protocol and its detailed description are provided in \Cref{sec:complete:algorithm}.

\subsection{The basic privacy preserving protocol}

\begin{figure*}[!ht]
\centering
\resizebox{\textwidth}{!}{
\begin{tikzpicture}[>=stealth,scale=1]

 \tikzstyle{client}=[circle,draw=blue!70,fill=blue!20,minimum size=8mm]
 \tikzstyle{Exec}=[rectangle,draw=green!70,fill=green!20,minimum width=14mm,minimum height=8mm]
 \tikzstyle{Res}=[rectangle,draw=red!0,fill=red!0,minimum width=14mm,minimum height=8mm]
 \tikzstyle{Agg}=[rectangle,draw=red!70,fill=red!20,minimum width=14mm,minimum height=8mm]


\node[client] (ci) at (-3,2.5) {$c_i$};
\node[Exec] (train) at (-15,0) {$\texttt{Train}$};
\node[Res] (wi) at (-13,-0.5) {$g_i^{\tau}$};
\node[Res] (hi) at (-7.5,-0.5) {$h_i^{\tau} = g_i^{\tau} + A \cdot s_i^{\tau} + e_i^{\tau} $};
\node[Exec] (hide) at (-11,0) {$\texttt{DP and mask}$};

\node[Exec] (PVSS) at (-5,0) {$\SSShare$};
\node[Res] (r1) at (-2, -1) {$s_{(i,1)}^{\tau}$};
\node[Res] (r2) at (-2, 0) {$s_{(i,2)}^{\tau}$};
\node[Res] (r3) at (-2, 1) {$s_{(i,3)}^{\tau}$};
\node[Exec] (enc1) at (1, -1) {Encrypt};
\node[Exec] (enc2) at (1, 0) {Encrypt};
\node[Exec] (enc3) at (1, 1) {Encrypt};
\node[Agg] (a1) at (6, 0) {$\mathbf{A_1}$};
\node[Agg] (a1bis) at (9, -1) {$\mathbf{A_1}$};
\node[Agg] (a2) at (9, 0) {$A_2$};
\node[Agg] (a3) at (9, 1) {$A_3$};
\node[Exec] (dec1) at (12, -1) {Decrypt};
\node[Exec] (dec2) at (12,0) {Decrypt};
\node[Exec] (dec3) at (12,1) {Decrypt};
\node[Res] (r1b) at (14, -1) {$s_{(i,1)}^{\tau}$};
\node[Res] (r2b) at (14, 0) {$s_{(i,2)}^{\tau}$};
\node[Res] (r3b) at (14, 1) {$s_{(i,3)}^{\tau}$};

\draw[msgBold,blue!70] (a1.north) to[bend right=10] (ci);
\path (a1.north) to[bend right=20] node[pos=0.5,above,inner sep=1pt,yshift=4pt] {$w_i^{\tau-1}$} (ci);

\draw[msgBold,dashed,gray] (ci.west) to[bend right=10] (train);
\path (ci.west) to[bend right=20] node[pos=0.5,above,inner sep=1pt,yshift=4pt] {$w_i^{\tau-1}$} (train);

\draw[msgBold,dashed,gray] (train.east) to (hide.west);

\draw[msgBold,dashed,gray] (hide.east) to (PVSS.west);

\draw[msgBold,dashed,gray] (PVSS.east) to (r1.west);
\draw[msgBold,dashed,gray] (PVSS.east) to (r2.west);
\draw[msgBold,dashed,gray] (PVSS.east) to (r3.west);

\draw[msgBold,dashed,gray] (r1.east) to (enc1.west);
\draw[msgBold,dashed,gray] (r2.east) to (enc2.west);
\draw[msgBold,dashed,gray] (r3.east) to (enc3.west);

\draw[msgBold,dashed,blue!70] (enc1.east) ..  controls (a1) .. (a1bis.west);
\draw[msgBold,dashed,blue!70] (enc2.east) ..  controls (a1) .. (a2.west);
\draw[msgBold,dashed,blue!70] (enc3.east) ..  controls (a1) .. (a3.west);


\draw[msgBold,dashed,gray] (a1bis.east) to (dec1.west);
\draw[msgBold,dashed,gray] (a2.east) to (dec2.west);
\draw[msgBold,dashed,gray] (a3.east) to (dec3.west);

\draw[msgBold,dashed,gray] (dec1.east) to (r1b.west);
\draw[msgBold,dashed,gray] (dec2.east) to (r2b.west);
\draw[msgBold,dashed,gray] (dec3.east) to (r3b.west);


\node[fill=white,circle,draw=black,inner sep=2pt,font=\footnotesize\bfseries] at (1,2.7) {1};
\node[fill=white,circle,draw=black,inner sep=2pt,font=\footnotesize\bfseries] at (-15,1) {2};
\node[fill=white,circle,draw=black,inner sep=2pt,font=\footnotesize\bfseries] at (-11,1) {3};
\node[fill=white,circle,draw=black,inner sep=2pt,font=\footnotesize\bfseries] at (-5,1) {4};
\node[fill=white,circle,draw=black,inner sep=2pt,font=\footnotesize\bfseries] at (1,1.8) {5};
\node[fill=white,circle,draw=black,inner sep=2pt,font=\footnotesize\bfseries] at (12,1.8) {6};

\end{tikzpicture}
}
\caption{\textbf{Privacy preserving update sharing:} Aggregator shares its model with a client (1). client trains the model on local data (2), adds noise $e_i^\tau$ to the update and masks it with $s_i^{\tau}$ (3). Shares are produced using a $(n_a, n_a-t_a)$-SS scheme (4). Shares are encrypted and sent to their recipient, using $a_1$ as a proxy (5). Aggregators receive and decrypt their shares (6).}
\label{fig:LWE}
\end{figure*}

The basis of our protocol is an adaptation of the FedAvg scheme \cite{MMRHA17} to the replicated aggregators case. It is calibrated for the optimal Byzantine resilience $n_a\ge 3t_a+1$.
Our protocol is described in \Cref{alg:LWE:client} for the client and in \Cref{alg:lwe:aggregator} for the aggregator. The protocol is initiated through the operation $\mathsf{Train}()$, invoked by aggregators. Then, communication between clients and aggregators relies on two message types: $\trainmsg$ messages and $\updatemsg$ messages.
$\trainmsg$ messages are used to activate clients and share the current model of an aggregator with the clients it coordinates. The $\updatemsg$ messages are used by each client to answer to a $\trainmsg$ message.

As outlined in \Cref{fig:LWE}, at the beginning of the round $\tau$, client $c_i$ receives a global model $\widehat{W}_{\texttt{Proposed}}^\tau$ from an aggregator. It performs local steps of gradient descent algorithm using its local data, thus obtaining a model update in the form of a gradient $g_{i}^{\tau}$. However, it wants this gradient to be protected from reconstruction and membership inference attacks. To do so, $c_i$ clips its gradient $g_i^\tau$ such that $\bar{g}_{i}^{\tau}\gets g_{i}^{\tau}/\max(1, {||g_{i}^{\tau}||_2}/{C})$. Then, $c_i$ hides $\bar{g_i}^\tau$ by computing $h_i = \bar{g}^\tau_i +A\cdot s_i + e_i$,  where $A$ is a public matrix, $s_i$ is the masking value, and $e_i$ is the Gaussian noise that enables differential privacy. The use of $A$ makes it possible to reduce the size of the mask $s_i$ without decreasing security. We denote by $N_s$ the size of $s_i$, which depends on the security required, and we denote by $N_g$ the size of $g_i^\tau$, where $N_s$ is smaller than $N_g$ by orders of magnitude. 
We refer to~\cite{SSVRCN22} for the evaluation of the size of the parameters. 
The amplitude of the noise $e_i$ is computed such that $(\alpha,\epsilon)$-RDP~\cite{M17} is guaranteed only if the noisy gradient $\bar{g}_i^\tau +e_i$ of $c_i$ is combined with enough other updates. We denote this number by $\rho$, where $\rho<k$.
Thus, we have $e_i \sim \mathcal{N}(0,({\sigma^2}/{\rho})\mathbf{I})$, and the sum of the $\rho$ noises follow the distribution
$\mathcal{N}(0, {\sigma^2}\mathbf{I})$, where $\sigma^2 = ({T\cdot C^2 \cdot  \alpha})/({2\cdot \epsilon{\texttt{max}}}) \Leftrightarrow \epsilon_{\texttt{max}}(\alpha) =  ({T\cdot C^2 \cdot  \alpha})/({2\cdot \sigma^2})$, with $T$ the maximum number of times a client’s gradient is added to a global model, and $\alpha$ and $\epsilon_{\texttt{max}}$ the necessary RDP parameters required to reach the desired privacy guarantees. This noise ensures that our protocol is $(\alpha, \epsilon_{\texttt{max}})$-RDP~\cite{SSVRCN22}.

Then, $c_i$ builds a set of $n_a$ shares $s_{(i,j)}, \forall a_j\in \mathcal{A}$ of the mask $s_i$ using a $(n_a,n_a-t_a)-$SS scheme and distributes the shares along with the masked and noisy gradient to its coordinator $a_j$. This aggregator acts as a proxy between the clients it coordinates and the final recipients of the shares.
To protect privacy, each share is encrypted with the key of a symmetric encryption scheme of its final recipient.
The use of a proxy removes the need for aggregators to confirm that other aggregators received their shares for specific clients’ gradients. Any other solution for such confirmation would involve expensive communication.
When a coordinator receives enough masked gradients and encrypted shares, it chooses a set $\mathcal{S}$ of $\rho$ clients, and broadcasts their encrypted shares to the other aggregators using a $\aggregatemsg$ message. The aggregators then decrypt and aggregate the shares. Once the aggregator $a_k$ has aggregated the share $\tilde{s}_k=\sum_{c_j\in \mathcal{S}} s{(j,k)}$, it sends it back to the coordinator $a_i$ using an $\disseminatemsg$ message. The coordinator computes the cluster-level aggregated gradient as: $$\SSReconstruct(\{\tilde{s}_{k}\}_{\forall a_k\in \mathcal{A}}), \quad
\widehat{H} = \sum_{j\in \mathcal{S}} h_j, 
\quad\text{ and } \widehat{G} = \widehat{H} - A \cdot \widehat{s}.
$$

Once $a_i$ has computed the cluster-level gradient, it can begin the inter-cluster aggregation stage.
During this stage, aggregators share their cluster-level aggregated gradients with other aggregators using $\gathermsg$ messages. Once the aggregator $a_i$ receives aggregated gradients from at least $n_a-t_a$ aggregators, it averages those gradients and reconstructs the global model.
The aggregator uses this global model as its $\widehat{W}_{i}^{\tau+1}$ value in the next round of training.

Unlike prior secure aggregation protocols, our protocol makes an innovative use of SS to achieve secure aggregation.
Our solution does not require clients to participate in the reconstruction of $s_i$;
instead, this task is delegated to the aggregators themselves.
This design significantly reduces communication overhead with a potentially large number of clients,
while SS ensures that an individual gradient $g_i^\tau+e_i$ is never reconstructed by Byzantine aggregators without the participation of $n_a-2t_a>t_a+1$ correct aggregators. 
Thus, those correct aggregators are entitled to verify that potential Byzantine aggregators do not reconstruct $g_i^\tau +e_i$ if not aggregated with $\rho -1$ others.

\begin{algorithm}
\small
\nl\Function{$\CreateMaskedModel(g^{\tau}_{i})$}{
        $s \gets \mathsf{GenerateRandomSecret}()$; \
        $e \gets \mathsf{DrawRandomNoise}()$; \\
        $\bar{g}_{i}^{\tau}\gets g_{i}^{\tau}/\max(1, \frac{||g_{c_i}^{\tau}||_2}{C})$;\\
        $h \gets \bar{g}^{\tau}_i + A\cdot s + e$;\\
        $\texttt{SecretShares} \gets (n_a, n_a-t_a)$-$\SSShare(s)$;\label{line:LWE:cli:sec:shares:begin}
        $\EncSecretShares\gets \emptyset^{n_a}$;\\
        \For{$a_\ell \in \mathcal{A}$}{
            $\EncSecretShares[\ell] \gets \Encrypt((c_i, \texttt{SecretShares}[l], \Sign(($\\ \nl\ \ \ \ $\tau, c_i, \texttt{SecretShares}[l]), \SKSig_{c_i})), \SKEnc_{(a_\ell, c_i)})$;\label{line:LWE:cli:sec:shares:end}
        }
        \return $(h,\EncSecretShares)$;
    }
\nl\WhenReceived{$\trainmsg<\tau, \widehat{W}^{\tau}_{\text{proposed}}>$}{
        $a_j \gets \assigned(\tau, c_i)$;\quad
        $g^{\tau}_i \gets  \nabla F_{i}(\widehat{W}^{\tau}_{\text{proposed}}, \xi^{\tau}_i)$;\\
        $(h, \sigma_{h},\EncSecretShares) \gets \CreateMaskedModel(g^{\tau}_{i})$;\\
        \send $\updatemsg\langle \tau, h, \sigma_{h}, \EncSecretShares \rangle$ to aggregator $a_j$;\\
    }
\caption{Privacy preserving training scheme (for $c_i$).}
\label{alg:LWE:client}
\end{algorithm}

\begin{algorithm}
\small
\WhenReceivedFromClient{$\updatemsg\langle \tau, h_j^{\tau}, \EncSecretShares^{\tau}_j \rangle$}{
        $\MaskedUpdates^{\tau}_i[j] \gets h_j^{\tau}$;\quad
        $\EncSecretShares^{\tau}_i[j] \gets \EncSecretShares^{\tau}_j$;\\
        $\mathcal{S}_i^{\tau} \gets \mathcal{S}_i^{\tau} \cup \{c_j\}$;\label{line:lwe:agg:update:gather:end}\\
        \If{$|\MaskedUpdates^{\tau}_i| \ge \rho$ \cAnd no $\aggregatemsg$ sent}{
            $\mathcal{S}_i^{' \tau} \gets \select(\mathcal{S}_i^{\tau})$;\label{line:lwe:agg:update:selection}\\
            \lFor{$c_j \in \mathcal{S}_i^{\tau}$}{
                $\widehat{H}_i^{\tau}\gets \widehat{H}_i^{\tau} + \MaskedUpdates_i^{\tau}[j]$
            }
            \For{$a_k \in \mathcal{A}$\label{line:lwe:agg:prepare:agg:beg}}{
                $\texttt{TmpShares} \gets \EncSecretShares^{\tau}_i[\ell][k], \forall \ell \in \mathcal{S}_i^{\tau}$;\\
                $\send$ $\aggregatemsg\langle \tau +E , \mathcal{S}_i^{'\tau}, \texttt{TmpShares}\rangle$ to $a_k$\label{line:lwe:agg:prepare:agg:end};
            }
            }
        }
\WhenReceivedFromAggregator{$\aggregatemsg\langle \tau, \mathcal{S}_j^{'\tau}, \texttt{TmpShares}\rangle$}{
    \lIf{not $|\mathcal{S}_j^{'\tau}| = \rho$ }{\return}
        $\texttt{sumShares} \gets \{0\}^{N_s}$;\\
        \For{$\texttt{encShare} \in \texttt{TmpShares}$\label{line:lwe:agg:decrypt:and:sum:beg}}{
            $(c_k, \texttt{share}, \texttt{sig}) \gets\Decrypt(\texttt{encShare}, \SKEnc_{(a_i, c_k)})$;\\
            \lIf{not $\Verify((\tau,c_k, \texttt{share}), \PKSig_{c_k})$}{\return}
            $\mathcal{S}_{\texttt{tmp}} \gets \mathcal{S}_{\texttt{tmp}} \cup c_k$;
            $\texttt{sumShares} \gets \texttt{sumShares} \oplus \texttt{share}$\label{line:lwe:agg:decrypt:and:sum:end};
        }
            $\send$ $\disseminatemsg\langle\tau, \texttt{sumShares}\rangle$ to $a_j$

}

\WhenReceivedFromAggregator{$\disseminatemsg\langle\tau, \texttt{sumShares}\rangle$}{
    $\texttt{SumShareSet}^{\tau}_i \gets \texttt{SumShareSet}^{\tau}_i \cup \{ \texttt{SumShares}\}$;\\
    \If{$|\texttt{SumShareSet}^{\tau}_i| \ge n_a-t_a$ \cAnd no $\gathermsg$ sent}{

        $\widehat{g}^{\tau}_i \gets \widehat{H}_i^{\tau} - A \cdot  (n_a, n_a-t_a)$-$\SSReconstruct(\texttt{SumShareSet})$; \label{line:lwe:agg:unmask}\\

        \broadcast $\gathermsg\langle \tau, \widehat{g}^{\tau}_i\rangle$ to processes in $\mathcal{A}$;
    }
}

\WhenReceived{$\gathermsg\langle \tau, \widehat{g}^{\tau}_j\rangle$}{
    \WaitUntil{$\gathermsg\langle \tau-1, \star\rangle$ is received from $a_j$\label{line:lwe:agg:verify:participation}}{
    $\FinalSelection_i^{\tau} \gets \FinalSelection_i^{\tau} \cup \{\widehat{g}^{\tau}_j\}$;\\
    \If{ $|\FinalSelection_i^{\tau}|\ge n_a-t_a$ and no $\trainmsg\langle \tau, \star \rangle$ sent}{
        $\widehat{G}_i^{\tau} \gets \frac{1}{\rho \cdot |\FinalSelection|}\sum_{k=1}^{|\FinalSelection|} \FinalSelection_i^{\tau}[k]$;\\
        $\widehat{W}^{\tau+1}_i = \widehat{W}^{\tau}_i - \gamma^{\tau} \widehat{G}_{a_i}^{\tau}$;\\
        \broadcast $\trainmsg\langle\tau+1, \widehat{W}^{\tau+1} \rangle$ to all clients in $\mathcal{C}$.
    }
    }
}
\caption{Privacy preserving training scheme (for $a_i$).}
\label{alg:lwe:aggregator}
\end{algorithm}

\subsection{Assignment} \label{sec:assignment}

Our system’s privacy requires that each aggregator compute a unique, noisy aggregate such that each individual gradient is hidden among a set of $\rho-1$ other gradients.

However, if each aggregator were allowed to independently include its own set of clients, then we might end up with $n_a$ different sets of aggregated updates that may intersect. If two of those aggregates differ by only one client’s update, it is easy to subtract those two aggregates to find the individual value of this client, thus violating privacy.
One natural idea to solve this problem is to require aggregators to \textit{agree} on a unique set of clients to activate. This step is known as Agreement on a Common Subset (ACS), and has to be solved using a consensus algorithm. Yet, consensus is impossible to achieve deterministically in asynchronous systems with faults \cite{FLP85}.

To enforce non-intersection of client sets in a fair manner and without consensus, we deterministically partition the set of clients at each aggregation step using a public, predefined shuffling function.
This partition function $\assign$ takes as input the round number and outputs a deterministic partition of the set of clients, assigning each client to a unique aggregator. An aggregator $a_i$ is thus assigned the cluster $\mathcal{S}_i^\tau\subset \mathcal{C}$ at round $\tau$, where $|\mathcal{S}_i^\tau| = k = {n_c}/{n_a}$. \footnote{To ease the rest of the discussions and without loss of generality, we assume that $n_a \mid n_c$. This condition can be relaxed in a real-world implementation, such that all clusters have the same size $\lfloor{n_c}/{n_a}\rfloor$ except one cluster, which is larger.}

The assignment algorithm, albeit deterministic, assigns clients to aggregators uniformly at random in expectation. In other words, in expectation, each client will be assigned to each aggregator during the execution of the protocol. Furthermore, if a client is assigned to aggregators with the same set of clients over the rounds and if this set of clients contains too many ($k-\rho-1$) crashed clients, then the correct clients may never participate, thus biasing the resulting model. To avoid this problem, we force the assignment algorithm to shuffle the clients in the sets $\mathcal{S}_i$.

Formally, the assignment algorithm can be defined as follows:

\begin{definition}{\textbf{Assignment algorithm.}}
Let $\assign: \mathbb{N} \rightarrow \mathcal{C}^{k\times n_a}$ be a deterministic function that takes as input a round number $\tau$ and outputs $\{\mathcal{S}^\tau_1, \cdots, \mathcal{S}^\tau_{n_a}\}$ a partition of $\mathcal{C}$ of size $n_a$, where $|\mathcal{S}^\tau_i| = k$, $\forall i \in \{1, \cdots, n_a\},\ \forall \tau \in \mathbb{N}$ and with $k = {n_c}/{n_a}$. 
Furthermore, we define a function $\assigned: \mathbb{N} \times \mathcal{C} \rightarrow \mathcal{A}$ that takes as input the round $\tau$ and the identity of a client, and outputs $a_i$, the identity of the client's coordinator at step $\tau$. 
The $\mathsf{Assign}$ function fulfills the two following properties:
\begin{itemize}
\item \textbf{Random aggregator assignment.} $\forall \mathcal{S}^* \in \assign(\tau)$ \\$=$ $\{\mathcal{S}^\tau_1, \cdots, \mathcal{S}^\tau_{n_a}\}$, $\forall c_i \in \mathcal{C}, 
Pr[c_i \in \mathcal{S}^*]$ $= {1}/{n_a}$.
\item \textbf{Random set of clients.} $\forall c_i, c_j \in \mathcal{C}, c_j\ne c_i$ and $\forall \tau \in \mathbb{N}$, if $\assigned(\tau, c_i) = a_k$ then $Pr[\assigned(\tau, c_j)$ $=a_k] = ({k-1})/({n_c-1})$.
\end{itemize}
\end{definition}

Such an assignment algorithm can be easily instantiated using a hash function whose input is the round and produces an output of size $n_c\times n_a$ bits. One such implementation is the Ethereum swap-or-not shuffling algorithm \cite{HMR12}.

\paragraph{Byzantine equivocation}
The assignment algorithm solves the intersection problem for correct aggregators. However, if an aggregator $a_i$ is Byzantine, it may deviate from its algorithm and equivocate, i.e., it can choose two different sets $\mathcal{S}_i^{'\tau} \subseteq \assign(\tau)[i]$ and $\mathcal{S}_i^{''\tau} \subseteq \assign(\tau)[i]$ of size $\rho$ that differ in exactly one client. If $a_i$ manages to lure correct aggregators into aggregating and reconstructing the values associated with both subsets, it could learn the exact gradient of an individual client.

To avoid this equivocation, we carefully craft the SS scheme used to reconstruct the sum of the masks $s_j$ such that Byzantine aggregators cannot reconstruct two different sums of masks. In practice, we require that $n_a-t_a$ shares must be gathered to reconstruct a mask. Therefore, reconstructing two different sums of masks for a unique cluster would require getting two non-intersecting quorums of $n_a-2t_a$ correct processes to reconstruct different intra-cluster aggregates for this cluster, which is not possible since $n_a>3t_a$.

\subsection{Inclusion} \label{sec:inclusion}
After the assignment, aggregators have to choose a subset of $\rho<k$ clients' gradients in their cluster. However, if this inclusion, and thus the data included in the training, is biased, then the final model will be biased too. Such bias is introduced when aggregators include fast clients first due to asynchronous communications and heterogeneous data. To solve this problem, we introduce a ``debiasing'' client inclusion mechanism. This mechanism works by letting an aggregator $a_i$ wait for more client participations than required, and then aggregate gradients of the clients less considered in the previous rounds by $a_i$. 

The inclusion algorithm works best if aggregators wait for the maximum possible number of client participations in their assigned cluster each round. However, our failure assumption depends on the global set of clients, not on each cluster. Thus, to know how many clients an aggregator can wait for, we require that clients send an additional message to all aggregators \emph{after} finalizing the sending of the masked gradient and shares to their coordinator. This \pingmsg message contains a signature of the round number from the client. It allows aggregators to determine when they waited for the maximum number of clients they can expect, i.e., at least $n_c-t_c$ clients from all clusters participated overall. 
However, this mechanism could lead to interlocked aggregators, i.e., each aggregator sees the participation of clients from other clusters, and no aggregator received enough participations from their own clusters. To solve this problem, we add a broadcast communication phase meant to ensure that aggregators have a shared knowledge of clients that participated. This communication phase works by letting aggregators broadcast the identity of the $n_c-t_c$ clients that sent them \pingmsg messages, along with the signature contained in those messages for tamper-resistance. Thanks to the properties of intersecting quorums, a $n_c-t_c$ common core of participating clients is guaranteed as $n_a > 3t_a$ \cite{MR97}. Due to the reliable communication model, if a message is sent, it is not dropped. Furthermore, \pingmsg messages are sent \emph{after} sending the masked gradients. Thus, if an aggregator knows that a client sent a \pingmsg message, but did not receive its gradient yet, it knows it will eventually receive it.

We prove experimentally that, if this mechanism is well crafted, then the frequency at which clients are included is uniform. More precisely, if $t_c >{k}/{2}$, $n_c>4t_c+1$,  $k> 2\rho$ and $\rho > 1 + \sqrt{1+k}$, then we show that, in expectation, the $n_c-2t_c$ fastest clients are uniformly included by each correct aggregator in expectation.

\paragraph{Impact of inclusion on budget}
Interestingly, this inclusion algorithm also decreases the required amplitude of the noise clients add to their gradient. Indeed, we recall that the noise $e_i$ follows a distribution  $e_i \sim \mathcal{N}(0,({\sigma^2}/{\rho})\mathbf{I})$, with $\sigma^2 = ({T\cdot C^2 \cdot  \alpha})/({2\cdot \epsilon_{\texttt{max}}})$, where $T$ is the maximum number of times a client's update is included. In our case, this value is 
$\min(n_a$($ {(\tau_{\texttt{max}}\cdot \rho})/({n_c-t_c})+\maxDelta$),$\tau_{\texttt{max}}$), 
where $\tau_{\texttt{max}}$ is the maximum number of training rounds and $\maxDelta$ is a parameter generally set to $1$ described in \Cref{sec:edge:cases}. In other Gaussian noise based FL protocols, asynchronous settings imply that $T=\tau_{\texttt{max}}$ in the worst case.

\paragraph{Edge cases} The inclusion mechanism can induce two specific edge cases. First, a Byzantine aggregator could decide to bias its clients selection by not using the inclusion algorithm. We propose to use an inclusion fraud detection mechanism to alleviate any such Byzantine misbehaviour. Second, some correct aggregators may not be able to include $\rho$ clients in a round if too many clients in their cluster crash. To solve this issue, this aggregator uses the intra-cluster model of the other aggregators this round. In following rounds, this aggregator will be able to resume to the classical aggregation mechanism, thanks to the shuffling of the assignation mechanism. Due to space limitations, both edge-cases are detailed in \Cref{sec:edge:cases}.

\subsection{Certification}\label{sec:certification}

There exists a Byzantine behavior that may undermine convergence of the training. Byzantine aggregators may behave correctly during the aggregation, but at the beginning of a new round, they could provide the clients with global models that they arbitrarily choose, rather than the result of the intra-cluster and inter-cluster aggregations. 
Those values are used as the basis of clients' gradient descent algorithm,
thus this attack can undermine convergence for all correct aggregators.
To overcome this attack, our protocol certifies global models that are sent to clients. This certification comes in the form of a proof of correct aggregation certifying that at least one correct aggregator verified that each global model sent to clients was produced according to the protocol.

This certification mechanism requires two building blocks: threshold signatures and Publicly Verifiable Additionally Homomorphic Secret Sharing (PVAHSS). Threshold signatures  \cite{S00} are signatures schemes with two additional operations \CombineThre and \VerifyCombThre. A threshold signature scheme is defined for a set of signatories $S_\sigma$ and a threshold $x$. \CombineThre allows an actor to combine a set of $x$ signatures signed by $x$ different signatories in $S_\sigma$ and combine them in a unique cryptographic element which proves that $x$ signatories out of $|S_\sigma|$ endorsed a common message. The \VerifyCombThre algorithm is used to verify this combined signature.

Certification proceeds as follows. When a client is activated, it creates shares of its mask along with two proofs \VSSProof and \VSSPartialProof. The first proof is a commitment to the noisy gradient, while the \VSSPartialProof is a set of commitment to each share produced by the client. \VSSProof is embedded in the encryption of the shares such that each aggregator can receive them untampered. 
\VSSPartialProof{s} are sent to coordinators.
Once an aggregator received $\rho$ encrypted shares for a given cluster through a \aggregatemsg message, it decrypts them, sums them, and then threshold-signs the homomorphic sum of the \VSSProof{s} before sending everything back to the coordinator.
When the coordinator has received $n_a-t_a$ sums of shares, it verifies each \VSSPartialProof to detect any outlier. Then, it reconstructs the shares to obtain the sum of the masks and it combines the signatures of the \VSSProof{s}.
Thanks to the trust assumption, the combined signatures of the \VSSProof{s} proves that $n_a-2t_a$ correct aggregators participated in the reconstruction. Therefore, the combined signature can be used as a proof that the sum of the commitments was built according to the protocol, and that no equivocation occurred. 

Once this proof is constructed, aggregators have to go through the inter-cluster aggregation phase. This phase has to be modified to provide certification. As earlier, aggregators share their cluster-level aggregated gradients. However, they broadcast the aggregated gradients along with the proof of legitimate reconstruction created during the intra-cluster aggregation phase (i.e., the threshold signature on the sum of the \VSSProof). When an aggregator receives enough intra-cluster aggregated gradients and proofs, it broadcasts them for certification. Each aggregator will verify the proofs of correct reconstruction. If this verification passes, the aggregators aggregate the intra-cluster gradients, threshold-signs the result, and send back this signature to the aggregator that requested this certification. This aggregator can combine the threshold signatures once it received $n_a-t_a$ of them. This final threshold signature proves that, at each point in the intra-cluster aggregation phase and in the inter-cluster aggregation phase, at least one correct aggregator participated, and certified that the global model was built using clients' gradients. Thus, it can be used by clients as a proof that no Byzantine aggregator tampered the model they received.

\section{Evaluation}\label{sec:exp:analyzes}

This section evaluates our solution in federated learning scenarios. 
More precisely, we aim to answer the following questions:

\begin{itemize}
    \item How does the computational overhead induced by the cryptographic mechanisms impact actors of the protocol ?
    \item How does our inclusion mechanism performs in highly adversarial networks and data distributions ?
    \item How the parameters of the protocol affect convergence of the models when differential privacy is applied ?

\end{itemize}

We use FLDP \cite{SSVRCN22} as our baseline as it is the state of the art in terms of DP-based FL. Indeed, FLDP allows to add noise with an amplitude inversely proportional to the number of included client's updates. Therefore, they are equivalent to a central DP learning algorithm in the homogeneous communications delays model. Any other homogeneous delays-based baseline would exhibit equivalent or worst utility. 
Additionally, FLDP is state-of-the-art in term of computationally efficient privacy preserving federated training.

\subsection{Cryptographic mechanisms evaluation}\label{sec:computationnal:overhead}
In this section we isolate the cryptographic mechanisms of our protocol to evaluate their computational overhead.

\subsubsection{Experimental setup}
In this section, we use simulated gradients produced as random vectors of a parameterizable size with 4 point decimal precision. We implemented FLDP and our cryptographic protocol in python, using 128 bit security cryptographic building blocks. For encryption, we used AES128 \cite{pycryptodome}. For commitments, we used discrete log-based commitments in elliptic curve cryptography groups on the curve P256 \cite{P256, fastecdsa}. For threshold signatures, we used the thresholdified version of the BLS signature scheme \cite{BLS01, blsimplem}. We used a modulus of 100003 for the LWE scheme and 1002523 for the SS scheme. To achieve 128 bit security using the LWE scheme, we used a LWE parameter estimator \cite{APS15} which, given $\sigma$, the standard deviation of the noise and the modulus of the scheme, evaluates $N_s$, the size of the $s$ vector of the LWE scheme. This parameter being the most important in term of computation as it governs the number of element secretly shared, as well as the number of committed elements. In our experiments, $N_s$ varies between $410$ and $710$.
Except when those parameters are tested with multiple values, the experiments have been run with $n_c= 1000$, $t_c= 249$, $\rho=32$, $N_g = 25000$, and $\epsilon = 5$.  

\subsubsection{Evaluation}

\begin{figure*}
    \centering
    \begin{subfigure}{0.33\textwidth}
        \includegraphics[width=\linewidth]{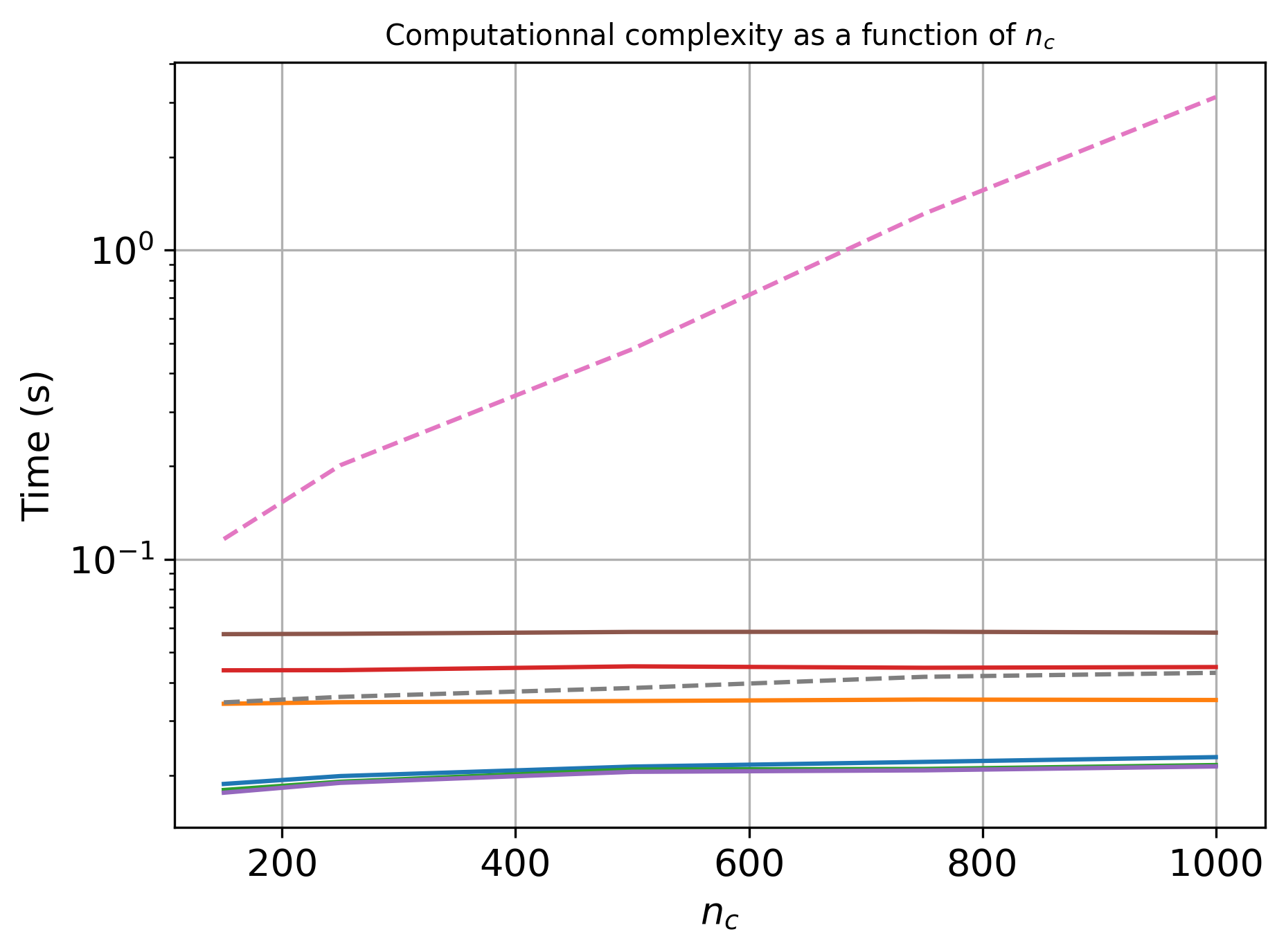}
    \end{subfigure}
    \begin{subfigure}{0.33\textwidth}
        \includegraphics[width=\linewidth]{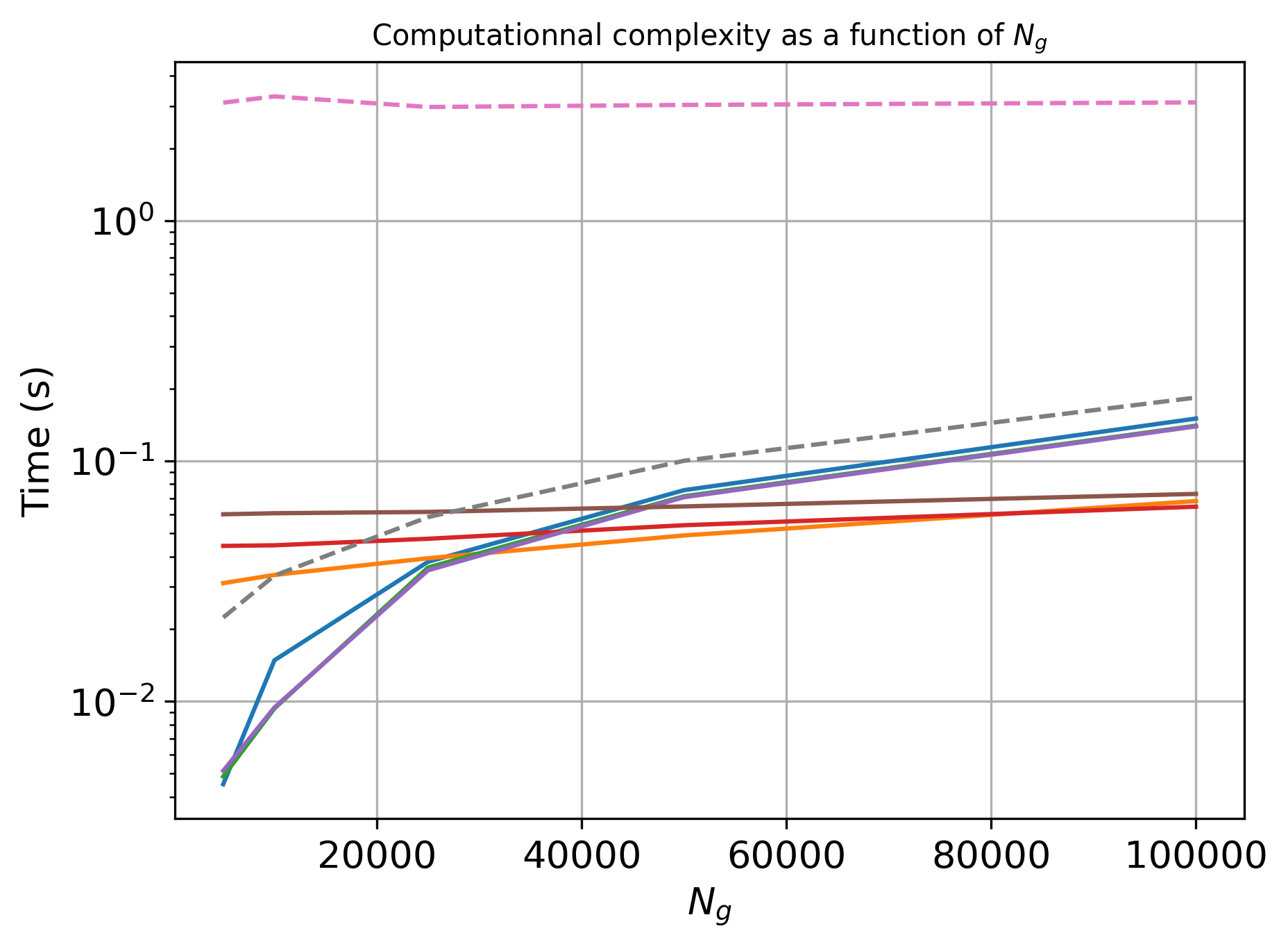}
    \end{subfigure}
    \begin{subfigure}{0.33\textwidth}
        \includegraphics[width=\linewidth]{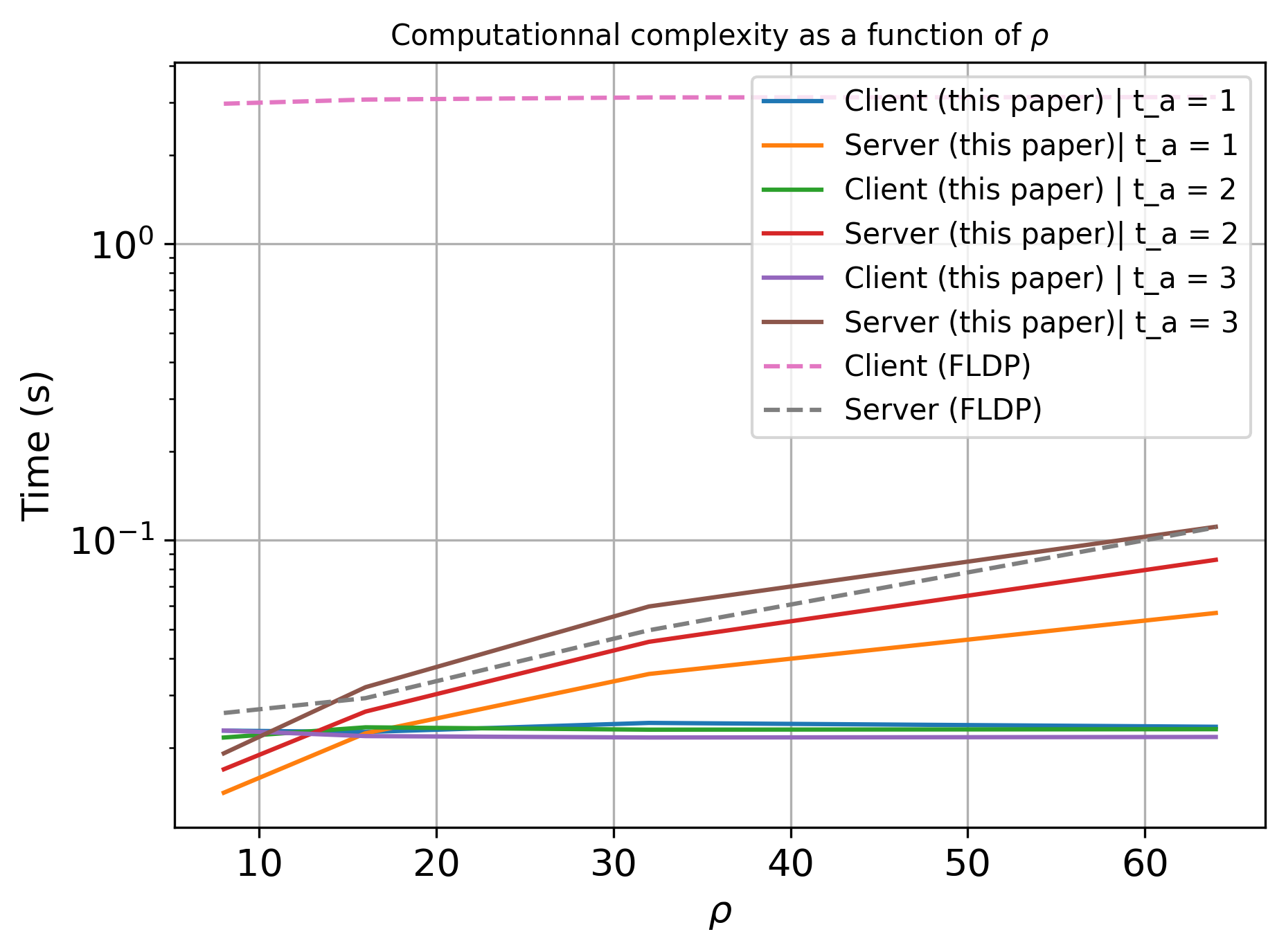}
    \end{subfigure}
    \caption{Computational overhead of the cryptographic scheme as a function of $n_c$ (left), $N_g$ (middle), and $\rho$ (right).}
    \label{fig:crypto}
\end{figure*}

\Cref{fig:crypto} shows the impact of the evolution of the parameters $\rho$, $N_g$ and $n_c$ on the computational overhead. We can see that, in our protocol, clients are only impacted by the $N_g$ parameter, whereas aggregators are mostly impacted by the $\rho$ parameter. Importantly, in those experiment, we see that the cryptographic overhead of our protocol is an order of magnitude lower than training computational overhead when the amount of data held by clients increases (\Cref{tab:training:time}). Furthermore, we see that state-of-the-art FLDP is highly impacted by the $n_c$ parameter, which undermines its scalability. On the other hand, our one-shot-client approach is not impacted by the number of clients, thus making it a good candidate for large scale deployments.

\begin{table}
    \centering
    \begin{tabular}{c|c|c}
        model & avg. \# data held by clients & avg. training time (s)\\
         \hline
        MNIST& 20 & 0.086\\
        MNIST & 30 & 0.119\\
        CIFAR-5m & 500 & 0.462 \\
        CIFAR-5m & 1000 & 0.922 \\
        CIFAR-5m & 2000 & 1.690 
    \end{tabular}
    \caption{Average client's training time with our protocol (without cryptographic mechanisms) as a function of the average amount of data held by clients.}
    \label{tab:training:time}
\end{table}

\subsection{Accuracy of training in adversarial scenarios}
In this section, we evaluate two complementary aspects of our
protocol: (i)~the inclusion mechanism's ability to counteract selection bias
under asynchronous network, and (ii)~the privacy-accuracy trade-off
enabled by our DP scheme. 
Byzantine scenarios and additional experimental analyzes are deferred
to~\Cref{sec:additionnal:expe}.

\subsubsection{Experimental setup}


We implement our experiments in two phases. We use the Multi-Agent
eXperimenter (MAX)~\cite{MAX} framework to generate per-agent communication
traces, where message delays follow probabilistic distributions.
Unless stated otherwise, we use three Gamma distributions modeling $n_a$
aggregators, $n_c - 2t_c - 1$ fast clients, and $2t_c + 1$ slow clients. Gamma distributions can model network delays and fit closely to real data \cite{WAKLMNW24}.
The traces are then consumed by Python scripts that perform model training and
aggregation according to the protocol described in this paper.

The protocol is evaluated on the MNIST datasets comprising 70,000 28 $\times$ 28 grayscale image partitioned into 10 classes and on the CIFAR-5m dataset \cite{CIFAR5m}, a CIFAR-10 like task comprising of 6 million 32 $\times$ 32 RGB images partitionned into 10 classes. Models used are presented in detail in \Cref{sec:additionnal:expe}. To summarise, we used a 26,000 parameters model for the MNIST dataset and a 550,000 parameters model for the CIFAR-5m dataset.

In order to take into account the heterogeneous nature of FL, we use the popular Dirichlet method \cite{JSHVKAC2024} to generate heterogeneity of data distribution between clients. 
Data heterogeneity is simulated by splitting the data samples of each class between the clients with Dirichlet distributions with parameter $\alpha_{dir}=0.5$ to attain a high heterogeneity setting while ensuring with high probability that all clients have at least one data sample.

Finally, when applying DP, we target an $(\epsilon,\delta)$-DP guarantee with
$\delta=10^{-5}$. We convert to $(\alpha,\epsilon_{\mathrm{RDP}})$-RDP by
minimizing $\alpha$, then calibrate $\sigma$ to achieve the worst-case
$(\epsilon,\delta)$-DP guarantee over all possible client selection sequences.

\subsubsection{Inclusion mechanism evaluation in highly adversarial networks and data distribution}

\begin{figure*}
    \centering
        \includegraphics[width=\linewidth]{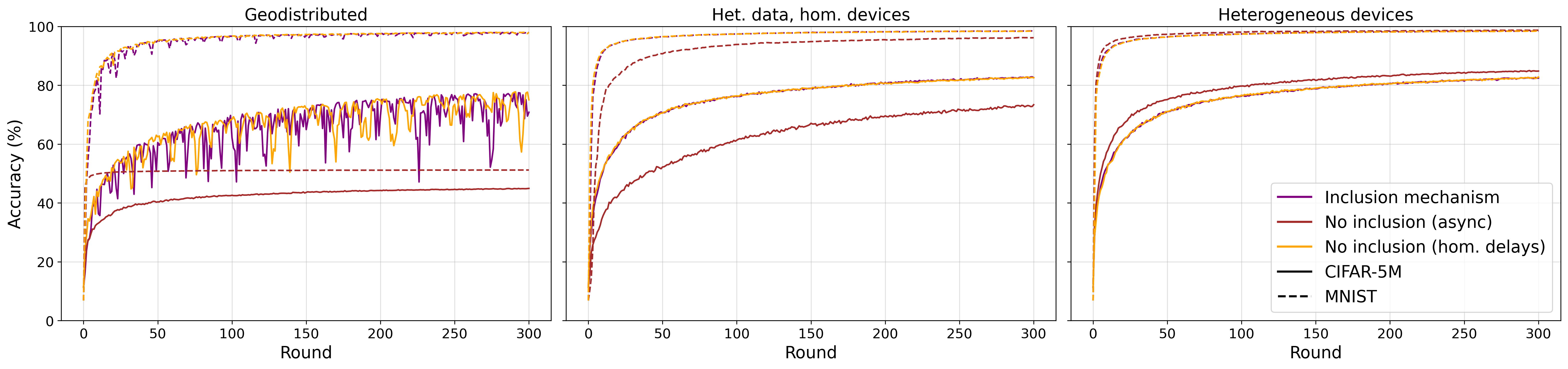}
    \caption{Accuracy of MNIST and CIFAR-5m under three heterogeneity scenarios. Each subfigure compares our protocol with the inclusion mechanism, without inclusion in an asynchronous network, and without inclusion with homogeneous delays.}
    \label{fig:exp:inclusion}
\end{figure*}

\begin{figure*}
    \centering
    \begin{subfigure}{0.32\textwidth}
        \includegraphics[width=\linewidth]{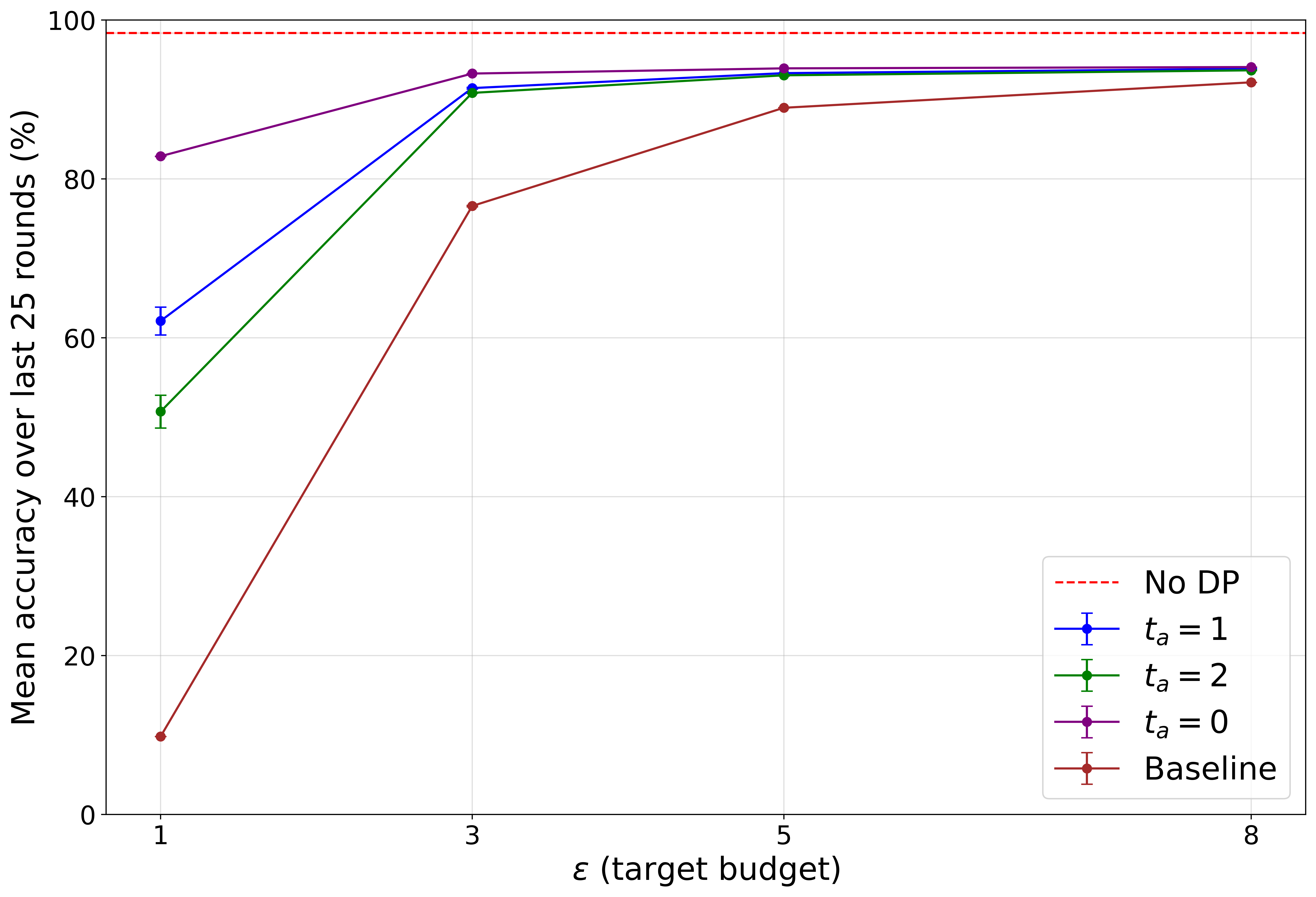}
    \end{subfigure}
    \begin{subfigure}{0.32\textwidth}
        \includegraphics[width=\linewidth]{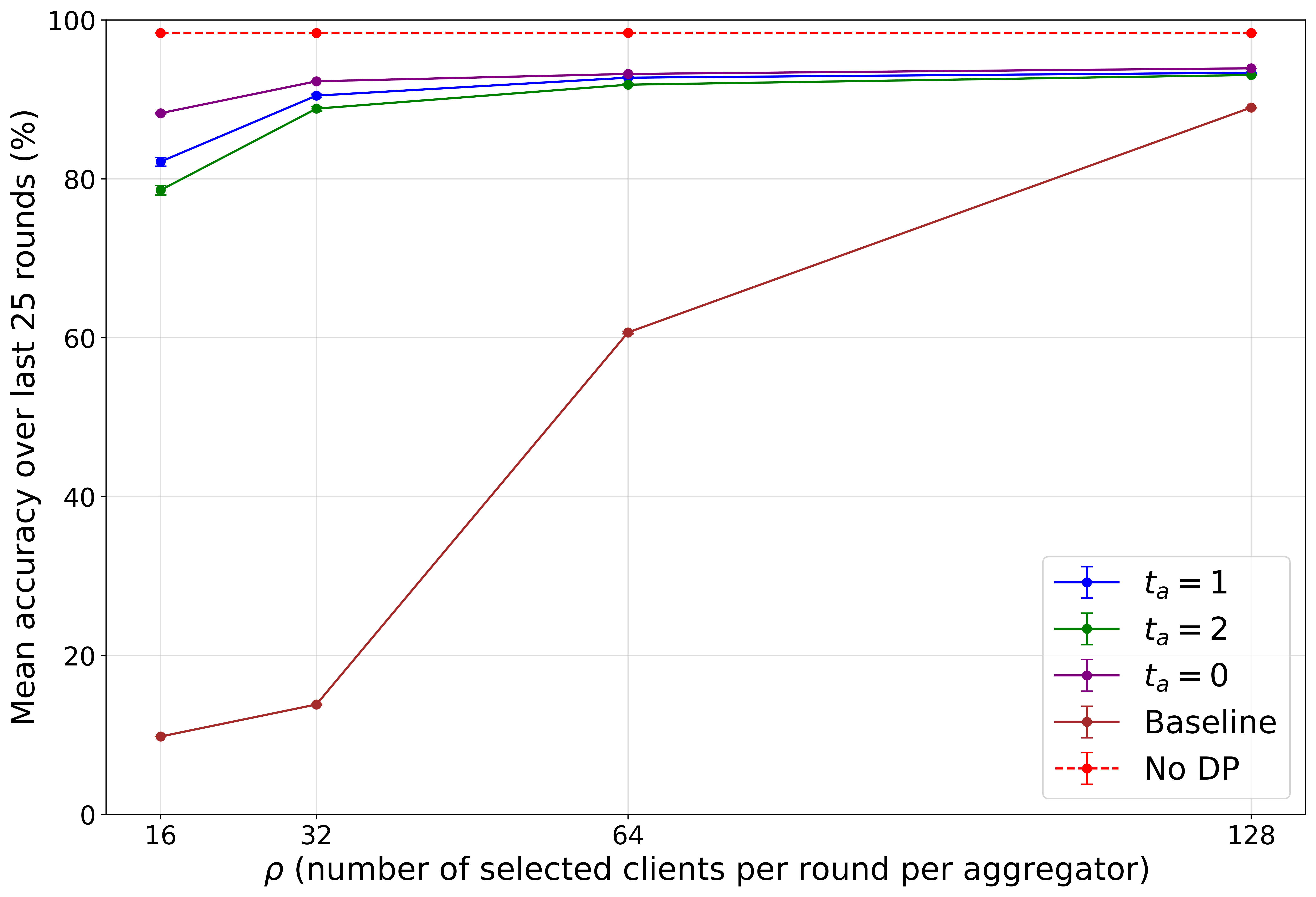}
    \end{subfigure}
    \begin{subfigure}{0.32\textwidth}
        \includegraphics[width=\linewidth]{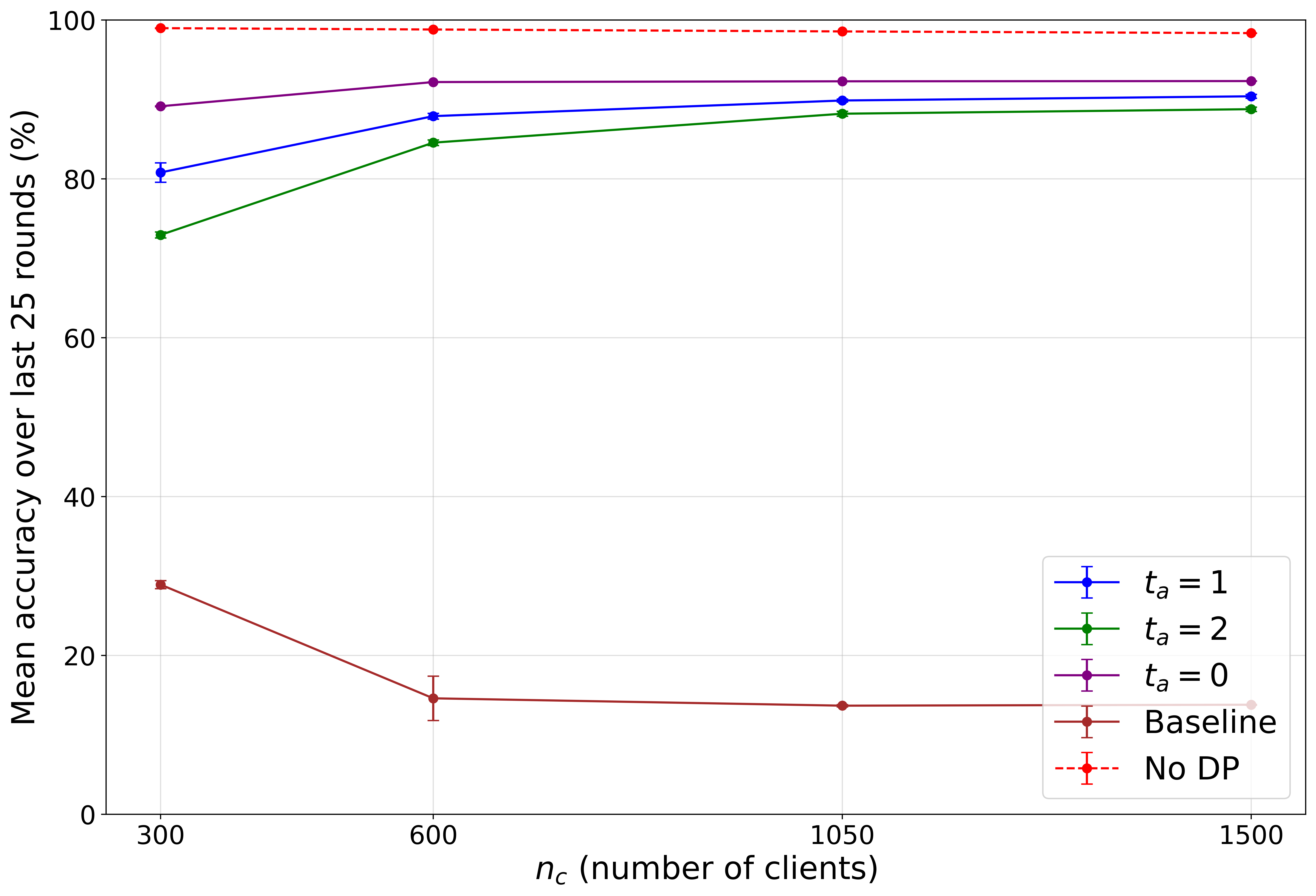}
    \end{subfigure}
    \begin{subfigure}{0.32\textwidth}
        \includegraphics[width=\linewidth]{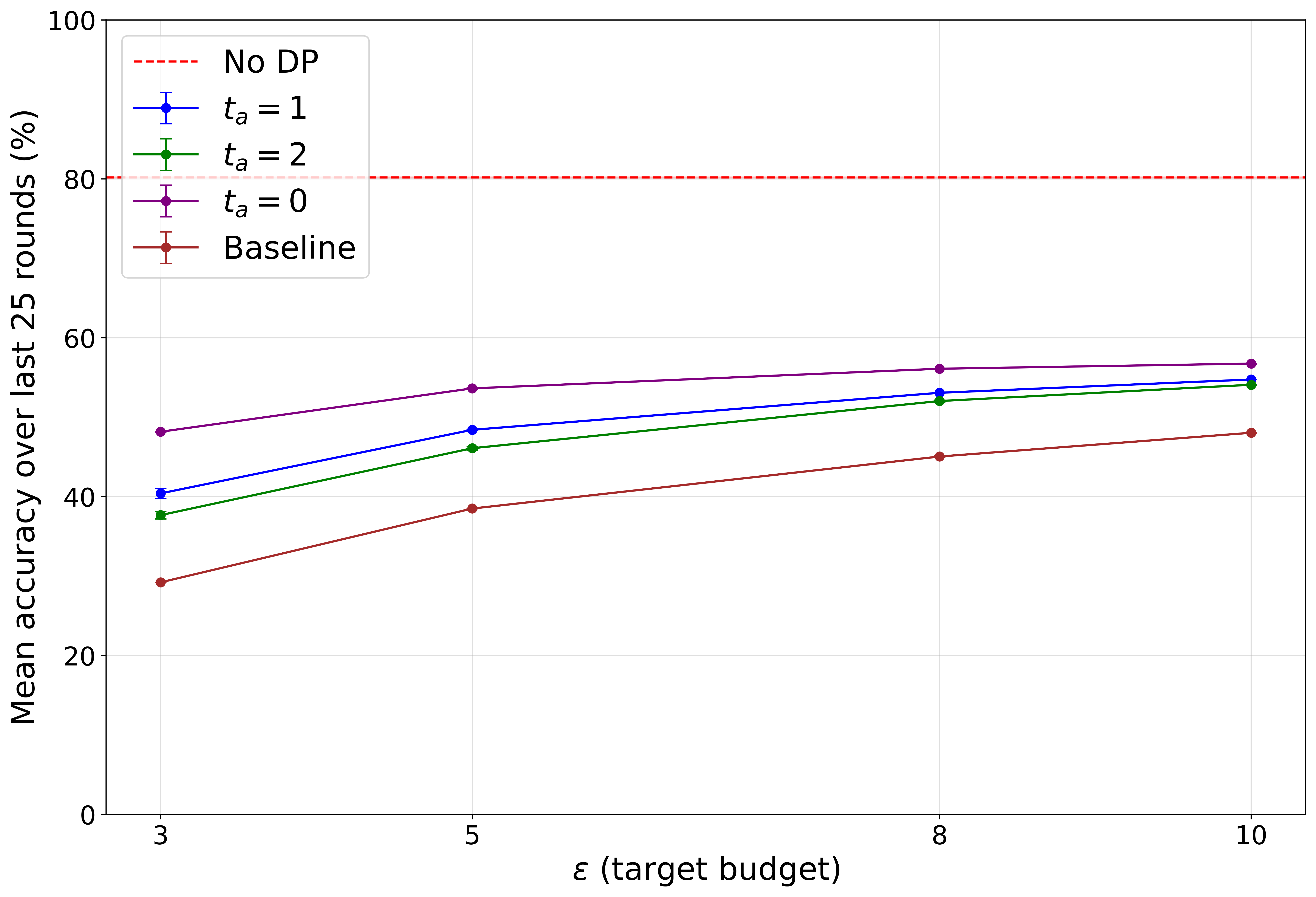}
    \end{subfigure}
    \begin{subfigure}{0.32\textwidth}
        \includegraphics[width=\linewidth]{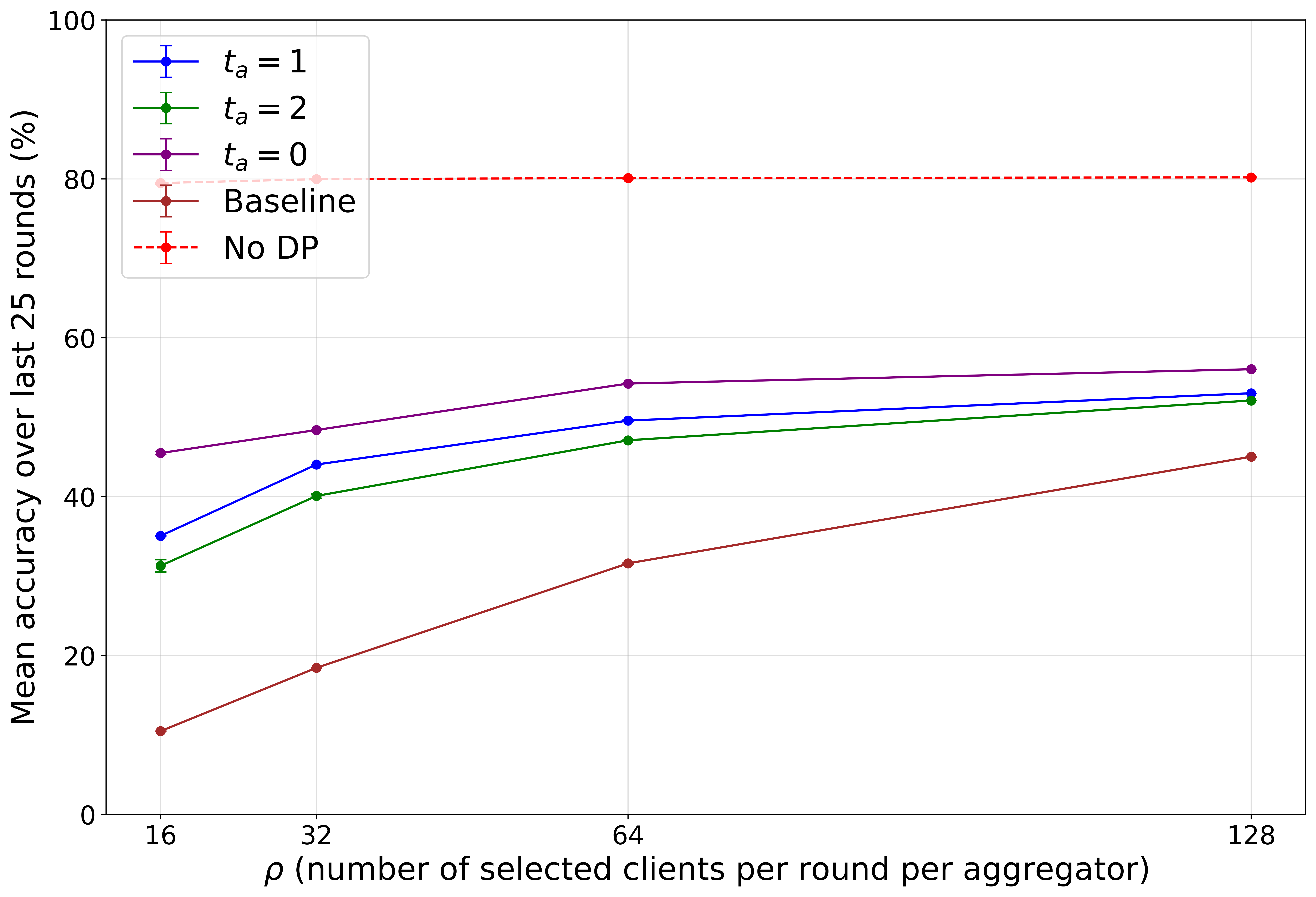}
    \end{subfigure}
    \begin{subfigure}{0.32\textwidth}
        \includegraphics[width=\linewidth]{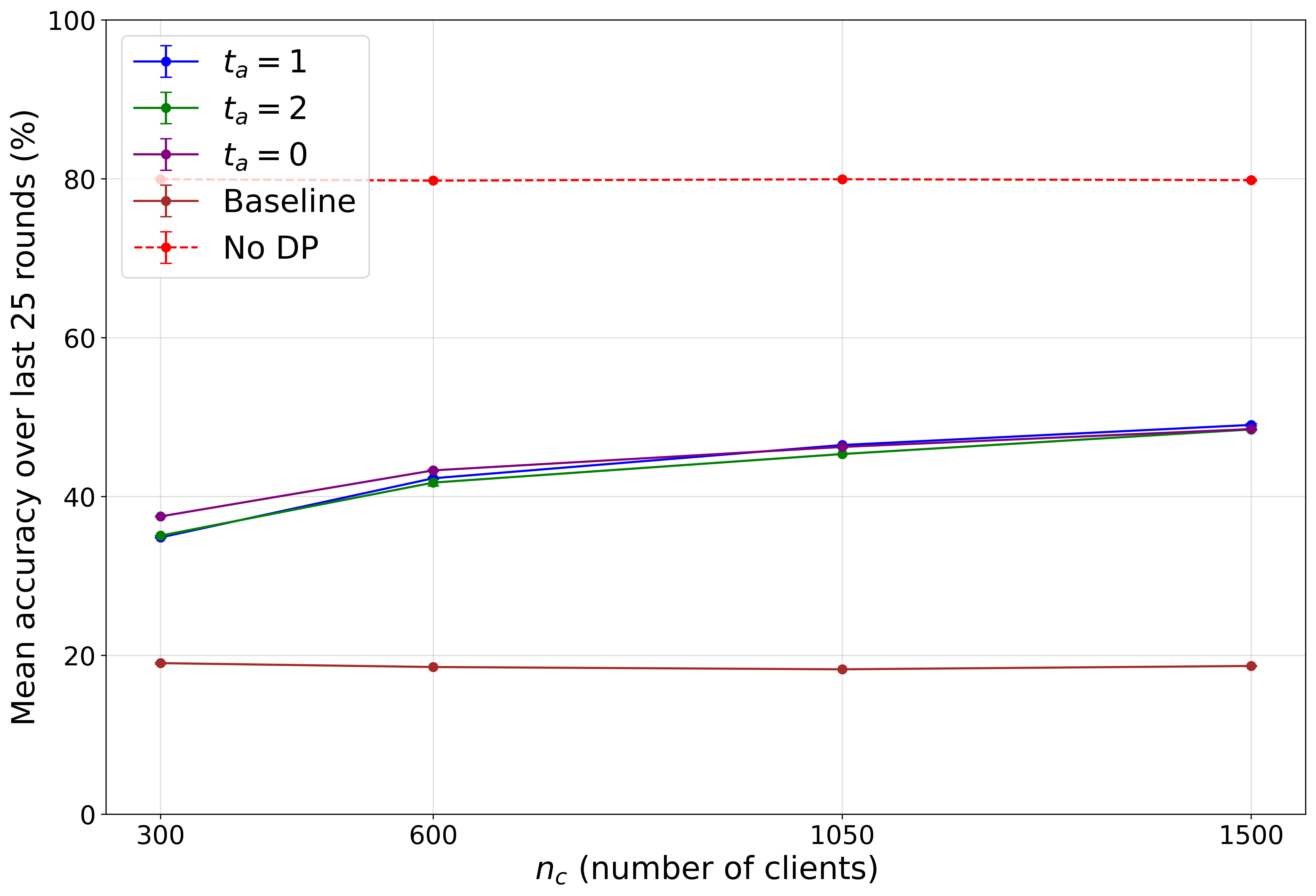}
    \end{subfigure}
    \caption{Mean accuracy over the last 25 rounds for MNIST (top) and CIFAR-5m (bottom)
        as a function of $\epsilon$ (left), $\rho$ (center), and $n_c$ (right),
        for our protocol with $t_a \in \{0,1,2\}$ vs.\ the
        baseline~\cite{SSVRCN22} (no inclusion).
        Error bars show std dev across repetitions; for $t_a>0$, aggregator
        scores are averaged per run before computing the mean and std over
        repetitions.
        Results averaged over 4 repetitions (MNIST) and 2 repetitions (CIFAR-5m).}
    \label{fig:performance}
\end{figure*}

In \Cref{fig:exp:inclusion}, we evaluate the inclusion mechanism across three scenarios that differ in the
correlation between client delays and data distribution.
For each, we compare three configurations:
(i)~our protocol \emph{with} inclusion under heterogeneous delays — 2 Gamma distributions for fast and slow clients,
(ii)~without inclusion under highly heterogeneous delays — fast clients are
systematically over-represented, and
(iii)~without inclusion but with \emph{homogeneous} delays — all clients
draw from the same Gamma distribution, so no group is structurally favored;
this serves as an upper-bound reference for what inclusion can recover.
All experiments use $n_c=1500$ clients and a single aggregator
($t_a=0$, $\rho=128$), without DP.

\paragraph{Scenario 1 — Correlated data and delay (geo-distributed).}
The first half of clients is fast and holds only classes $\{0,\ldots,4\}$;
the second half is slow and holds only classes $\{5,\ldots,9\}$. This models settings where geographic
position determines both the data type and network latency of a client (e.g.,
regional sensor networks or cross-silo FL across distant data centers).

Without the inclusion mechanism, the aggregator converges to $\approx50\%$
accuracy on MNIST and $\approx45\%$ on CIFAR-5m — learning only half the label
space. With inclusion, the aggregator reaches $98.0\%$ on MNIST and
$\approx72\%$ on CIFAR-5m, closely matching the unbiased reference.
On CIFAR-5m, both the inclusion and unbiased-reference configurations exhibit
significant accuracy oscillations (std $\approx 10\%$), which are much less
pronounced on MNIST (std $< 1\%$). The oscillations are slightly more marked
with inclusion (max drop $11.7$ pp vs.\ $9.5$ pp). 
This oscillation does not appear in the no-inclusion
asynchronous configuration, which avoids it by ignoring slow clients
entirely — at the cost of learning half the label space.

\paragraph{Scenario 2 — Data-induced delay.}
Half the clients are fast because they hold few data samples ($10\%$ of the
total), while the other half are slow because they hold many ($90\%$),
incurring longer training and transmission times per round.
This models settings where data volume is the bottleneck — e.g., cross-silo FL
between organizations of vastly different database sizes but similar hardware.

The inclusion mechanism reduces convergence time to 95\% accuracy from 158 to 24
rounds on MNIST (6.6$\times$), and improves final accuracy by $9.2$ pp on
CIFAR-5m ($73.4\%$ vs.\ $82.6\%$). The effect is more pronounced on CIFAR-5m
because the harder task amplifies the impact of excluding heavy-data clients.

\paragraph{Scenario 3 — Device-induced delay.}
Half the clients are fast, resource rich servers holding $90\%$ of the data,
while the other half are slow, low-end devices holding only $10\%$.
Here the delay is a property of the device, not of the data volume: slow
clients happen to have little data, not the other way around.
This models settings such as FL between enterprise servers and consumer IoT
devices.

In this scenario, the no-inclusion asynchronous configuration converges
slightly faster than our protocol, while our protocol matches the homogeneous
reference. This is the expected outcome: slow clients hold only $10\%$ of the
data, so the aggregator already captures a representative sample by favoring
fast clients.

\Cref{fig:exp:inclusion} shows these results on both MNIST and CIFAR-5m.
The two datasets exhibit consistent qualitative behavior, confirming the
generality of the mechanism beyond the specific task.

\subsubsection{Differential privacy and accuracy trade-off evaluation}\label{sec:exp:dp:accuracy}

\Cref{fig:performance} displays the mean accuracy over the last $25$ rounds of training while varying the parameters $\epsilon$, $\rho$, or $n_c$. Across all experiments, the baseline consistently underperforms our       
protocol                                                                  
under tight DP budgets.                                                 
Larger $\rho$ consistently improves accuracy, and higher fault tolerance  
$t_a$ trades accuracy for Byzantine resilience due to the additional noise
it requires.  
These experiments illustrate that thanks to our inclusion mechanism, we are able to introduce less noise for the same differential privacy guarantees, resulting in a faster convergence than our baseline in all cases.

\section{Conclusion}
\label{sec:ccl}
In this paper, we introduce the first privacy-preserving and fair federated learning protocol for asynchronous networks with fully Byzantine aggregators. Our protocol replicates the aggregator server and clusters clients around a coordinator server, with the clustering reshuffled at each round to cope with potential Byzantine aggregators. Reshuffling also aids convergence by mixing client updates across aggregators in the absence of per-round consensus. The protocol adapts lightweight secure masking techniques and differential privacy, while guaranteeing one-shot client communication and no client interdependency. It also employs verification mechanisms to certify the correctness of the aggregated model. Moreover, to operate in asynchronous networks, we devise a fair inclusion mechanism. This mechanism mitigates bias induced by network asynchrony and reduces the amount of noise added to the model.
Experimental results on MNIST and CIFAR-5m datasets validate our protocol, showing its ability to converge even when the data is strongly partitioned across clients. We also show that our inclusion mechanism reduces the noise added by clients to their updates, thus increasing the utility of each individual contribution, whereas FLDP \cite{SSVRCN22} under the same conditions fails to converge. Interestingly, our experiments also reveal that the additional cryptographic mechanisms used do not increase computationnal time compared to other approaches, as we leverage aggregator replication to reduce the overall amount of computation. 

\newpage
\printbibliography

\appendix

\section{Open Science}
The projects we built to test both the accuracy and the computational cost of our protocol are available at \cite{implem}. The repository is composed of two folders. The first folder contains the material for accuracy and parameter evaluation, while the second contains the material for evaluating the computational cost of the cryptographic primitives used throughout our protocol.

The CIFAR-5m dataset, which was used to provide each client with more samples than the classical CIFAR-10 dataset, is available at \cite{CIFAR5m}.

\section{Detailed Complexity Analysis}
\label{app:complexity}

This section provides a detailed analysis of the complexity figures reported 
in Table~\ref{tab:comparisonIntro}, first for the protocol presented in this paper, then for the other protocols presented in the table.

\subsection{Our Protocol}

\paragraph{Client-side.}
A client's computation consists of the following operations:
\begin{itemize}
    \item Generating the length-$N_s$ secret vector $s$ and the length-$N_g$ 
    noise vector $e_i$: $\mathcal{O}(N_g + N_s)$.
    \item Computing Shamir secret shares of $s$ for each of the $n_a$ aggregators, 
    including commitments on the $\mathcal{O}(n_a)$ polynomial coefficients: 
    $\mathcal{O}(N_s n_a)$. Note that packed secret sharing as used in 
    FLDP~\cite{SSVRCN22} cannot be applied here, since the number of shares 
    $n_a$ is intended to be small relative to $N_s$.
    \item Computing the matrix-vector product $A \cdot s$ for mask generation, 
    where $A \in \mathbb{Z}^{N_g \times N_s}$: $\mathcal{O}(N_g N_s)$.
    \item Computing commitments on the masked update $w_i + e_i$ of length $N_g$: 
    $\mathcal{O}(N_g)$.
\end{itemize}
Total client computational complexity is $\mathcal{O}(N_s n_a + N_g N_s)$,
dominated by the mask generation matrix multiplication.

Client communication consists of:
\begin{itemize}
    \item Sending the masked update with commitments to one aggregator: 
    $\mathcal{O}(N_g)$.
    \item Sending ping messages to the $n_a$ aggregators $O(1*n_a)$
    \item Sending one share of $s$ plus commitments on the polynomial coefficients 
    to each of the $n_a$ aggregators: $\mathcal{O}(N_s n_a)$.
\end{itemize}
Total client bits complexity is $\mathcal{O}(N_g + N_s n_a)$,
and client message complexity is $\mathcal{O}(n_a)$.
Crucially, all client-side metrics depend on $n_a$ rather than $n_c$, 
ensuring excellent scalability when $n_a \ll n_c$.

\paragraph{Server-side.}
Each aggregator's computation consists of:
\begin{itemize}
    \item Intra-cluster aggregation of model updates for its $\mathcal{O}(n_c/n_a)$ 
    clients: $\mathcal{O}(N_g \cdot n_c/n_a)$, summing across all $n_a$ aggregators 
    gives $\mathcal{O}(N_g n_c)$.
    \item Inter-cluster aggregation among the $n_a$ aggregators: 
    $\mathcal{O}(N_g n_a)$.
    \item Reconstruction and verification of the secret shares of $s$ for each 
    client in its cluster: for each of the $\mathcal{O}(n_c/n_a)$ clients, 
    reconstruction of the length-$N_s$ vector requires $\mathcal{O}(n_a^2)$ 
    operations per coordinate, giving 
    $\mathcal{O}(n_c/n_a \cdot N_s \cdot n_a^2) = \mathcal{O}(N_s n_c n_a)$ in total.
    \item Matrix-vector product $A \cdot \sum s$ for the aggregated mask: 
    $\mathcal{O}(N_g N_s)$.
\end{itemize}
Total server computational complexity is 
$\mathcal{O}(N_g n_c + N_g n_a + N_s n_c n_a + N_g N_s)$.

Server communication consists of:
\begin{itemize}
    \item Broadcasting commitments on $w_i + e_i$ for its $\mathcal{O}(n_c/n_a)$ 
    clients to the $n_a$ aggregators: $\mathcal{O}(N_g n_c)$.
    \item Sending intra- and inter-cluster model updates to the $n_a$ aggregators: 
    $\mathcal{O}(N_g n_a)$.
    \item Forwarding secret shares of $s$ for its cluster clients to the $n_a$ 
    aggregators: $\mathcal{O}(N_s n_c)$.
    \item Sending client participation lists to the $n_a$ aggregators: 
    $\mathcal{O}(n_c n_a)$.
    \item Broadcasting the final aggregated model to all $n_c$ clients: 
    $\mathcal{O}(N_g n_c)$.
\end{itemize}
Total server bits complexity is 
$\mathcal{O}(N_g(n_c + n_a) + n_c(N_s + n_a))$,
and server message complexity is $\mathcal{O}(n_a + n_c)$.

In practice, rather than each aggregator pushing the final model to all $n_c$ 
clients --- which would multiply the $\mathcal{O}(N_g n_c)$ broadcast cost by 
$n_a$ --- aggregators instead notify clients that the model is available, and 
clients fetch it from any $t_a + 1$ aggregators of their choice. Since 
$t_a + 1 = \mathcal{O}(n_a)$, the client fetch cost is $\mathcal{O}(N_g n_a)$. The server notification cost is $\mathcal{O}(n_c)$. 
Asymptotically, this optimization does not change the complexity bounds, 
but significantly reduces the practical communication burden on aggregators.

\subsection{Baseline Protocols}

\paragraph{Bonawitz et al.~\cite{BIKMMPRSS17}.}
Each client communicates with all $n_c$ clients for pairwise mask negotiation, 
giving client bits complexity $\mathcal{O}(N_g + n_c)$, message complexity 
$\mathcal{O}(n_c)$, and computational complexity $\mathcal{O}(n_c(N_g + n_c))$. 
The server aggregates all masked updates and reconstructs dropout seeds, 
giving server bits complexity $\mathcal{O}(n_c(N_g + n_c))$, message complexity 
$\mathcal{O}(n_c)$, and computational complexity $\mathcal{O}(N_g n_c^2)$. 
The quadratic terms in $n_c$ make this impractical at scale.

\paragraph{Bell et al.~\cite{bell2020secure}.}
Replacing the complete graph with a $k$-regular graph of degree $\mathcal{O}(\log n_c)$ 
reduces client bits complexity to $\mathcal{O}(N_g + \log n_c)$, message complexity 
to $\mathcal{O}(\log n_c)$, and computational complexity to 
$\mathcal{O}(\log n_c (N_g + \log n_c))$. Server bits complexity becomes 
$\mathcal{O}(n_c(N_g + \log n_c))$, message complexity $\mathcal{O}(n_c)$, 
and computational complexity $\mathcal{O}(n_c(\log^2 n_c + N_g \log n_c))$.

\paragraph{FLDP~\cite{SSVRCN22}.}
Client bits complexity is $\mathcal{O}(N_g + N_s + n_c)$, message complexity 
$\mathcal{O}(n_c)$, and computational complexity $\mathcal{O}(n_c \log n_c + N_g N_s)$, 
where the $n_c \log n_c$ term arises from the MPC protocol for aggregating masks. 
Server bits complexity is $\mathcal{O}(N_g n_c + N_s)$, message complexity 
$\mathcal{O}(n_c)$, and computational complexity 
$\mathcal{O}(N_g(n_c + N_s) + n_c \log n_c)$.

\paragraph{FlexScaAgg~\cite{flexible24}.}
The original protocol uses exactly two entities (server and initiator). 
Each client computes an E2 commitment on its length-$N_g$ gradient and sends 
its masked gradient to the server, giving client bits complexity $\mathcal{O}(N_g)$, 
message complexity $\mathcal{O}(1)$, and computational complexity $\mathcal{O}(N_g)$.
The server receives masked gradients and commitments from all $n_c$ clients and 
interacts with the initiator to compute the aggregation result, giving server bits 
complexity $\mathcal{O}(N_g n_c)$, message complexity $\mathcal{O}(n_c)$, 
and computational complexity $\mathcal{O}(N_g n_c)$.

In Table~\ref{tab:comparisonIntro}, we generalize FlexScaAgg to $n_a$ 
non-colluding entities rather than fixing $n_a = 2$, in order to enable 
a fair comparison with our protocol. Under this generalization, each client 
must send its masked gradient and commitments to each of the $n_a$ entities, 
giving client bits complexity $\mathcal{O}(N_g n_a)$, message complexity 
$\mathcal{O}(n_a)$, and computational complexity $\mathcal{O}(N_g n_a)$. 
Server bits complexity becomes $\mathcal{O}(N_g(n_c + n_a))$, message complexity 
$\mathcal{O}(n_a + n_c)$, and computational complexity $\mathcal{O}(N_g(n_c + n_a))$.

\subsection{Asymptotic Simplification}

Under the practical assumptions $n_a = \mathcal{O}(\log n_c)$ and 
$N_s \log n_c = o(N_g)$ --- which holds in practice since the LWE mask size 
$N_s$ is typically a few hundred to a few thousand dimensions while $N_g$ 
reaches millions of parameters in modern FL deployments --- the complexities 
simplify as follows. The term $N_s n_c n_a = N_s n_c \log n_c$ satisfies 
$N_s \log n_c = o(N_g)$, hence $N_s n_c n_a = o(N_g n_c)$ and is absorbed 
into $\mathcal{O}(N_g n_c)$. Similarly, $N_g n_a = N_g \log n_c = o(N_g n_c)$ 
and $n_c n_a = n_c \log n_c$ are both dominated by $N_g n_c$. 
This yields simplified complexities matching those discussed in 
Section~\ref{sec:relatedwork}.
\section{Complete algorithm and edge cases management}

\subsection{Edge cases of the inclusion mechanism} \label{sec:edge:cases}

In this section, we detail two edge cases of the inclusion algorithm that were discussed in \Cref{sec:inclusion}.

\paragraph{Wasted clusters}
The assignment algorithm assigns clients to aggregators in an expected uniformly random manner. Thus, each round, some aggregators have a probability to be assigned to clients that have crashed (or slow enough to appear as crashed). As our privacy preservation mechanism requires aggregators to include a fixed number ($\rho$) of client updates to aggregate each round, when more than $k-\rho -1$ crashed clients are assigned to an aggregator $a_i$, this aggregator will not be able to aggregate clients' values this round.

The solution for the aggregator is simply to inform other aggregators that it does not participate this round, and to wait for the next round to start. Thanks to the expected random shuffling of assigned clients, in the next rounds, $a_i$ will be associated with new clients, and, in expectation, it will receive at least $\rho$ model updates.
If a correct aggregator detects that it may be assigned to a ``wasted'' cluster, i.e., if it did not receive $\rho$ gradients from its assigned clients while it already received $n_a - t_a$ \clientunificationmsg messages from the other aggregators, then it informs the others that it will not participate during this round by sending a \wastedmsg message. This aggregator will still help the other aggregators in the reconstruction of their cluster-level gradient, but it will not send any \aggregatemsg message for its own cluster-level's gradient. A ``wasted'' aggregator at round $\tau$ still produces a model at round $\tau$ using the averaged gradients of other aggregators during the inter-cluster aggregation phase.
Furthermore, once a \wastedmsg message is received by $a_i$ from an aggregator $a_j$, no aggregation request nor cluster-level gradient will be waited from it. Therefore, this mechanism does not impact non-wasted aggregators when they train, and as shown in \Cref{sec:additionnal:expe}, it does not impact overall training accuracy.

\paragraph{Inclusion fraud detection}
Byzantine processes can voluntarily bias their client inclusion by including in priority some clients rather than others, thereby biasing their own models due to data heterogeneity, but also biasing others aggregators' models in the inter-cluster aggregation phase. To solve this problem, we add a fraud detection mechanism. 
If an aggregator $a_i$ is detected by another aggregator $a_j$ as voluntarily biasing its inclusion of clients during round $\tau$, then $a_j$ will locally mark $a_j$ as \emph{blamed}. A correct aggregator does not help a blamed aggregator to reconstruct and aggregate its gradients during the intra-cluster aggregation stage.
If a correct aggregator blames another aggregator $a_k$, either enough correct aggregators blame $a_k$, or $a_k$ proposed two different sets of included clients. In the first case, the correct aggregators that detect the bias will not participate in the reconstruction of the update. Thus, $a_k$ will not be able to reconstruct its cluster-level gradient, hence, it will not be able to introduce bias in the final result. In the second case, either one of the set does not introduce bias, and enough correct aggregators received it, thus $a_k$ will be seen as a correct aggregator and its unbiased cluster-level gradient will be reconstructed, or one of the sets is biased and as earlier, $a_k$ cannot participate anymore this round.

To detect a fraudulent client inclusion, we compute the maximum difference between the least frequently included and the most frequently included clients ($\Delta(\lambda)$), allowing a tolerance of $\maxDelta$. If an aggregator $a_i$ includes clients with an absolute difference that is higher than $\maxDelta$, then, when it asks for summation of those clients' gradients to the other correct aggregators, those aggregators, that share the same view on which clients $a_i$ included in previous rounds thanks to the $n_a>3t_a$ threshold, will all locally \emph{blame} $a_i$. Thus, no correct aggregator will help reconstruction $a_i$ intra-cluster update.

However, the inclusion algorithm we use is stochastic. Thus, the blaming mechanism can also impact correct aggregators as the probability they include biased clients in their sets is non zero. To solve this problem, correct aggregators detect on their own set of included clients if they can be perceived as fraudulent. If they do, they declare themselves ``AUTO-BLAMED'' and wait for the following round to be able to participate again, just as if their cluster was wasted. 

\subsection{Complete algorithm and detailed description}\label{sec:complete:algorithm}

\begin{algorithm*}
\smallskip
\small
\nonl\Init{$A \gets $ a public matrix of size $n\times m$ used to hide the update $w_i$;\\
$C \gets$ Clipping parameter.}
\nl\Function{$\CreateMaskedModel(g^{\tau}_{i})$}{
        $s \gets \mathsf{GenerateRandomSecret}()$; \quad
        $e \gets \mathsf{DrawRandomNoise}()$; \label{line:LWE:cli:gen:mask:noise}\\
        $\bar{g}_{i}^{\tau}\gets g_{i}^{\tau}/\max(1, \frac{||g_{c_i}^{\tau}||_2}{C})$;\label{line:clipping}\\
        $h \gets \bar{g}^{\tau}_i + A\cdot s + e$;\label{line:LWE:cli:masking}\\
        $\texttt{SecretShares} \gets (n_a, n_a-t_a)$-$\SSShare(s)$;\label{line:LWE:cli:secret:share}\quad\\
        $\VSSPartialProof[j] \gets \commit(\texttt{SecretShare}[j]), \ \forall j \in \{1, \cdots, n_a\}$;\label{line:cli:vsspartial:proof}\\
        $\VSSProof \gets \commit( g^{\tau}_i + e)$;\label{line:cli:vss:proof}\\
        $\EncSecretShares\gets \emptyset^{n_a}$;\\
        \For{$a_\ell \in \mathcal{A}$}{
            $\EncSecretShares[\ell] \gets \Encrypt((\tau, c_i, \texttt{SecretShares}[l], \VSSProof, \Sign(($\\ \nl\ \ \ \ $\tau, c_i, \texttt{SecretShares, \VSSProof}), \SKSig_{c_i})), \SKEnc_{(a_\ell, c_i)})$;\label{line:LWE:cli:encrypt}
        }
        $\sigma_{h} \gets \Sign(\tau, h, \SKSig_{c_i})$;\label{line:cli:h:sign}\\
        \return $(h,\sigma_h, \EncSecretShares, \VSSProof, \VSSPartialProof)$;
    }
\WhenReceived{$\trainmsg<\tau, \widehat{W}^{\tau}_{\text{proposed}}, \sigma^\tau_{\widehat{W}^{\tau}_{\text{proposed}}}>$\label{line:receive:model}}{
        $a_j \gets \assigned(\tau, c_i)$;\label{line:cli:assigned}\\
        $g^{\tau}_i \gets  \nabla F_{i}(w^{\tau-1}_i, \xi^{\tau}_i)$;\label{line:compute:gradient}\\
        $(h, \sigma_{h},\EncSecretShares, \VSSProof, \VSSPartialProof) \gets \CreateMaskedModel(g^{\tau}_{i})$;\\
        \send $\updatemsg\langle \tau, h, \sigma_{h}, \Sign(\tau, \SKSig_{c_i}), \EncSecretShares, \VSSProof, \VSSPartialProof \rangle$ to aggregator $a_j$;\\
        \broadcast $\pingmsg\langle \tau, \Sign(\tau, \SKSig_{c_i})\rangle$ to all aggregators in $\mathcal{A}$ except $a_j$.
    }
\caption{Asynchronous privacy preserving federated learning algorithm with aggregator fault resilience (for client $c_i$).}
\label{alg:async:client}
\end{algorithm*}

In this section, we provide the complete privacy preserving learning algorithm with replicated aggregators. This algorithm, is presented in eight parts in Algorithms \ref{alg:async:client}, \ref{alg:wasted:detection:uniform}, \ref{alg:blaming:uniform}, \ref{alg:select:async}, \ref{alg:prepare:agg:full}, \ref{alg:async:aggregator:1} and \ref{alg:async:aggregator:2}.

\subsubsection{Detailed client's algorithm description}
\Cref{alg:async:client} presents client's main algorithm. In this algorithm, the main function is triggered when the client $c_i$ receives a $\trainmsg$ message at round $\tau$. This message includes the initial model if $\tau = 0$ or the model computed during the previous round $\tau-1$ and a proof of correct computation of this model otherwise (line \ref{line:receive:model}). $c_i$ first registers its coordinator this round using the $\assigned$ function (line \ref{line:cli:assigned}).
Then, it computes a gradient on the model it received and on its local data (line \ref{line:compute:gradient}) and it invokes the \CreateMaskedModel~function. 

The \CreateMaskedModel~function has two parts. The first part of the function aims at hiding the client's gradient, while the second part is used to provide material to unmask this hidden client's gradient. In the first part of the \CreateMaskedModel~function, $c_i$ first generates $s$, the mask of the gradient and $e$, the DP noise (line \ref{line:LWE:cli:gen:mask:noise}). It then clips the gradient (line \ref{line:clipping}), i.e., it limits the maximum norm of the gradient to the clipping parameter $C$, such that DP can be ensured. Then, the noise and the mask are added to the clipped gradient (line \ref{line:LWE:cli:masking}). 

In the second part of the \CreateMaskedModel~function, $c_i$ creates $n_a$ shares of its mask $s$ (line \ref{line:LWE:cli:secret:share}). Those shares will later be summed by the aggregators, and the sum of the shares will be interpolated to retrieve the sum of the masks of $\rho$ clients. This is the main improvement that was proposed by FLDP \cite{SSVRCN22} over other secret sharing-based FL algorithms \cite{bell2020secure}, as it reduces the size of the vector we are working on. As shown in \Cref{sec:exp:analyzes}, $s$ has a size $N_s$ between $10^2$ and $10^3$ for 128 bit security and is independent of the size $N_g$ of $g$. Therefore, we can work and exchange mostly values related to $s$ rather than $g$ and reduce the overall communication and computation complexity of the protocol. Once the shares have been computed, $c_i$ produces the two proofs \VSSProof~and \VSSPartialProof~(lines \ref{line:cli:vsspartial:proof} and \ref{line:cli:vss:proof}). The first one is an additionally homomorphic commitment to $c_i$'s unmasked (but noisy) gradient, that will be used by the aggregators to ensure that cluster-level gradients are computed accordingly to the protocol. The second one is an additionally homomorphic commitment to each share. It will be used by $a_j$, the coordinator of $c_i$'s cluster, to ensure that no aggregator misbehave by sending a wrong sum of client's shares. Those proofs makes it possible to ensure that the Byzantine aggregators cannot undermine neither liveness nor safety of the protocol.

Once the proofs have been computed, $c_i$ encrypts the data destined to each aggregator with their shared symmetric encryption key (line \ref{line:LWE:cli:encrypt}). For each aggregator $a_l$, $c_i$ encrypts the round number $\tau$, its identity, the secret share $\texttt{SecretShare}[l]$ of the mask $s$ destined to $a_l$, and \VSSProof. $c_i$ adds a signature of those elements to avoid tampering. This encrypted message will be sent to $a_l$ using the coordinator $a_j$ as a proxy. Therefore, $a_j$ may perform attacks such as replay attacks. To avoid this behaviour, the round number and the client's identity are encapsulated in the encrypted message. The secret share, on the other hand, is encrypted to ensure that $a_j$ is not able to reconstruct the unaggregated $s$ value, which would break user's privacy.

$c_i$ signs $h$ to avoid a Byzantine coordinator from selecting an arbitrary $h$ during the inter-cluster phase (line \ref{line:cli:h:sign}). This additional signature is required as the \VSSProof~only ensures that $s$ is well reconstructed, but no information is given on the unmasked model. Therefore, aggregators can verify integrity of the $h$ value shared by the aggregator using $\sigma_h$ and they can legitimately compute the unmasked sum of gradients.

$h$, $\sigma_h$, $\VSSProof$ and $\VSSPartialProof$ are sent to $a_j$, the coordinator of $c_i$'s cluster for round $\tau$, using an $\updatemsg$ message. Additionally, $c_i$ sends a signature of $\tau$ to $c_i$ to act as a $\pingmsg$ message that $a_j$ will be able to share with other aggregators during the unification phase. Finally, $c_i$ sends a $\pingmsg$ message to each aggregator to help them detect if they can begin the intra-cluster aggregation phase, or if they are wasted. 

\subsubsection{Detailed aggregator's algorithm description}

\Cref{alg:async:aggregator:1} and \Cref{alg:async:aggregator:2} present2 aggregator's main algorithm. The aggregator $a_i$ begins by invoking the $\mathsf{Train}()$ operation, which sends the first $\trainmsg$ message to the clients, with an initial model that can either be randomly selected, or that can be a parameter of the algorithm (line \ref{line:agg:initial:train}).

Afterward, $a_i$ waits for $\pingmsg$ and $\updatemsg$ messages. Once it receives a $\pingmsg$ message, it stores the associated signature of the round number to use it during the unification phase (line \ref{line:agg:ping:store:1}). $\pingmsg$ messages makes it possible for $a_i$ to know when $n_c-t_c$ clients participated, thus it allows $a_i$ to know if it can still wait more message from its cluster's clients, or if it has to declare itself ``WASTED''.

When $a_i$ receives an $\updatemsg$ message from client $c_j$, it first stores all the element that it received (lines \ref{line:agg:update:store:begin} to \ref{line:agg:update:store:end}). Importantly, the $\updatemsg$ message is also recorded as a $\pingmsg$ (line \ref{line:agg:update:store:end}). It contains a $\sigma_\pingmsg^{\tau}$ element that is a signature of the round number by $c_j$. This signature is important as $c_j$ do not send  a $\pingmsg$ message to its cluster's coordinator. Hence, it allows $a_i$ to count the clients that already participated in and outside its cluster.

Once this recording is done, $a_i$ checks if it can wait for more messages from clients (line \ref{line:agg:update:ping:check}) by checking if it received at least $n_c-t_c$ $\pingmsg$ messages. If it is the case, $a_i$ begins the unification phase (line \ref{line:agg:update:unification:phase}) by sending a $\clientunificationmsg$ message. The goal of this message is to share with other aggregators the list of participating clients. Because $\pingmsg$ messages are sent after $\updatemsg$ messages by clients and because clients cannot be Byzantine, once an aggregator $a_k$ knows a client from its cluster sent a $\pingmsg$ message to another aggregator, it knows that it will eventually receive the $\updatemsg$ message from this client and that it can wait for it. This unification phase makes it possible to avoid inter-locked situations where each aggregator sees the $\pingmsg$ messages of other aggregator's clusters, but none of them (or not enough to reach the $\rho$ threshold) are from its cluster. When $a_i$ receives a $\clientunificationmsg$ message, it verifies the set of signature contained in the message, and it records each unknown signature in its own $\pinglist$ (lines \ref{line:agg:unification:begin} to \ref{line:agg:unification:end}). 

Otherwise, if $a_i$ did not receive enough $\pingmsg$ messages, it waits for the additional clients' participations that are ensured to eventually arrive thanks to our fault model.

Afterward, in both cases (lines \ref{line:agg:prepare:agg:1} or \ref{line:agg:prepare:agg:2}), $a_i$ verifies if it received enough $\clientunificationmsg$ messages, and if it did, it executes the $\PrepareAggregation$ function.

The goal of the $\PrepareAggregation$ function, presented in \Cref{alg:prepare:agg:full}, is to perform the inclusion mechanism on the participations $a_i$ received, and to prepare the material necessary for each aggregator to perform the intra-cluster aggregation. First $a_i$ verifies if it received enough ($\rho$) participations to require an aggregation using the $\wasteddetection$ function (line \ref{line:agg:wasted:detection:prep:agg}). If it does not, it broadcasts a $\wastedmsg$ message so that other aggregators do not wait for its $\aggregatemsg$ message. If $a_i$ received enough participations on the other hand, it gathers them (lines \ref{line:agg:prep:agg:gather:deb} to \ref{line:agg:prep:agg:gather:end}). Then, it performs the inclusion on this set of client (line \ref{line:agg:prep:agg:inclusion}) and verifies if the resulting set fulfills the expected uniform inclusion of client thanks to the $\blaming$ function (line \ref{line:agg:auto:blaming:detection}). This function checks if the most included client and the least included client of $a_i$ have a difference of participation of more than $\maxDelta$. This function ensure that no Byzantine process bias the inclusion mechanism. If the function returns $\ttrue$, then correct aggregators may see $a_i$ as a Byzantine aggregator. It therefore send an $\autoblamingmsg$ message, saying it cannot participate this round. $\autoblamingmsg$ messages are treated as $\wastedmsg$ messages.
Otherwise, $a_i$ can proceed. It homomorphically sums the $\VSSProof$ that are associated to participations of included clients for later verification (line \ref{line:agg:prep:agg:full}). Finally it prepares each encrypted share for each aggregator, and sends it using a $\aggregatemsg$ message (lines \ref{line:agg:prep:agg:deb:send} to \ref{line:agg:prep:agg:end:send}).
\begin{algorithm}
\small
\Function{$\wasteddetection()$}{
    $\ParticipatingClients \gets \emptyset$;\\
    \For{$(c_k, \star, \star) \in \pinglist_i^{\tau}$}{ 
        \lIf{$c_k \in \assign(\tau)[i]$}{ 
            $\ParticipatingClients \gets \ParticipatingClients \cup c_k $
        }
    }
    
    \If{$|\ParticipatingClients| < \rho $ and at least $n_a-t_a$ \clientunificationmsg messages have been received}{
        \return \ttrue
    }
    \return \ffalse
}

\caption{Function to detect ``wasted'' clusters (for $a_i$)} 
\label{alg:wasted:detection:uniform}
\end{algorithm}

\begin{algorithm}
\small
\Function{$\blaming(\lambda)$}{
    $\lambda' \gets$ the $n_c-2t_c$ most frequent values in $\lambda$\; 
    \lIf{ $\Delta(\lambda')>\maxDelta$}{
        \return \ttrue
    }
    \return \ffalse
}

\caption{Blaming function where $\Delta(\lambda')$ outputs the maximum frequency difference in $\lambda'$ (for $a_i$)}
\label{alg:blaming:uniform}
\end{algorithm}

\begin{algorithm}
\small
\Function{$\select(\lambda, S)$}{
    $S_{\texttt{sorted}}\gets$ $S$ sorted in ascending order from fewer to most appearances in $\lambda$;\\
    \lIf{$S\ge \rho$}{
        \return the $\rho$ first elements of $S_{\texttt{sorted}}$
    }
    \return \ffalse
}
\caption{Inclusion function (code for $a_i$).} 
\label{alg:select:async}
\end{algorithm}

When $a_i$ receives a $\aggregatemsg$ message from $a_j$, it first waits for the previous $\aggregatemsg$ or $\wastedmsg$ or $\autoblamingmsg$ message from $a_j$ (line \ref{line:agg:sumshare:wait}). This makes it possible to know which clients $a_j$ selected in previous rounds, such that the inclusion fraud detection mechanism can work. Afterward, $a_i$ verifies that $a_j$ client inclusion is correct (lines \ref{line:agg:sumshares:verify:inclusion:deb} to \ref{line:agg:sumshares:verify:inclusion:end}), i.e., it first check that $a_j$ included exactly $\rho$ client's updates, then it checks if the inclusion mechanism was properly performed using the $\blaming$ function. If those steps pass, $a_i$ updates the variable  $\Lambda_i[j]$ that stores the frequency of inclusions of clients by $a_j$ (line \ref{line:agg:sumshare:lambda:update}).  
Then, $a_i$ decrypts the shares of clients selected by $a_j$ using their shared symmetric key (line \ref{line:agg:sumshare:decyrpt}). It sums the decrypted shares and the associated $\VSSProof$ (lines \ref{line:sumshare:sum:vss:proof} and \ref{line:sumshare:sum:shares}). Finally, it threshold-signs the sum of $\VSSProof$ and sends everything back to $a_j$ using a $\disseminatemsg$ message (lines \ref{line:sumshare:sign:vss:proof} and \ref{line:sumshare:send:disseminate}).

The goal of the $\disseminatemsg$ message is for the aggregators to collect sum of shares of their cluster, and combine those sum of shares to unveil the sum of unmasked gradients of their selected clients in their cluster.
The first action $a_i$ takes when it receives an $\disseminatemsg$ message from $a_j$ is to verify the $\sigma_\texttt{proof}$ signature, which is the threshold signature by $a_j$ of $\VSSProof$, and that will be used in the following phase by aggregators to verify that the intra-cluster phase was conducted following the protocol. Those $\VSSProof$ signatures proves compliance to the protocol as $a_i$ is required to aggregate at least $n_a-t_a$ of them, which implies that at least one correct aggregator did participate in the intra-cluster phase. Furthermore, $a_i$ verifies that the commitment to the sum of shares sent by $a_j$ is equal to the sum of commitments to shares (the $\VSSPartialProof$) that was received from clients. This ensures that $a_j$ summed the shares and did not provide arbitrary elements which would modify the reconstructed gradient (line \ref{line:agg:intra:cluster:verification:proofs}). Then, it stores the sum of shares and the signature of the proofs until it receives $n_a-t_a$ valid $\disseminatemsg$ messages (line \ref{line:agg:condition:intra:cluster:na:ta}). At this point, $a_i$ reconstruct the sum of the masks $s$ (line \ref{line:intra:cluster:reconstruct:s}) and combines the signatures to obtain a constant size proof of the good reconstruction of the sum of $s$ values (line \ref{line:intra:cluster:sig:gather}). Those elements are finally broadcast to aggregators using a $\gathermsg$ message, along with the masked gradients $h$ of included clients ($\MaskedUpdates_i^{\tau}$), and the signature by the clients of those masked gradients ($\SigMaskedUpdates_i^{\tau}$).

The aggregators now have to collect the sum of the unmask gradients of other aggregators clusters, sum those different gradients and request a certification from other aggregators.
This is done when $a_i$ receives an $\gathermsg$ message from $a_j$. It verifies the signature $\sigma_{\widehat{s}_j^{\tau}}$ that certifies that at least one correct aggregator participated in the intra-cluster phase, and thus that this phase was conducted accordingly with the protocol, and that no two different intra-cluster gradients exist for $a_j$'s cluster (line \ref{line:agg:gather:sig:verify}). Indeed, if $a_j$ is able to build two different signatures  $\sigma_{\widehat{s}_j^{\tau}}$ and $\sigma_{\widehat{s'}_j^{\tau}}$ for two different intra-cluster gradients $\widehat{s}_j^{\tau}$ and ${\widehat{s'}_j^{\tau}}$, it means $a_j$ was able to find two sets of $n_a-t_a$ aggregators willing to sign those two different intra-cluster gradients. However, because correct aggregators will not sign two such different intra-cluster gradients, if we assume $n_a -2t_a$ correct aggregators signed the intra-cluster gradient $\widehat{s}_j^{\tau}$, then only $n_a -(n_a-2t_a)=2t_a < n_a-t_a$ (as $n_a>3t_a)$ will sign the second inter-cluster gradient ${\widehat{s'}_j^{\tau}}$, thus the second aggregated threshold signature cannot exist, and a Byzantine aggregator cannot equivocate.
$a_i$ also verifies the signatures of the sum of masked gradients to ensure no tampering was performed by $a_j$.
Once those verifications passed, $a_i$ stores the received elements until it receives the maximum number of such $\gathermsg$ messages it can expect. $a_i$ can expect $n_a-t_a$ messages, but it has to take into account wasted and auto-blamed aggregators, and remove them from the count.
Therefore, once $a_i$ received $ n_a-t_a - \wastedcnt_i^{\tau}$ valid $\gathermsg$ messages, it can compute its inter-cluster update (line \ref{line:agg:gathermsg:compute:final:update}) and send all the material that was used to compute and verify the inter-cluster update to the other aggregators for certification using a $\certifymsg$ message.

The goal of the certification phase is to verify that the inter-cluster gradient was computed following the protocol. Namely, it aims at verifying the same elements as in the previous phase, and to threshold sign the model that has been produced in the previous phase. The certification conducted during the certification phase thus proves that the whole protocol was respected.
When the certification has passed, the threshold signature of the final model is sent to the aggregator that requested it using a $\finalizemsg$ message. 

When $a_i$ receives a $\finalizemsg$ message, it saves the signature that it contains. Once it received $n_a-t_a$ such signatures, it can aggregate them. This aggregated signature proves that at least one correct aggregator conducted the certification process, and that the model signed was computed following the whole protocol. This model can thus be used by clients in the following round $\tau+1$ to produce a new update.

\begin{algorithm}
\small
\Function{$\PrepareAggregation()$}{
    \If{$\wasteddetection() = \ttrue$ \label{line:agg:wasted:detection:prep:agg}}{\broadcast $\wastedmsg\langle\tau\rangle$ to processes in $\mathcal{A}$;\\
    \return;}
    $\ParticipatingClients\gets \emptyset$\label{line:agg:prep:agg:gather:deb}\;
    \For{$(c_k,\star,\star) \in \pinglist$ }{ \lIf{$ck \in \assign(\tau)[i]$}{$\ParticipatingClients\gets \ParticipatingClients \cup c_k$;\label{line:agg:prep:agg:gather:end}}}
    $\mathcal{S}_i^{\tau} \gets$ $\select( \Lambda_i^{\tau-1}[i],\ParticipatingClients)$;\label{line:agg:prep:agg:inclusion}\\
    \For{$c_k \in \ParticipatingClients$}{
        $\Lambda_{\texttt{tmp}} \gets \Lambda_i^{\tau-1}[i][k] + 1$;
    }
     \If{$\blaming(\Lambda_{\texttt{tmp}})=\ttrue$\label{line:agg:auto:blaming:detection}}{\broadcast $\autoblamingmsg\langle\tau\rangle$ to processes in $\mathcal{A}$;\\
     \return;}
    \Else{
    \WaitUntil{$\updatemsg$ messages are received from each client in $\mathcal{S}_i^\tau$}{
    \For{$c_j \in \mathcal{S}_i^{\tau}$}{
        $\VSSProof_i^{\tau} \gets \VSSProof^{\tau}_i \oplus \texttt{ProofStore}_i^{\tau}[j]$;\label{line:agg:prep:agg:full}
    }
    \For{$a_k \in \mathcal{A}$\label{line:agg:prep:agg:deb:send}}{
        $\texttt{TmpShares} \gets \emptyset$;\\
        \For{$c_\ell \in \mathcal{S}_i^{\tau}$}{
            $\texttt{TmpShares} \gets \texttt{TmpShares} \cup \EncSecretShares^{\tau}_i[\ell][k]$; 
        }
        $\send$ $\aggregatemsg\langle \tau, \mathcal{S}_i^{\tau}, \texttt{TmpShares}\rangle$ to $a_k$;\label{line:agg:prep:agg:end:send}
        }
        }
        }
}
\caption{Prepare aggregation function (code for $a_i$)} 
\label{alg:prepare:agg:full}
\end{algorithm}

\begin{algorithm*}
\small
\nonl\Init{$A \gets $ a public matrix of size $n\times m$ used to hide the update $w_i$;\\
$C \gets$ Clipping parameter.}
\Operation{$\mathsf{Train}()$}{
    \broadcast $\trainmsg\langle 0, W^0, \emptyset\rangle$ to clients;\label{line:agg:initial:train}
}
\WhenReceivedFromClient{$\pingmsg\langle\tau, \sigma_{\tau}\rangle$}{
    $\pinglist_i^{\tau} \gets \pinglist_i^{\tau} \cup (c_j, \tau, \sigma_{\tau})$;\label{line:agg:ping:store:1}
}
\WhenReceivedFromClient{$\updatemsg\langle \tau, h_j^{\tau}, \sigma_{h_j^{\tau}},\EncSecretShares^{\tau}_j, \sigma^{\tau}_{\texttt{PING}}, \VSSProof_j^{\tau}, \VSSPartialProof_j^{\tau} \rangle$}{
        $\MaskedUpdates^{\tau}_i[j] \gets h_j^{\tau}$;\label{line:agg:update:store:begin}\\
        $\SigMaskedUpdates^{\tau}_i[j]\gets \sigma_{h_j^{\tau}}$;\\
        $\EncSecretShares^{\tau}_i[j] \gets \EncSecretShares^{\tau}_j$;\\
        $\texttt{PartialProofStore}_i^{\tau}[j] \gets \VSSPartialProof$;\\
        $\texttt{ProofStore}_i^{\tau}[j] \gets \VSSProof$;\\
        $\pinglist_i^{\tau} \gets \pinglist_i^{\tau} \cup (c_j, \sigma^{\tau}_{\texttt{PING}})$;\label{line:agg:update:store:end}\\
        \If{$|\pinglist_i^{\tau}| \ge n_c-t_c$ \cAnd $\clientunificationmsg$ or $\aggregatemsg$ have not been sent \label{line:agg:update:ping:check}}{
            \broadcast $\clientunificationmsg\langle \tau, \pinglist_i^{\tau} \rangle$\label{line:agg:update:unification:phase}
        }
        \If{$n_a -t_a$ valid \clientunificationmsg messages have been received \cAnd at least $\rho$ \trainmsg have been received this round}{
            $\PrepareAggregation()$;\label{line:agg:prepare:agg:1}
        }
}
\WhenReceivedFromAggregator{$\clientunificationmsg\langle\tau, \pinglist_j^{\tau} \rangle$}{
    \If{$|\pinglist_j^\tau|\ge n_c-t_c$\label{line:agg:unification:begin}}{
        \For{$(c_k, \sigma^{\tau}_{\texttt{PING}})\in \pinglist_j^{\tau}$}{
            \lIf{$\Verify(\sigma^{\tau}_{\texttt{PING}}, \tau, \PKSig_{c_k})= \ffalse$}{\return}
        }
        $\pinglist_i^{\tau} \gets \pinglist_i^{\tau} \cup \pinglist_j^{\tau}$;\label{line:agg:unification:end}\\
    }
    \If{$n_a -t_a$ valid \clientunificationmsg messages have been received \cAnd at least $\rho$ \trainmsg messages have been received this round}{
            $\PrepareAggregation()$;\label{line:agg:prepare:agg:2}
    }
}

\WhenReceivedFromAggregator{$\aggregatemsg\langle {\tau}, \mathcal{S}_j^{\tau}, \texttt{TmpShares}\rangle$}{
    \WaitUntil{$\aggregatemsg\langle {\tau}, \star, \star\rangle$ or $\wastedmsg\langle\tau\rangle$ or $\autoblamingmsg\langle\tau\rangle$ is received from $a_j$ or $\tau=0$ \label{line:agg:sumshare:wait}}{
    \If{$|\mathcal{S}_j^{\tau}| = \rho$ \cAnd $\mathcal{S}_j^{\tau} \subseteq \assign^{\tau}[j]$\label{line:agg:sumshares:verify:inclusion:deb}}{
        $\lambda_{\texttt{tmp}} = \Lambda_i[j]$;\\
        \For{$c_k \in \mathcal{S}_j^{\tau}$}{
                $\Lambda_{\texttt{tmp}} \gets \Lambda_i^{\tau-1}[j][k] + 1$;
            }
        \lIf{$\blaming(\Lambda_{\texttt{tmp}}=\ttrue)$\label{line:agg:sumshares:verify:inclusion:end}}{\return}
        $\Lambda_i^{\tau-1}[j]\gets \Lambda_{\texttt{tmp}}$;\label{line:agg:sumshare:lambda:update}\\
        $\texttt{sumShares} \gets \{0\}^{N_s}$;\\
        $\texttt{sumProofs} \gets \{0\}^{N_s}$;\\
        \For{$\texttt{encShare} \in \texttt{TmpShares}$}{
            $(\tau, c_k, \texttt{share},\VSSProof, \texttt{sig}) \gets\Decrypt(\texttt{encShare}, \SKEnc_{(a_i, c_k)})$;\label{line:agg:sumshare:decyrpt}\\
            \lIf{not $\Verify((c_k, \tau, \texttt{share},\VSSProof), \PKSig_{(c_i})= \ttrue$}{\return}
            $\mathcal{S}_{\texttt{tmp}} \gets \mathcal{S}_{\texttt{tmp}} \cup c_k$;\\
            $\texttt{sumProofs} \gets \texttt{SumProofs} \oplus \VSSProof$;\label{line:sumshare:sum:vss:proof}\\
            $\texttt{sumShares} \gets \texttt{sumShares} \oplus \texttt{share}$\label{line:sumshare:sum:shares};
        }
        $\aggcnt_i^{\tau} \gets \aggcnt^{\tau}_i +1$;\\
        \If{$\mathcal{S}_{\texttt{tmp}} = \mathcal{S}_j^{\tau}$}{
            $\sigma_{\texttt{proof}} \gets \SignThre(\texttt{SumProofs}, \SKThreSig_i)$\label{line:sumshare:sign:vss:proof}\;
            $\send$ $\disseminatemsg\langle\tau, \texttt{sumShares}, \sigma_{\texttt{proof}}\rangle$ to $a_j$;\label{line:sumshare:send:disseminate}
        }
        \lElse{
        \return
        }
    }
    
}
}
\WhenReceivedFromAggregator{$\wastedmsg\langle\tau\rangle$ \cOr $\autoblamingmsg\langle\tau\rangle$}{
    \If{no \wastedmsg has been received from $a_j$}{
    $\wastedcnt_i^{\tau} \gets \wastedcnt_i^{\tau} +1$;
    }
}
\rememberlines{asyncAlg}
\caption{Privacy preserving asynchronous federated learning algorithm with aggregator fault resilience (for aggregator $a_i$). (Part I)}
\label{alg:async:aggregator:1}
\end{algorithm*}

\begin{algorithm*}
\small
\resumenumbering{asyncAlg}
\WhenReceivedFromAggregator{$\disseminatemsg\langle\tau, \texttt{sumShares}, \sigma_{\texttt{proof}}\rangle$}{
    \If{$\commit(\texttt{sumShares})=\commit(\sum_{k \in \mathcal{S}_i^\tau}\texttt{PartialProofStore}_i^{\tau}[k][j]^\tau))$ \cAnd $\VerifyThre(\sigma_{\texttt{proof}}, \VSSProof_i^{\tau}, \PKThreSig_j)$ \label{line:agg:intra:cluster:verification:proofs}}{
        $\StoredSumShares^{\tau}_i \gets \StoredSumShares^{\tau}_i \cup \{\texttt{SumShares}\}$;\
        $\StoredSumProofs^{\tau}_i \gets \StoredSumProofs^{\tau}_i \cup \{\sigma_\texttt{proof}\}$
    } 
    \If{$|\StoredSumShares^{\tau}_i| \ge n_a-t_a$ \cAnd no $\gathermsg$ has been sent \label{line:agg:condition:intra:cluster:na:ta}}{
        $\widehat{s}_i^{\tau} \gets  (n_a, n_a-t_a)$-$\VSSReconstruct(\texttt{SumShareSet})$;\label{line:intra:cluster:reconstruct:s}\\
        $\sigma_{\widehat{s}_i^{\tau}} \gets \CombineThre(\StoredSumProofs_i^{\tau})$;\label{line:intra:cluster:sig:gather}\\
        \broadcast $\gathermsg\langle \tau, \MaskedUpdates_i^{\tau}, \SigMaskedUpdates_i^{\tau},\widehat{s}_i^{\tau}, \sigma_{\widehat{s}_i^{\tau}}\rangle$ to processes in $\mathcal{A}$;
    }
}

\WhenReceivedFromAggregator{$\gathermsg\langle \tau, \MaskedUpdates_j^{\tau},\SigMaskedUpdates_j^{\tau}, \widehat{s}_j^{\tau}, \sigma_{\widehat{s}_j^{\tau}}\rangle$}{
    \lIf{$\VerifyCombThre(\sigma_{\widehat{s}_j^{\tau}},\commit(\widehat{s}_j^{\tau}))= \ffalse$ \cOr $\Verify(\SigMaskedUpdates_j^{\tau}[\ell],(\tau, \MaskedUpdates_j^\tau[\ell]), \PKSig_{c_\ell})=\ffalse \forall \ell \in \mathcal{C}$ \label{line:agg:gather:sig:verify}}{\return}
    $\FinalSelection_i^{\tau} \gets \FinalSelection_i^{\tau} \cup \{(a_j,\widehat{s}_j^{\tau}, \sigma_{\widehat{s}_j^{\tau}}, \MaskedUpdates_j^\tau,\SigMaskedUpdates_j^\tau)\}$;\\
    \If{ $|\FinalSelection_i^{\tau}|\ge n_a-t_a - \wastedcnt_i^{\tau}$ and no $\certifymsg$ has been sent}{
        $\widehat{G}_i^{\tau} \gets \frac{1}{\rho|\FinalSelection_i^{\tau}|} \sum_{(\star,\hat{s}_\ell^{\tau}, \star, \MaskedUpdates, \star) \in \FinalSelection_i^{\tau}} (\sum_{ h \in \MaskedUpdates} h)- A\cdot \hat{s}_\ell^{\tau}$;\\
        $\widehat{W}^{\tau+1}_i = \widehat{W}^{\tau}_i - \gamma^{\tau} \widehat{G}_{a_i}^{\tau}$;\label{line:agg:gathermsg:compute:final:update}\\
        \broadcast $\certifymsg\langle \tau, \FinalSelection_i^{\tau}, \widehat{W}^{\tau}_i \rangle$
    }
}
\WhenReceivedFromAggregator{$\certifymsg\langle \tau, \FinalSelection_j^{\tau}, \widehat{W}^{\tau}_j,\sigma^{\tau}_{\widehat{W}^{\tau}_j}\rangle$}{
    \WaitUntil{$\certifymsg\langle \tau, \star, \star, \star\rangle$ is received from $a_j$ or $\tau = 0$}{
    \For{$(a_\ell,\hat{g}_\ell^{\tau}, \sigma_{\widehat{g}_\ell^{\tau}}, \MaskedUpdates_\ell^\tau, \SigMaskedUpdates_\ell^\tau) \in \FinalSelection_j^{\tau}$}{
            \lIf{$\VerifyCombThre(\sigma_{\widehat{g}_\ell^{\tau}}, \commit(\widehat{g}_\ell^{\tau}))= \ffalse$ \cOr $\VerifyCombThre(\sigma^{\tau}_{\widehat{W}^{\tau}_j}\rangle, \widehat{W}^{\tau}_j)= \ffalse$ \cOr $\Verify(\SigMaskedUpdates_\ell^{\tau}[k],(\tau, \MaskedUpdates_\ell^\tau[k]), \PKSig_{c_k})=\ffalse \forall k \in \mathcal{C}$}{\return}
        }
        $\widehat{G}^{\tau} \gets \frac{1}{\rho|\FinalSelection_j^{\tau}|} \sum_{(\star,\hat{s}_\ell^{\tau}, \star, \MaskedUpdates, \star) \in \FinalSelection_i^{\tau}} (\sum_{ h \in \MaskedUpdates} h)- A\cdot \hat{s}_\ell^{\tau}$;\\
        $\widehat{W}_j^{\tau+1} = \widehat{W}^{\tau}_j - \gamma^{\tau} \widehat{G}^{\tau}$;
        $\sigma_W \gets \SignThre(\widehat{W}^{\tau+1}_j,\SKThreSig_i)$;\\
        $\send$ $\finalizemsg\langle \tau, \sigma_{W} \rangle$ to $a_j$
    }
    }

\WhenReceivedFromAggregator{$\finalizemsg\langle \tau, \sigma_W \rangle$}{
    \If{a $\certifymsg(\langle \tau, \star\rangle)$ has been sent to $a_j$ \cAnd $\Verify(\sigma_W, \widehat{W}_i^{\tau})=\ttrue$}{
        $\sigSet_i \gets \sigSet_i \cup \sigma_W$;\\
        \If{$|\sigSet_i|\ge n_a-t_a$ and no \trainmsg has been sent for round $\tau$}{
            $\sigma_{\widehat{W}^{\tau+1}_i}\gets \CombineThre(\sigSet_i)$;\\
            \broadcast $\trainmsg\langle\tau+1, \widehat{W}^{\tau+1}_i, \sigma^{\tau+1}_{\widehat{W}^{\tau+1}_i}\rangle$ to all clients.
        }
    }
}
\caption{Asynchronous federated learning algorithm with aggregator fault resilience in the fully asynchronous communication model (for aggregator $a_i$). (Part II)}
\label{alg:async:aggregator:2}
\end{algorithm*}

\subsection{Model sharing optimisation}

In the previous section, the algorithm presented assumes all the aggregators send their models to all the clients each round. This behaviour is intended to avoid a situation where a Byzantine aggregator does not send any model at round $\tau$, and some crashed clients are assigned to correct aggregators. In this case, correct aggregators may never receive $n_c-t_c$ $\pingmsg$ messages, thus impacting liveness of the protocol. By sending models to all clients, correct aggregators ensure that they will receive $\pingmsg$ from all correct clients. However, this behaviour is expensive in term of bit complexity, as models are heavy.

To reduce the bit complexity, we can change the model sharing mechanism from a ``push'' mechanism, where aggregators broadcast their models, to a ``pull'' mechanism, where aggregators advertise clients that a new model is available through a $\texttt{ModelAvailable}$ message, and clients then require this model to some specific aggregators.

More precisely, when an aggregator begins a new round, it still sends the new model to the clients in its cluster, but it only sends a $\texttt{ModelAvailable}\langle\tau\rangle$ message to clients outside of its cluster. When a client receives such message, and if it did not already begin training for round $\tau$, it requests $t_a+1$ aggregators of its choice to send their models. Therefore, the client is ensured to eventually receive a model for round $\tau$, while reducing bit complexity.

\section{Additional Experimental Details and Results}
\label{sec:additionnal:expe}

\begin{figure*}[h]
    \centering
    \includegraphics[width=\linewidth]{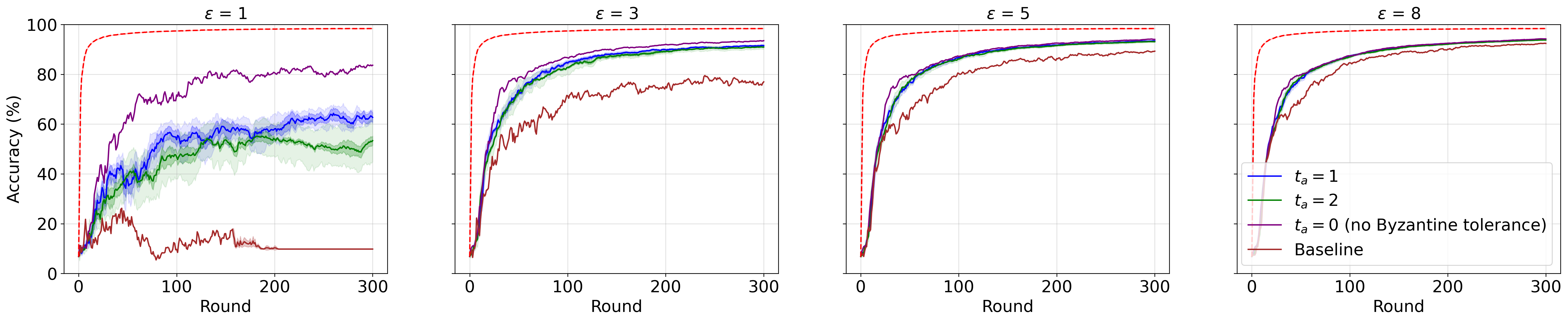}\\[6pt]
    \includegraphics[width=\linewidth]{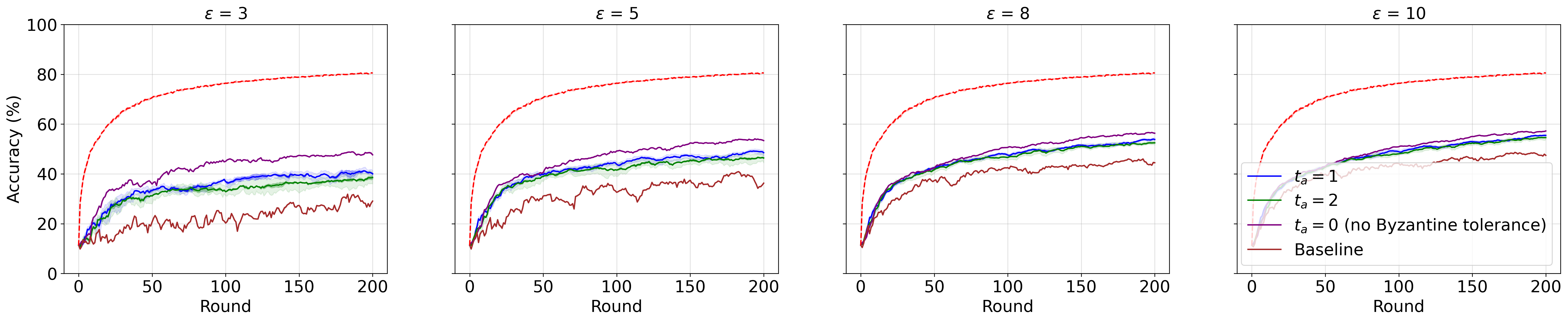}
    \caption{%
        Accuracy over rounds for $\epsilon \in \{1, 3, 5, 8\}$ on MNIST
        and $\epsilon \in \{3, 5, 8, 10\}$ on CIFAR-5m
        ($\rho=128$, $N=1500$). Each sub-panel corresponds to one value of
        $\epsilon$. Our protocol converges across the full range
        on MNIST; the baseline fails to make progress at $\epsilon \leq 3$.
        The no-DP reference converges near-instantly in all panels.
    }
    \label{fig:exp:accuracy:epsvari}
\end{figure*}

\begin{figure*}
    \centering
    \includegraphics[width=\linewidth]{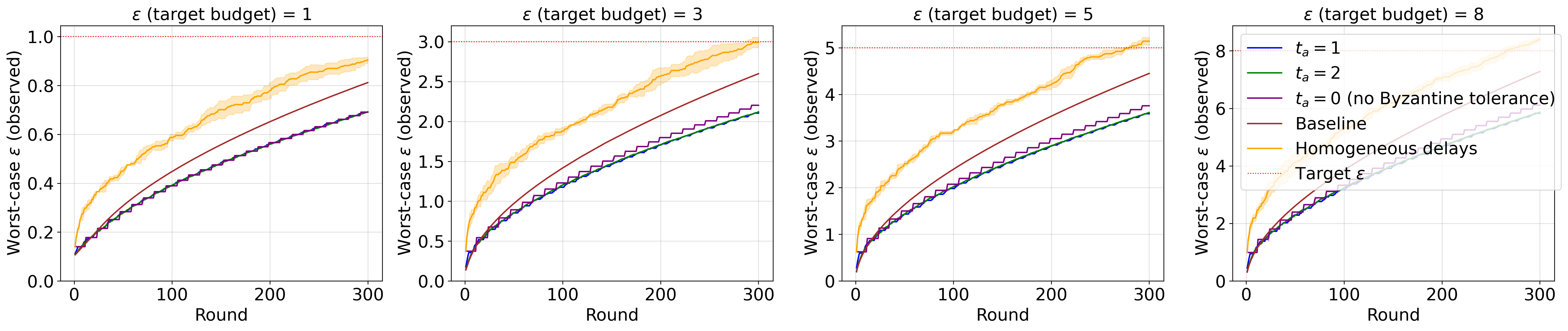}\\[6pt]
    \includegraphics[width=\linewidth]{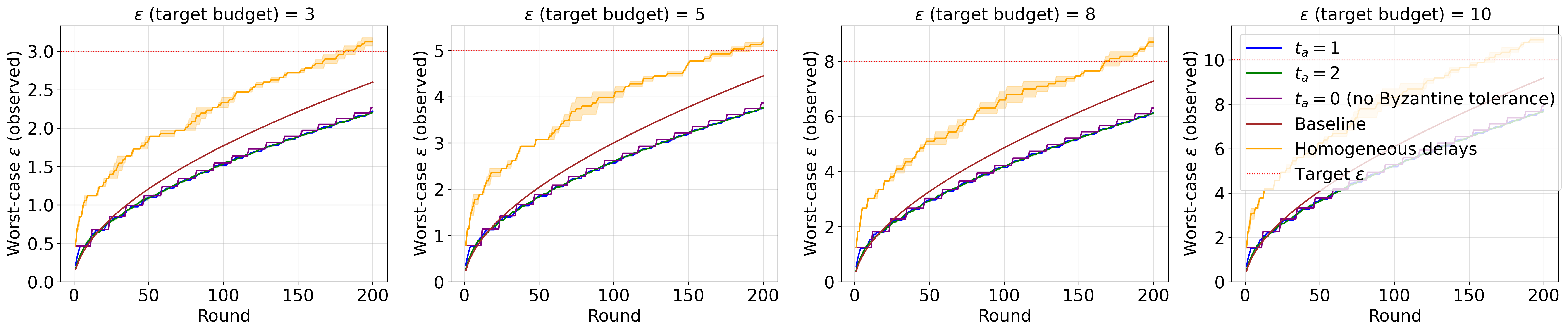}
    \caption{%
        Actual worst-case $\epsilon$ accumulated over rounds for the same
        experiment on MNIST (top) and CIFAR-5m (bottom).
        The dashed horizontal line marks the target $\epsilon$ for each panel.
        Both our protocol and the baseline remain within their respective
        budgets; the homogeneous-delays configuration exceeds the target at
        small $\epsilon$ (visible as curves crossing the dashed line).
    }
    \label{fig:exp:epsilon:epsvari}
\end{figure*}

\begin{figure*}
    \centering
    \includegraphics[width=\linewidth]{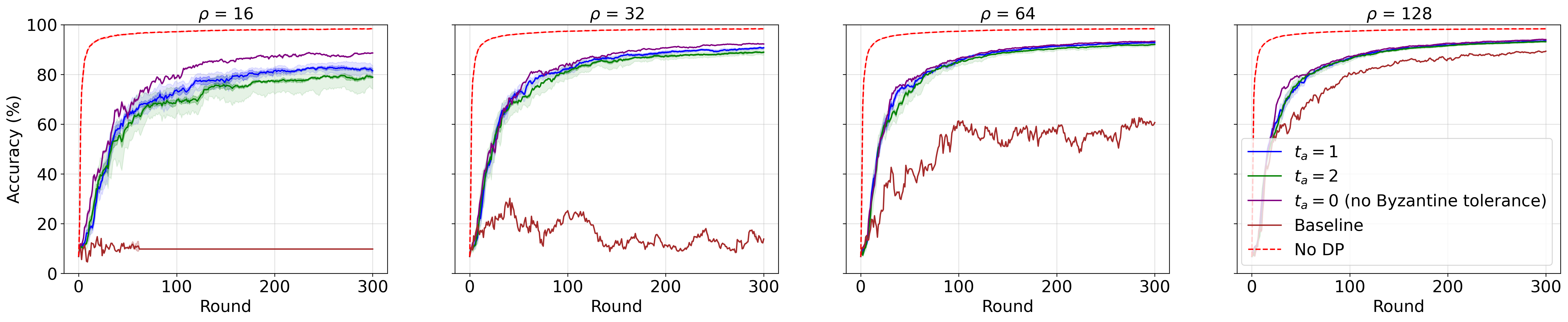}\\[6pt]
    \includegraphics[width=\linewidth]{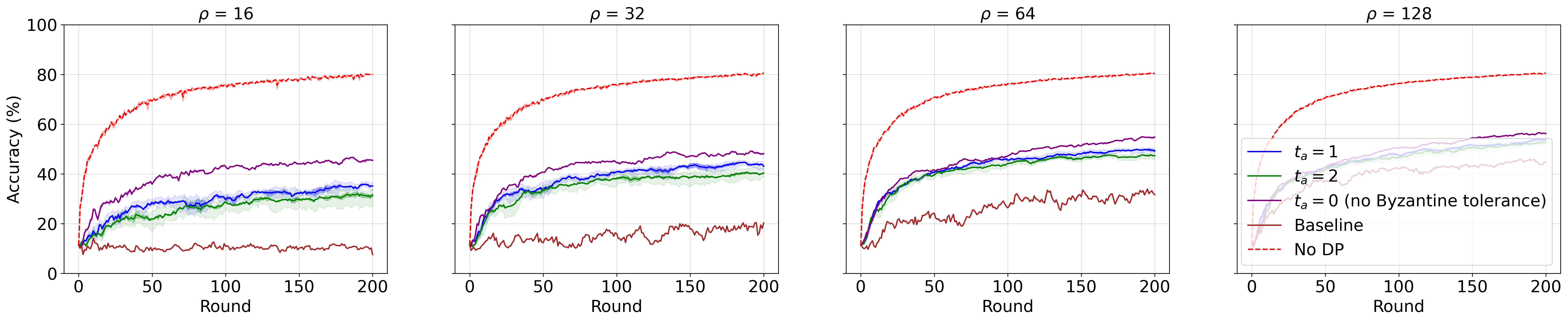}
    \caption{%
        Accuracy over rounds for $\rho \in \{16, 32, 64, 128\}$
        ($\epsilon=5$, $N=1500$) on MNIST (top) and CIFAR-5m (bottom).
    }
    \label{fig:exp:accuracy:rhovari}
\end{figure*}

\begin{figure*}
    \centering
    \includegraphics[width=\linewidth]{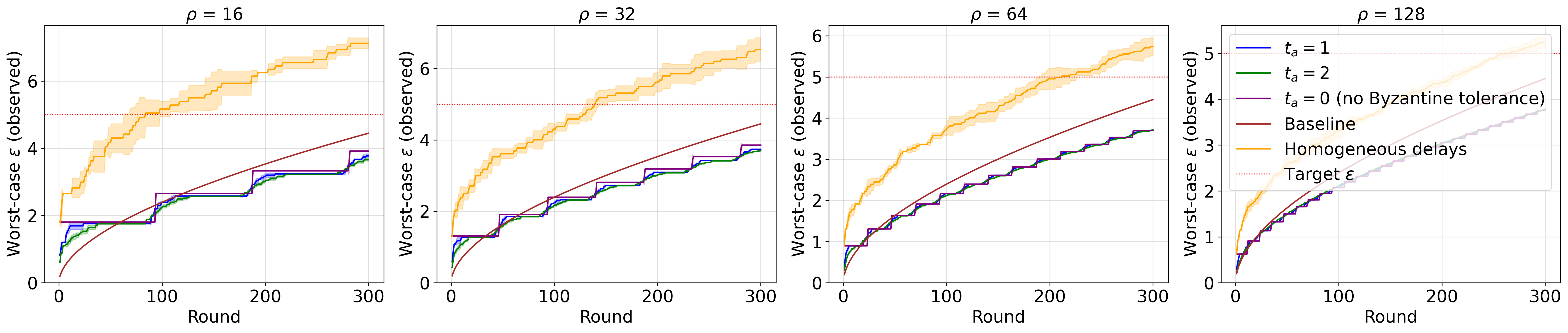}\\[6pt]
    \includegraphics[width=\linewidth]{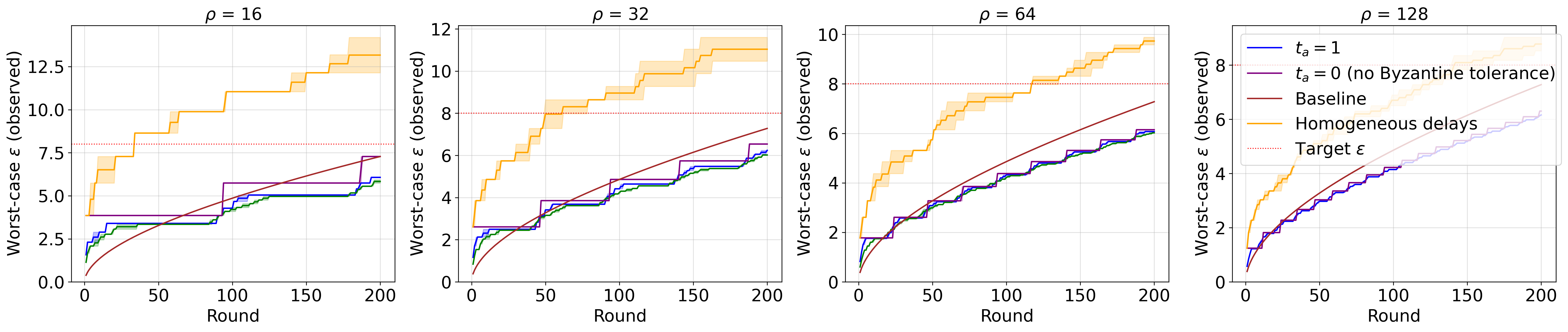}
    \caption{%
        Actual $\epsilon$ consumed over rounds for varying $\rho$
        on MNIST (top) and CIFAR-5m (bottom).
    }
    \label{fig:exp:epsilon:rhovari}
\end{figure*}

\begin{figure*}
    \centering
    \includegraphics[width=\linewidth]{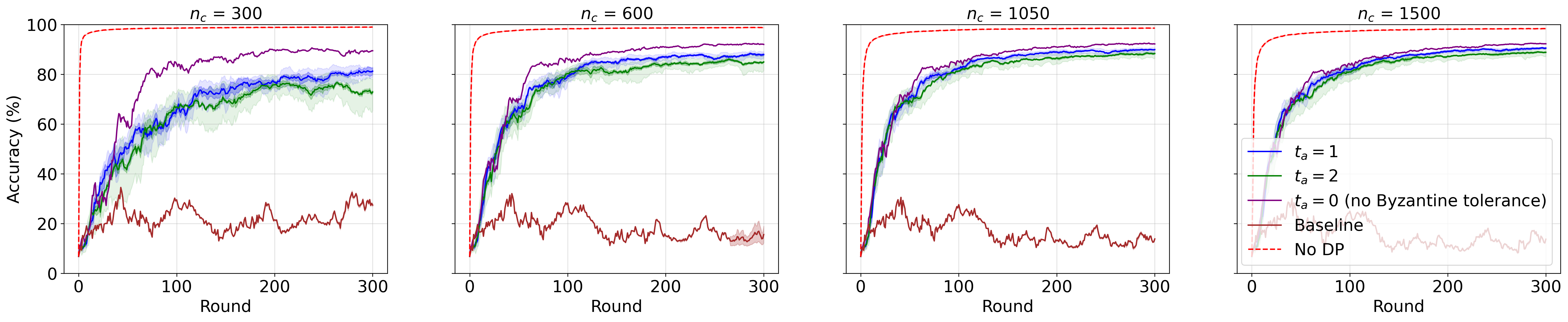}\\[6pt]
    \includegraphics[width=\linewidth]{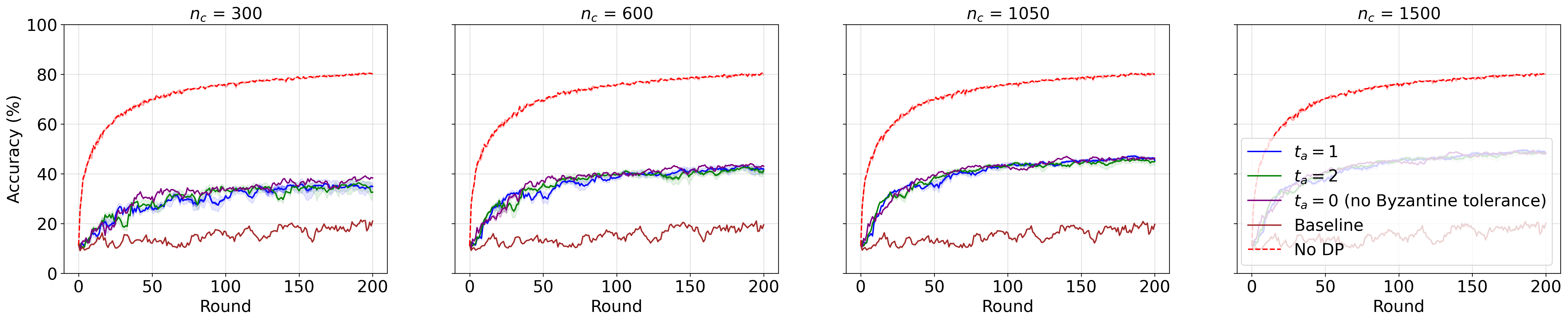}
    \caption{%
        Accuracy over rounds for $n_c \in \{300, 600, 1050, 1500\}$
        ($\epsilon=5$, $\rho=32$) on MNIST (top) and CIFAR-5m (bottom).
    }
    \label{fig:exp:accuracy:nvari}
\end{figure*}

\begin{figure*}[h]
    \centering
    \includegraphics[width=\linewidth]{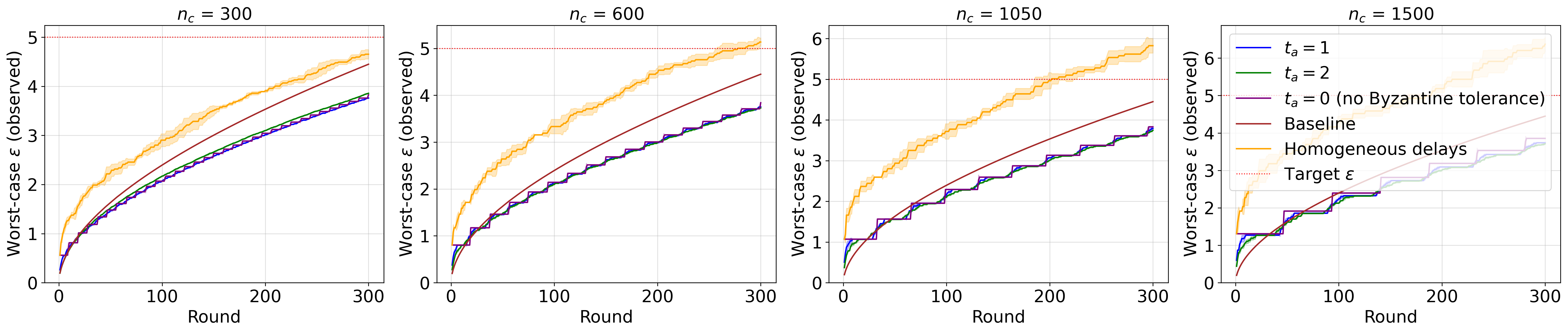}\\[6pt]
    \includegraphics[width=\linewidth]{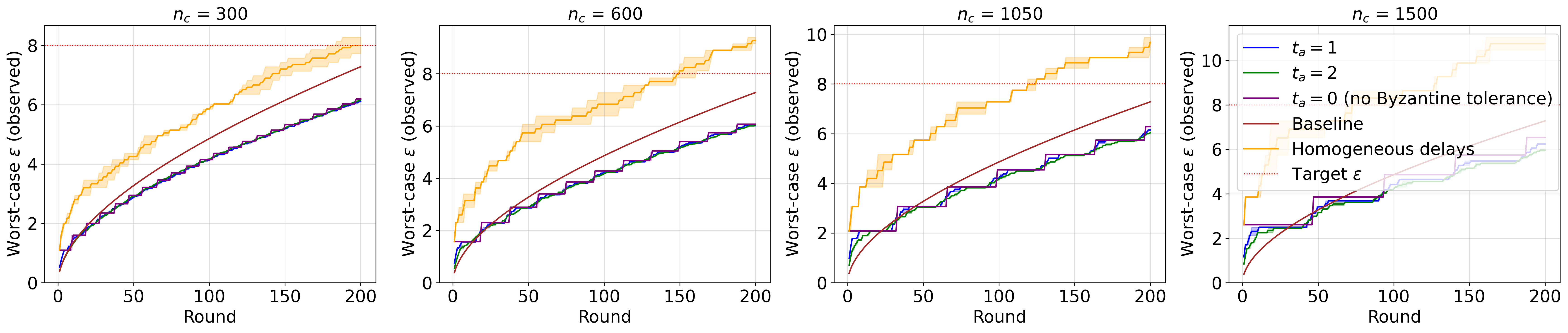}
    \caption{%
        Actual $\epsilon$ consumed over rounds for varying $n_c$
        on MNIST (top) and CIFAR-5m (bottom).
    }
    \label{fig:exp:epsilon:nvari}
\end{figure*}

\begin{figure*}
    \centering
    \includegraphics[width=\linewidth]{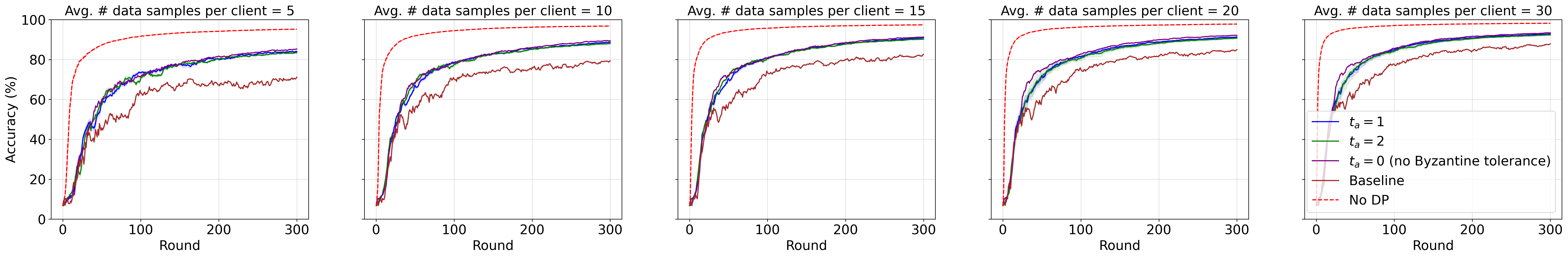}\\[6pt]
    \includegraphics[width=\linewidth]{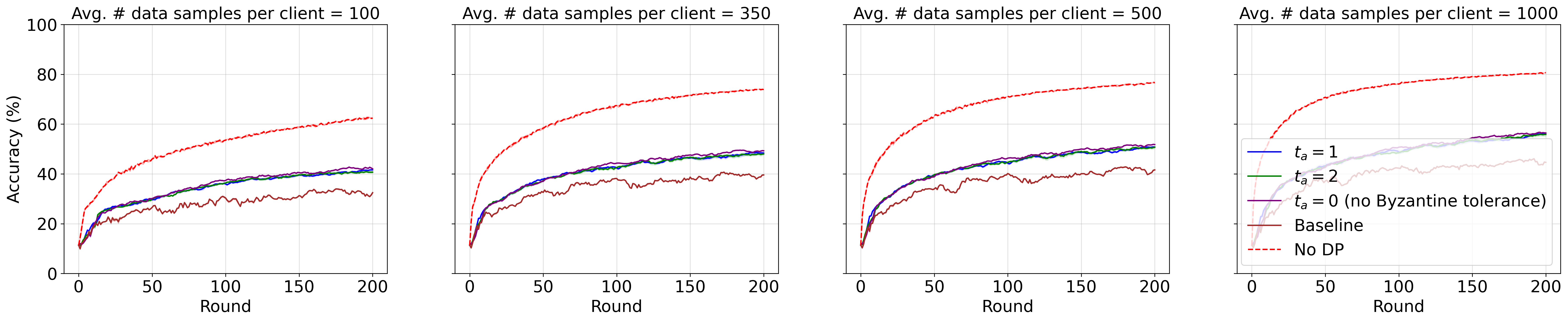}
    \caption{
        Top: accuracy over rounds for $d_q \in \{5, 10, 15, 20, 30\}$ samples per
        client on CIFAR-5m ($\epsilon=8$, geo-distributed split, $\rho=128$).
        Bottom: accuracy over rounds for $d_q \in \{100, 500, 1000\}$ samples per
        client on CIFAR-5m ($\epsilon=8$, geo-distributed split, $\rho=128$).
    }
    \label{fig:exp:datavari:cifar}
\end{figure*}

\begin{figure*}
    \centering
    \begin{subfigure}{0.33\textwidth}
        \includegraphics[width=\linewidth]{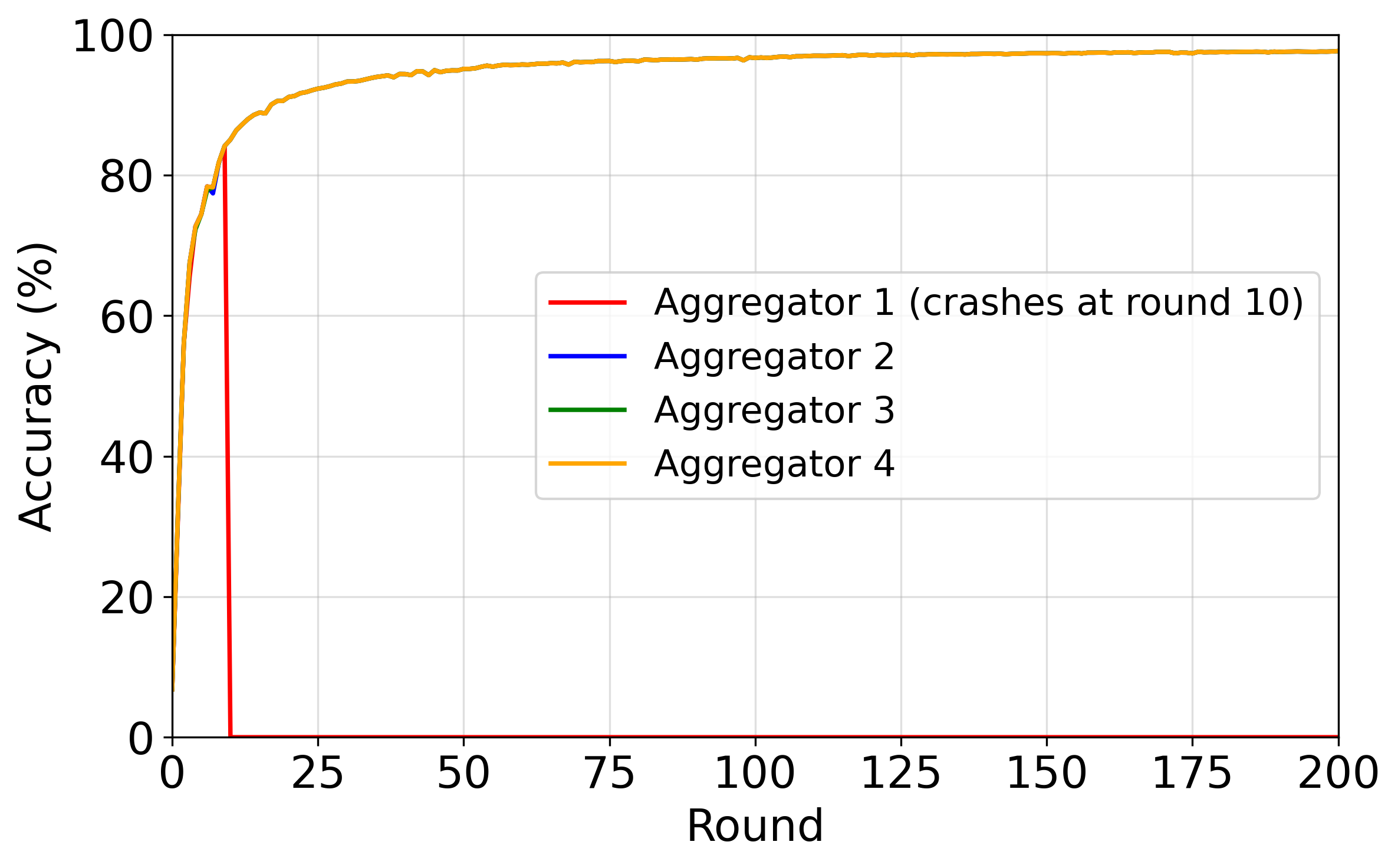}
    \end{subfigure}
    \begin{subfigure}{0.33\textwidth}
        \includegraphics[width=\linewidth]{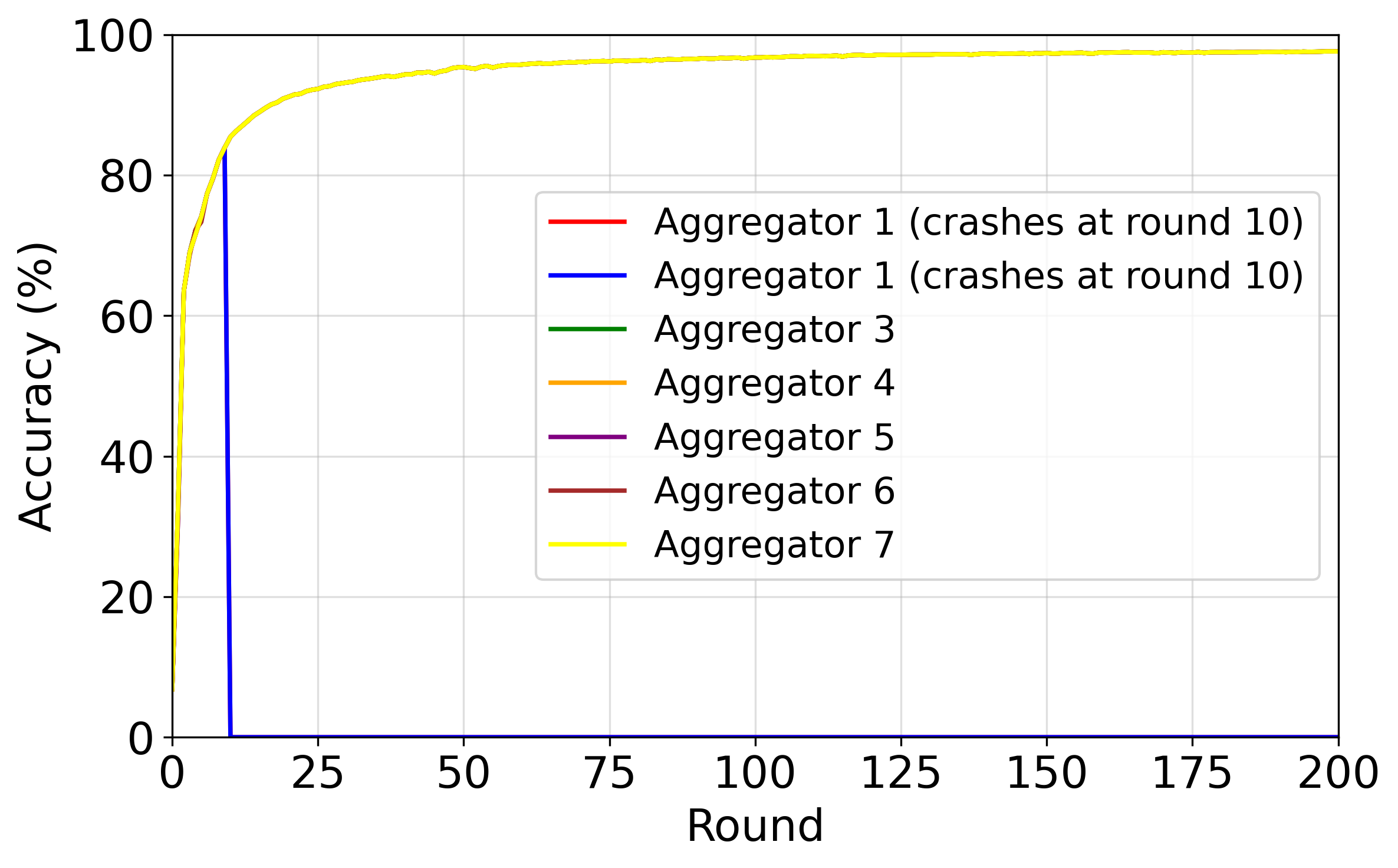}
    \end{subfigure}
    \begin{subfigure}{0.33\textwidth}
        \includegraphics[width=\linewidth]{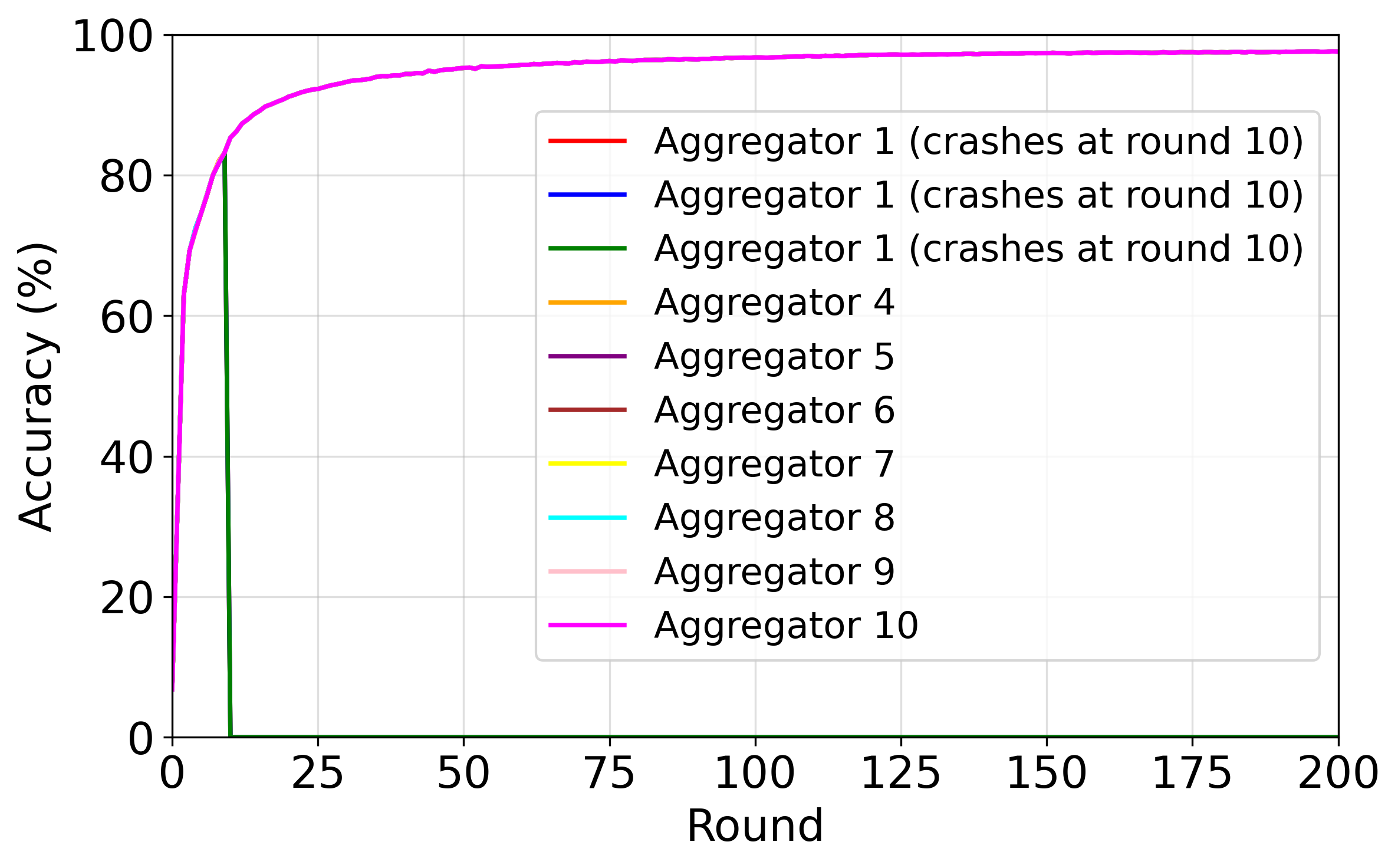}
    \end{subfigure}
    \caption{Training of the MNIST dataset with simulation of crash from $1$ aggregator (left), $2$ aggregators (center) and $3$ aggregators (right).}
    \label{fig:crash:test}
\end{figure*}

\begin{figure*}
    \centering
    \begin{subfigure}{\textwidth}
        \includegraphics[width=\linewidth]{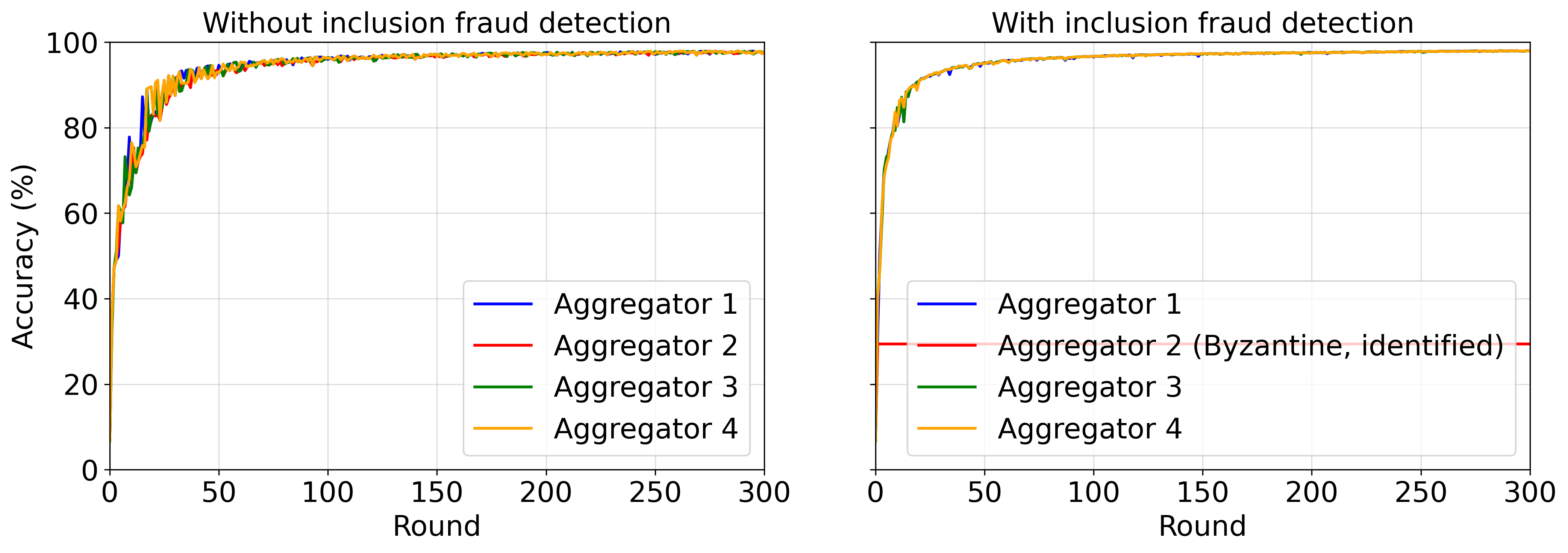}
    \end{subfigure}
    \begin{subfigure}{\textwidth}
        \includegraphics[width=\linewidth]{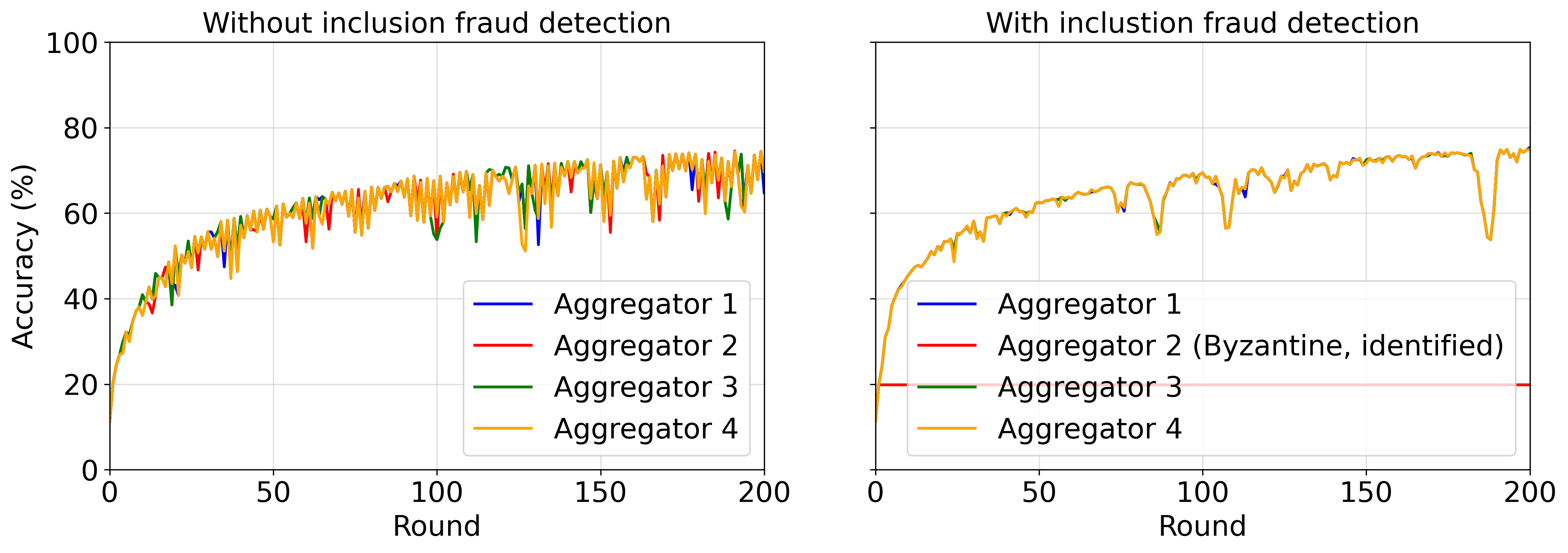}
    \end{subfigure}
    \caption{Impact of aggregator's Byzantine behaviour on the accuracy without the fraud detection mechanism (left) and with the fraud detection mechanism (right). Results on the MNIST dataset (top) and the CIFAR-5m dataset (bottom).}
    \label{fig:blaming:test}
\end{figure*}

\subsection{Model Architectures}

We use two convolutional neural networks.

\textbf{CNN-MNIST} ($\approx$27,000 parameters): two convolutional blocks ($3\times3$
kernels, 8 and 16 output channels respectively), each followed by $2\times2$
max-pooling, then two fully-connected layers (784$\to$32$\to$10). ReLU
activations throughout.

\textbf{CNN-CIFAR} ($\approx$550,000 parameters): three pairs of convolutional layers
($3\times3$, $3\to32\to32\to64\to64\to128\to128$ channels), each followed by
GroupNorm (8 groups) and $2\times2$ max-pooling after each pair, then two
fully-connected layers (2048$\to$128$\to$10). GroupNorm replaces BatchNorm
because clients may train with batch size~1, which makes BatchNorm statistics
degenerate; GroupNorm is unaffected by batch size.

\subsection{Full Convergence Trajectories and Privacy Budget Consumption}
\label{sec:exp:dp:additionnal}

The figures in this section complement the main body's maximum accuracy
summaries of \Cref{sec:exp:dp:accuracy} by showing, for each of the three experiments, the full
accuracy-over-rounds curves and the actual privacy budget consumed per round.
Each figure pairs results for MNIST (top) with result for CIFAR-5m (bottom).
In all accuracy plots, solid lines are means over repetitions, inner bands
($\pm$std) capture run-to-run variability, and outer bands (for $t_a>0$)
span the fastest to the slowest aggregator within a single run.

\subsubsection{Varying the Privacy Budget $\epsilon$}

\Cref{fig:exp:accuracy:epsvari} shows how accuracy trajectories evolve across
$\epsilon$ values. For large $\epsilon$ (low noise), all configurations
converge quickly to similar final accuracy. As $\epsilon$ decreases, the
performance gap widens sharply. 
We see that increasing the number of aggregators increases the fault tolerance, but decreases the performance and results in a lower accuracy, due to the additional noise required for Byzantine fault tolerance.
On CIFAR-5m, the range
shown is $\epsilon \in \{3,5,8,10\}$, where the model is already difficult
enough that $\epsilon=1$ yields no useful training for any configuration.
The bands for $t_a>0$ configurations widen at low $\epsilon$: the stochastic
nature of which aggregator first accumulates enough well-aligned gradient
updates causes significant spread between the fastest and slowest aggregator
within a single run.

\paragraph{Budget consumption}\Cref{fig:exp:epsilon:epsvari} shows the realized $\epsilon$ budget consumption per
round. A few key observations require careful interpretation.

Both our protocol and the baseline ($t_a=0$, no inclusion) remain
within their respective target budgets throughout the 300-round window for mnist (respectively 200 rounds for cifar) —
by design, since each is calibrated independently for its own worst case.
Our protocol displays lower realized $\epsilon$ than the baseline
(e.g., $\approx3.05$ vs.\ $\approx3.54$ after 200 rounds at $\epsilon=5$),
but this gap is partly an artifact of the experimental setup:
client crashes are not simulated. Therefore, clients are included less aggressively
than in the worst possible case.
The baseline, by contrast, has its worst case realized naturally by the
biased selection of fast clients.
The correct reading is therefore not that our protocol ``wastes'' less budget,
but that \emph{both protocols meet the stated $(\epsilon,\delta)$-DP guarantee},
and neither consumes its full budget in practice.

The homogeneous-delays configuration uses the same noise calibration as our
protocol, since it was designed to share the same DP accounting.
However, \emph{without the inclusion mechanism}, nothing prevents certain
clients from being selected disproportionately, even when all
delays are drawn from the same distribution.
This causes the realized $\epsilon$ to grow faster than calibrated, and for
small $\rho$ the budget is violated well before 300 rounds for MNIST (respectively 200 rounds for cifar)
(e.g., $\epsilon\approx6.25$ after 200 rounds at $\rho=16$, hitting the
$\epsilon=5$ target at round~77).
This is not an indictment of homogeneous delays as a scenario — it shows
that \emph{the inclusion mechanism is what makes the DP guarantee hold}:
it is the mechanism that enforces the uniform selection probability
$\approx\rho/N$ that the noise calibration assumes.
In a normal settings, the calibration of noise for the homogenous delays should be the same as for the baseline without selection, which would occurs poor performance in accuracy.

\subsubsection{Varying the number $\rho$ of included clients}

\Cref{fig:exp:accuracy:rhovari} illustrates the effect of the parameter $\rho$ on convergence. Larger $\rho$ consistently accelerates
convergence for all configurations, since more gradient updates are
incorporated per round. For our protocol, we see that the number of rounds to convergence to $80\%$ accuracy with the MNIST dataset increases
at every tested value: 102 rounds for $\rho=16$ and 57 rounds for $\rho=128$. The baseline only converges to 80\% accuracy with $\rho=128$ and fails
at all smaller values, confirming that it cannot recover from the noise for small $\rho$ parameters. The $t_a=2$ configuration shows the strongest
sensitivity to $\rho$: at $\rho=16$, convergence to 80\% accuracy requires nearly 300 rounds,
while at $\rho=128$ it reaches 80\% accuracy in 66 rounds. 

\paragraph{Budget consumption} \Cref{fig:exp:epsilon:rhovari} shows that $\rho$ has a notable effect on
budget consumption, particularly for the homogeneous-delays configuration.
At $\rho=16$, homogeneous delays hits $\epsilon=5$ at round~77
($\epsilon\approx6.25$ after 200 rounds), a severe budget violation.
The violation shrinks as $\rho$ grows: at $\rho=64$ it hits the target at
round~198, and at $\rho=128$ the budget is just barely respected
($\epsilon\approx4.52$ after 200 rounds). The interpretation is that with
small $\rho$, the random fluctuations in client selection cause the effective
sampling rate to be dominated by a few over-selected clients; as $\rho$
grows, the selection becomes empirically more uniform, mitigating the issue.
Our protocol and the baseline both remain within budget for all tested
$\rho$, confirming the earlier reading: both are calibrated for their own
worst case and the guarantees hold.

\subsubsection{Varying the number of clients $n_c$}

\Cref{fig:exp:accuracy:nvari} confirms the robustness of our protocol to
population size. For $t_a=0$, convergence speed to 80\% accuracy remains stable  on MNIST at
62--69 rounds across $n_c\in\{300,600,1050,1500\}$, showing that the protocol's
convergence is governed by the fraction of clients selected per round
($\rho/n_c=32/n_c$), not the absolute number of clients. The baseline fails to
converge at any $n_c$ value, highlighting that the selection bias cannot be
resolved by adding more clients alone. The $t_a=2$ variant is the only
configuration sensitive to $n_c$: with $n_c=300$, the model does not converge to 80\% accuracy within 200-round. 

\paragraph{Budget consumption }\Cref{fig:exp:epsilon:nvari} shows budget consumption as $n_c$ varies.
As in the previous figures, both our protocol and the baseline remain within
their respective targeted budgets for all $n_c$ values.
The differences in term of budget consumption across different $n_c$ values are small. At round 100, our protocol ranges from
$\approx2.07$ for $n_c=300$  and to $\approx2.40$ for $n_c=1500$; the baseline is
stable at $\approx2.41$ in any configuration. We conclude that $n_c$ has little effect on per-round privacy budget
once $\rho$ is fixed, since the noise calibration is dominated by the
worst-case inclusion behavior rather than by the amplification from a
specific amount of clients.
The homogeneous-delays curve shows analogous
budget pressure as in the experiments on the parameter $\rho$ for the same structural
reason: without the inclusion mechanism, uniform selection is not guaranteed.

\subsection{Influence of the amount of Data held by Clients}
\label{sec:exp:datavari}


This experiment varies the number of training samples per client $d_q$
on MNIST and CIFAR-5m ($\epsilon=8$, $N=1500$, $\rho=128$, geo-distributed split).

\Cref{fig:exp:datavari:cifar} shows that increasing per-client data volume ($d_q$)
consistently accelerates convergence for all configurations.
Crucially, our protocol maintains a clear advantage over the
baseline at every tested value of $d_q$: even at the lowest data volumes, the inclusion mechanism continues to reduce the effective noise
through DP amplification by subsampling, independently of how much data
each client possesses.
The gap between our protocol and the baseline is driven by the noise
multiplier ratio, not by the gradient quality, so varying $d_q$ does not
alter the relative ordering of the curves — it only shifts all curves
together toward faster or slower convergence.
This confirms that the advantage demonstrated in the main paper is not
specific to high-data regimes and holds robustly across data-scarce settings.

\subsection{Impact of aggregators crash}

One of the Byzantine behaviour that aggregators can exhibit is a simulation of crash, i.e., they cease to participate in the protocol. We tested this setup by letting $t_a$ aggregators simulate a crash after round $10$ with the MNIST dataset in a similar setting as the geodistributed setting presented in \Cref{sec:exp:analyzes}, with $n_c = 1500$, $rho=128$, $t_c = 374$, and no differential privacy. 

The results of this experimentation are shown in \Cref{fig:crash:test}. When an aggregator crashes, it cease to send messages to clients and aggregators, and we set its accuracy to $0$. As we can see, with $1$, $2$, or $3$ crash, the crash of an aggregator does not impact the final accuracy of the correct aggregators.

\subsection{Inclusion fraud detection impact on training}

\begin{figure}
    \includegraphics[width=\linewidth]{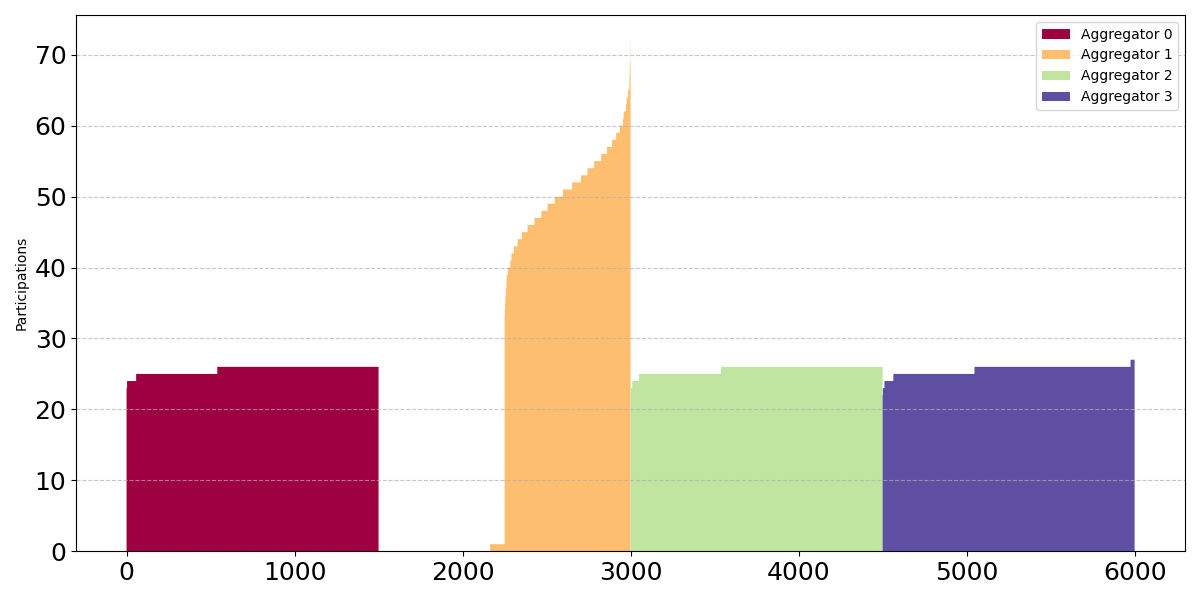}
    \caption{Number of times each aggregator included each client, sorted from least to most included client, with aggregator 1 exhibiting a Byzantine behaviour.}
    \label{fig:blaming:inclusion:summary}
\end{figure}

In this section, we want to experimentally analyze the inclusion fraud detection mechanism presented in \Cref{sec:edge:cases}.
To do so, we simulated a Byzantine behaviour, where an aggregator voluntarily bias its included clients to include in priority fastest clients. To increase the impact of this bias, we place this experimentation in the geodistributed setup presented in \Cref{sec:exp:analyzes}. Therefore, the Byzantine aggregator effectively tries to bias the model of all aggregators by only including clients with biased data (in this case, it only included clients with samples from the $5$ first classes). The difference between the clients included by correct and Byzantine aggregators can be seen in \Cref{fig:blaming:inclusion:summary}. This figure shows, for each aggregator, the number of time it included each client, sorted from the least included client to the most included client. It is clear that correct aggregators have an uniform inclusion policy whereas the Byzantine aggregator (aggregator 2) has a clearly biased inclusion policy.  

In \Cref{fig:blaming:test}, we show the difference between an execution with this type of Byzantine behaviour with and without the inclusion fraud detection mechanism both on the MNIST and CIFAR-5m dataset. The main impact of such Byzantine behaviour on the accuracy is visible as, without the fraud detection mechanism, accuracy lines are way more jagged than its fraud detection counterpart. This is due to the fact that the Byzantine aggregator draws everyone down. However, we cannot see a clear impact of the inclusion fraud detection mechanism on the maximum accuracy of the training. This is probably due to the fact that, in the scenario without fraud detection, correct aggregators average the behaviour of the Byzantine behaviour, whose impact is therefore reduced.
Overall, we can see that the inclusion fraud detection mechanism is a necessary mechanism as it makes it possible to have a smoother training.



\end{document}